\documentclass[]{aa}

\usepackage{graphicx,amssymb,amsmath,times,xcolor}
\usepackage{longtable}
\usepackage{bm} 
\usepackage{url}

\usepackage[varg]{txfonts}
\usepackage[T1]{fontenc}
\usepackage{natbib}
\bibpunct{(}{)}{;}{a}{}{,} 
\usepackage[normalem]{ulem}
\usepackage[colorlinks=true,     linkcolor=blue, citecolor=blue, filecolor=blue, urlcolor=blue]{hyperref}

\begin{document}

\def\simlt{\lower.5ex\hbox{\ltsima}}
\def\simgt{\lower.5ex\hbox{\gtsima}}

\def\farcm{\hbox{$\mkern-4mu^\prime$}}
\def\parcm{${}^{\prime}$\llap{.}}
\def\farcs{\hbox{$^{\prime\prime}$}~}
\def\parcs{${}^{\prime\prime}$\llap{.}}
\def\km{{\rm\,km}}
\def\kms{{\rm\,km\,s^{-1}}}
\def\mas{{\rm\,mas}}
\def\masyr{{\rm\,mas/yr}}
\def\kpc{{\rm\,kpc}}
\def\mpc{{\rm\,Mpc}}
\def\msun{{\rm\,M_\odot}}
\def\lsun{{\rm\,L_\odot}}
\def\rsun{{\rm\,R_\odot}}
\def\pc{{\rm\,pc}}
\def\cm{{\rm\,cm}}
\def\yr{{\rm\,yr}}
\def\au{{\rm\,AU}}
\def\g{{\rm\,g}}
\def\om{\Omega_0}
\def \ca {{\it ca.\/}}
\def \r {r$^{1/4}$ }
\def \magnitude {$^{\rm m}$}
\def\kr{${\cal K}_r$}
\def\kz{${\cal K}_z$}
\def\kzz{${\cal K}_z(z)$}
\def\mss{{\rm M}_\odot \rm pc^{-2}}
\def\msss{{\rm M}_\odot \rm pc^{-3}}
\newcommand{\fmmm}[1]{\mbox{$#1$}}
\newcommand{\scnd}{\mbox{\fmmm{''}\hskip-0.3em .}}
\newcommand{\scnp}{\mbox{\fmmm{''}}}
\newcommand{\mcnd}{\mbox{\fmmm{'}\hskip-0.3em .}}
\def\Aa{\; \buildrel \circ \over {\rm A}}
\def\AA{$\; \buildrel \circ \over {\rm A}$}
\def\yr{{\rm yr}}
\def\tmaxo{$T_{max}^{opt}$} 
\def\tmeano{$T_{mean}^{opt}$} 
\def\tminv{$T_{min}^{RV}$} 
\def\tmeanv{$T_{mean}^{RV}$} 
\def\tminvfe{$T_{min}^{RV(Fe)}$} 
\def\tmeanvfe{$T_{mean}^{RV(Fe)}$} 
\def\tminvhb{$T_{min}^{RV(H\beta)}$} 
\def\tmeanvhb{$T_{mean}^{RV(H\beta)}$} 

\def\gtsim{\;\lower.6ex\hbox{$\sim$}\kern-6.7pt\raise.4ex\hbox{$>$}\;}
\def\ltsim{\;\lower.6ex\hbox{$\sim$}\kern-6.9pt\raise.4ex\hbox{$<$}\;}
\def\hr{${}^{\hbox{\footnotesize h}}$}
\def\tm{${}^{\hbox{\footnotesize m}}$}
\def\ts{${}^{\hbox{\footnotesize s}}$}
\def\Ts{${}^{\hbox{\footnotesize s}}$\llap{.}}
\def\Deg{${}^\circ$\llap{.}}
\def\Min{${}^{\prime}$\llap{.}}
\def\Sec{${}^{\prime\prime}$\llap{.}}
\def\deg{${}^\circ$}
\def\min{${}^{\prime}$}
\def\sec{${}^{\prime\prime}$}
\def\hst{{\it HST\/}}
\def\gaia{{\it Gaia~}}
\def\bmv{\hbox{\it B--V\/}}
\def\bmr{\hbox{\it B--R\/}}
\def\bmi{\hbox{\it B--I\/}}
\def\vmi{\hbox{\it V--I\/}}
\def\umi{\hbox{\it U--I\/}}
\def\jmk{\hbox{\it J--K\/}}
\def\bmk{\hbox{\it B--K\/}}
\def\\gc#1{\hbox{NGC$\,$#1}}
\def\wcen{$\omega$ Cen~}
\def\trisv{$T_{ris}^{V}$}
\def\trisg{$T_{ris}^{g}$}
\def\trisr{$T_{ris}^{r}$}
\def\trisi{$T_{ris}^{i}$}
\def\trisz{$T_{ris}^{z}$}
\def\tris{$T_{ris}$}

\setlength{\tabcolsep}{3pt}

\newcommand{\rrl}{RR~Lyr\ae}
\newcommand{\GB}[1]{\textcolor{cyan!100}{#1$_{_{\mathbf{GB}}}$}}
\newcommand{\VFB}[1]{\textcolor{red!100}{#1$_{_{\mathbf{VFB}}}$}}

\title{On the use of field RR Lyrae as Galactic probes VII. light curve templates in the LSST photometric system\thanks{Corresponding author: vittorio.braga@inaf.it},\thanks{Tables 2, 4, 5, A1, A2, A3, A4, A5, A6 are only available in electronic form at the CDS via anonymous ftp to cdsarc.u-strasbg.fr (130.79.128.5) or via http://cdsweb.u-strasbg.fr/cgi-bin/qcat?J/A+A/},\thanks{The code needed to apply the light curve templates is publicly available at \url{https://github.com/vfbraga/RRL_lcvtemplate_griz_LSST}}}

\author{
V.~F.~Braga \inst{1,2}
\and M.~Monelli \inst{2,3,1}
\and M.~Dall'Ora \inst{4}
\and J.~P.~Mullen\inst{5}
\and R.~Molinaro\inst{4}
\and M.~Marconi \inst{4}
\and R.~Szab{\'o}\inst{6,7}
\and C.~Gallart\inst{2}
}
\institute{
INAF-Osservatorio Astronomico di Roma, via Frascati 33, 00040 Monte Porzio Catone, Italy
\and Instituto de Astrof\'isica de Canarias, Calle Via Lactea s/n, E38205 La Laguna, Tenerife, Spain
\and Departamento de Astrofísica, Universidad de La Laguna, Avenida Astrofísico Francisco Sánchez, s/n., 38200, San Cristóbal de La Laguna, Tenerife, Spain
\and INAF-Osservatorio Astronomico di Capodimonte, Salita Moiariello 16, 80131 Napoli, Italy
\and Department of Physics and Astronomy, Vanderbilt University, Nashville, TN 37240, USA
\and Konkoly Observatory, HUN-REN Research Centre for Astronomy and Earth Sciences, MTA Centre of Excellence H-1121 Konkoly Thege Mikl{\'o}s {\'u}t 15-17, Budapest, Hungary
\and ELTE E\"otv\"os Lor{\'a}nd University, Institute of Physics, H-1117, P{\'a}zm{\'a}ny P{\'e}ter s{\'e}t{\'a}ny 1/a, Budapest, Hungary}

\abstract
{The \textit{Vera C. Rubin} Observatory will start operations in 2025. During the first two years, too few visits per target per band will be available, meaning that mean magnitude measurements of variable stars will not be precise and thus, standard candles like RR Lyrae (RRL) will not be usable. Light curve templates (LCTs) can be adopted to estimate the mean magnitude of a variable star with few magnitude measurements, provided that their period (plus amplitude and reference epoch, depending on how the LCT is applied) is known. LSST will provide precise RRL periods within the first six months, allowing to exploit RRLs if LCTs were available. }
{We aim to build LCTs in the LSST bands to enhance the early science with LSST. Using them will provide a 1-2 years advantage with respect to a classical approach, concerning distance measurements.}
{We collected $gri$-band data from the ZTF survey and $z$-band data from DECam to build the LCTs of RRLs. We also adopted synthetic $griz$-band data in the LSST system from pulsation models, plus SDSS, \gaia and OGLE photometry, inspecting the light amplitude ratios in different photometric systems to provide useful conversions to apply the LCTs.}
{We have built LCTs of RRLs in the $griz$ bands of the LSST photometric system; for the $z$ band, we could build only fundamental-mode RRL LCTs. We quantitatively demonstrated that LCTs built with ZTF and DECam data can be adopted on the LSST photometric system. LCTs will decrease by a factor of at least two the uncertainty on distance estimates of RRLs, with respect to a simple average of the available measurements. Finally, within our tests, we have found a brand new behavior of amplitude ratios in the Large Magellanic Cloud.}
{}


\maketitle

\section{Introduction} \label{sect_intro}

The \textit{Vera C. Rubin} Observatory will see the first light at the beginning of 2025 and, for ten years, it will carry on the Legacy Survey of Space and Time (LSST), surveying the entire Southern sky in six passbands (\textit{ugrizy}). The wide field camera (3.5 square degrees) and the large primary mirror (8.4m diameter) in synergy, will be the keys for the unprecedented efficiency of the so-called Wide-Fast-Deep (WFD) survey within LSST. The WFD will deliver the most extended set of light curves for variable stars in terms of the following, combined properties: I) Epoch extension (10 years); II) Sky area (18,000 deg$^2$); III) Number of phase points (more than 800 in total); IV) Number of passbands (six, from \textit{u} to \textit{y}) and V) Depth: r$\sim$24.03/26.90 mag with 5$\sigma$ signal-to-noise, for the single exposure/10-years coadded image. The WFD is conceived to be a complement of the \gaia mission \citep{gaia_alldr} in the inner Bulge and in the outer Halo, concerning not only variability but also astrometry.

RR Lyrae (RRLs) are old, low-mass, core Helium-burning stars on the Horizontal Branch, pulsating regularly with periods of 0.2 -- 1.4 days \citep{soszynski2019,braga2021}. RRLs are present in every component of the Milky Way: Globular Clusters, Halo, Bulge, Thick Disk and, recently, several Thin Disk RRLs were also found \citep{prudil2020b,matsunaga2022}. RRLs obey to tight Period-Luminosity (PL) relations at effective wavelengths ranging from $\sim$600 nm to $\sim$4500 nm, thus including all passbands between, e.g., $R$ (or $r$) and $M$ (or $[4.5]$ in the Wide-field Infrared Survey Explorer [WISE] system) \citep{longmore86,catelan04,madore13,braga15}. This makes them reliable old-age standard candles, being therefore a complement, for distance estimates, to the young and more luminous Classical Cepheids \citep{leavitt08,madore1991,Skowronetal2019}. RRLs are also metallicity probes since they obey $P$ -- [Fe/H] -- $\phi_{31}$ relations, where $\phi_{31}$ is a parameter derived from the Fourier coefficient of the fit of the light curve \citep{jurcsik1996,mullen2022,kovacs2023}. Metallicity has an important role since it also affects PLs, therefore Period-Luminosity-Metallicity relations (PLZ) can be used also as iron abundance diagnostics once the distance is known \citep{martinezvazquez16a,braga16}. Period-Wesenheit relations (PW) and Period-Wesenheit-Metallicity relations (PWZ), where the Wesenheit is a pseudo-magnitude that is reddening-free by construction \citep{madore82} are used as distance indicators too. It is worth to note that some combination of magnitudes, especially towards bluer passbands, give metallicity-insensitive PWZs, as verified both theoretically and empirically \citep[][Accepted for publication in A\&A]{marconi15,marconi2022,ngeow2022,narloch2024} Thanks to all these properties, and being strictly old ($\gtrsim$9 Gyr, despite recent studies theorize that a minority of metal-rich, Thin Disk RRLs could be young, being the outcome of a merger \citealt{bobrick2024}), they can be used as tracers of the earliest stages of Galactic evolution \citep{fiorentino15a,iorio2021}.

To fully exploit RRLs as probes of the Galactic evolution, it is mandatory to estimate accurate mean magnitudes (<$mag$>) which, in turn, provide individual distance estimates by means of the PL relations. Being variable stars, a <$mag$> estimate requires a good sampling of the pulsation cycle to properly model the light curve. However, as shown by \citet{jones1996,braga2019}, light curve templates (LCTs) can be adopted to estimate accurate <$mag$> also having only one phase point observed (provided that the period, light amplitude [$Ampl$] and the reference epoch are well constrained). LCTs are, in short, analytical functions (or gridded points) that reproduce the typical shape of the light curve of a variable star within a specific range of periods/$Ampl$, pulsating in a specific mode. RRLs pulsate either in the Fundamental (F) or in the First Overtone (FO) mode, or both simultaneously (the so-called Double-Mode [DM]). Empirical evidence for the existence of Second Overtone (SO) RRLs is not clear \citep{kiss1999,rodriguez2003} and SO RRLs are not supported theoretically by pulsation models \citep{stothers1987}. Therefore, we ignored SO RRLs in this work. We also ignored DMs because 1) additional factors come into play (e.g., the difference in phase and the $Ampl$ ratio between the two modes); 2) concerning PLs and distance estimates, DMs were either discarded or treated as FO RRLs, since the dominating mode is, generally, the FO. To sum up, we focused only on F and FO LCTs. 

The empirical classification of RRLs into RRa, RRb and RRc, based on the shape of their light curves, dates back to \citet{bailey1902}. \citet{schwarzschild1940} associated the RRa and RRb stars (saw-tooth-shaped light curves, later unified into the RRab class) to the F mode and the RRc stars (sinusoidal light curves) to the FO mode. The pulsation mode is, therefore, the most important parameter to consider for LCTs. However, also RRLs pulsating in the same mode can have significantly different light curve shapes. The most reliable parameter to separate the different morphological classes of light curves is the pulsation period \citep{inno15,braga2019}.

The main aim for which the LCTs in this work are conceived is to enhance the science performed with LSST early data. By ``LSST early data'' we mean both the commissioning data, real-time calibrated images and Data Releases 1 and 2 (six and tewlve months of data, respectively). Observations from, e.g., DR1, will allow a period estimate with precision better than 0.05\% for typical RRL periods \citep{dicriscienzo2023}, but with uncertainties on <$mag$> up to 0.1-0.2 mag. This would translate into a distance uncertainty of up to 20\%, that is too large for any galactic archaeology analysis with RRLs. Using the LCTs from this work will mean to have an advantage of 1-2 years, compared with a more classical approach, concerning the distance determination of RRLs. This is going to be particularly interesting in the Bulge and in the outer Halo, where LSST will discover hundreds of thousands of new RRLs. We point out that the LCTs by \citet[][hereinafter, S10]{sesar2010}, although covering SDSS \textit{ugriz} passbands, are not fit for our purpose and this is the reason why we decided that new LCTs are necessary.

The paper is structured as follows: in Section~\ref{sect_data}, we present the dataset used to build the LCTs; the procedure adopted to build the LCTS is explained in Section~\ref{sect_template}. We discuss the accuracy of our LCTs and how to use them in Section~\ref{sect_validation} and report our conclusions in Section~\ref{sect_discussion}. We added three sections in appendix to discuss, in order, the shape of the LCTs, the comparison with S10 LCTs and the light amplitude ratios.

\section{Dataset}\label{sect_data}

In general, LCTs are built starting from normalized and overlapped light curves observed in a given passband. In our case, this is not possible since \textit{Rubin} Observatory has not started its operations yet, meaning that there is no existing observed data in the $ugrizy_{LSST}$ ($ugrizy_L$, in short) photometric system.

Therefore, as a database to build the LCTs, we decided to adopt light curves collected by other instruments, with passbands as similar as possible to the $ugrizy_L$ photometric system. The two empirical datasets that we have selected are: 1) Zwicky Transient Facility \citep[ZTF,][]{ztf,ztf2} light curves of RRLs in the $gri_{ZTF}$ bands ($gri_Z$, in short), selected from our catalog of $\sim$286,000 RRLs, mostly based on \gaia DR2+EDR3, plus other surveys \citep{fabrizio2019,fabrizio2021}; 2) The Dark Energy Camera \citep[DECAM,][]{decam_camera} survey of the Galactic Bulge by \citet{saha2019} in the $ugriz_{DECam}$ ($ugriz_D$, in short).

A critical question arises: can we really adopt ZTF and DECam passbands to derive LCTs to be used with LSST data? A detailed answer to this question will be given in Section~\ref{sect_conversion}.

\subsection{ZTF}\label{sect_ztf}


The extremely large 47\deg field of view of ZTF allowed this instrument to survey the sky very quickly, collecting hundreds of $gr_Z$-band, plus tens of $i_Z$-band phase points per target. From now on, the suffix $Z$ will refer to the ZTF photometric system; non-suffixed passband names refer will instead be used when referring to the LCTs and to passbands in general. As a starting sample, we have adopted our catalog of RRLs based on \gaia DR2+EDR3 plus other surveys like Catalina, ASASSN, PanSTARRS \citep[][and references therein]{fabrizio2019,fabrizio2021}. By using \gaia EDR3 coordinates for this catalog, we queried the ZTF DR17 database and retrieved a catalog of 35434/29510/9949 $g_Z/r_Z/i_Z$ light curves. To keep only the best sampled light curves to build the LCTs, we have rejected all the light curves with a value of \texttt{ngoodobsrel} (number of epochs within the current data release, without photometric quality issues) larger or equal than 80.

After this cut, the catalog narrowed down to 24925 RRLs in total. We have downloaded 23849/19146/2192 $g_Z/r_Z/i_Z$ light curves for these objects. The $g_Z/r_Z/i_Z$ light curves have, on average, 303/499/88 phase points. These numbers are the result of the rejection---already in the query step---of all the observations with \texttt{catflags} $\geq$32768 (i.e., measurements for which moon illumination or clouds might have hampered the photometric quality).



\subsection{DECam}\label{sect_decam}

Unfortunately, the ZTF survey did not collect images with passbands redder than \textit{i}. Nonetheless, for our purpose, it is crucial to push our LCTs at longer wavelengths, because LSST will acquire images also in $z$ and $y$. These are the most important filters to investigate the Bulge of the Milky way, since longer wavelengths are less affected by the extreme reddening in that region. Moreover, the PL relations have a higher slope (meaning that they are more precise distance indicators) at redder passbands.

For this reason, we complemented the ZTF sample described above with DECam time series of RRLs in the Bulge collected by \citet{saha2019}. They acquired tens of images in each of the $ugriz_D$ passbands (where the $D$ suffix refers to the DECam photometric system) and detected 474 RRab. Unfortunately, their sample does not include RRc since they did not search for them. This means that we will not be able to provide $z_D$-band LCTs for this pulsation mode. We also point out that we will not use $u_D$-band data to build $u$-band LCTs. First of all, the $u$ band is less important for distance determination, because the PL is flat and has a high dispersion at short wavelengths. Secondly, the steepness of the $u$-band light curve makes it difficult to provide good fits with the available sampling.

\section{Template building}\label{sect_template}

For the $gri$ LCTs, we adopted only the ZTF data for three reasons. 1) the contribution from DECam would be tiny: for the $gr$ bands, ZTF light curves have $\sim$10 times more phase points and the number of RRLs is $\sim$2 orders of magnitude larger. For the $i$ band, the ZTF RRL sample is $\sim$5 times larger than DECam one, and each light curve has around twice as many phase points. Overall, this means that, in the $gr$ and $i$ bands, the number of phase points in ZTF is three/one orders of magnitude larger than the DECam sample, respectively; 2) when possible, we prefer not to merge datasets from different instruments, to avoid transformations between photometric systems that would only introduce noise and, consequently, an intrinsic dispersion in the LCTs; 3) the ZTF and DECam catalogs of RRLs are qualitatively different, since they provide light curves of Halo and Bulge RRLs, respectively and it is well-known that the former are more metal-poor than the latter \citep[{[Fe/H]} --1.51$\pm$0.41 compared to $\sim$--1.0][]{fabrizio2021, walker1991a}. Since the shape of the light curve is notoriously dependent on the iron abundance at all wavelengths \citep{jurcsik1996,mullen2022}, we prefer not to merge the two datasets. Note, anyways, that we have compared the amplitudes of the ZTF and DECam light curves and found no significant difference (See ~\ref{sect:amplratio_appendix}) but the parameter that really depends on metallicity is $\phi_{31}$ \citep{jurcsik1996}. One might argue that metallicity should be taken more into account when building LCTs but there are two reasons for which we do not provide different LCTs for, e.g., different ranges of metallicities. First of all, only a tiny fraction ($\sim$1\%) of our RRLs has an iron abundance estimate. This would mean to sacrifice the excellent sampling that we have for the $g$ and $r$ bands and not having enough statistics to derive any $i$ and $z$ LCT. Secondly, our main aim is not to provide a complete taxonomy of the light curves of RRLs, but providing models that can be used to estimate <$mag$> and distances of RRLs that will be, mostly, new discoveries from LSST, and for which no estimate of [Fe/H] will be available, making it also impossible to use this information, even by having LCTs separated in [Fe/H] bins.

To build the LCTs, one has to 1) fold all the light curves by the proper period and adopt the same reference epoch ($T_0$) for the zero phase; 2) normalize the light curves; 3) merge all the normalized light curves in a given band and period bin into a single cumulated and normalized light curve (CNLCV); 4) derive an analytic fit of the CNLCV: this will be the analytic form of the LCT.

\subsection{Phasing of the light curves}\label{sect_phase}

As already discussed in \citet{braga2019,braga2021} we usually adopted, as a reference epoch for our LCTs, the epoch of <$mag$> on the rising branch of the \textit{V}-band light curve (\trisv), introduced by \citet{inno15}, instead of the more commonly used epoch of maximum light ($T_{max}$). In fact, \citet{inno15} and \citet{braga2021} demonstrated that \trisv provides a better anchor epoch with respect to $T_{max}$, because the CNLCVs have a smaller dispersion \citep[see Fig.2 and 8 in][]{braga2021}. We point out that the LCTs provided in the quoted papers ($JHK$-band light curves and radial velocity curve templates) were anchored to a reference epoch in the $V$ band because they are meant to be applied on scarcely sampled NIR or radial-velocity time series, by adopting ephemerides from well-sampled $V$-band data. They followed this approach because $V$-band surveys are larger and better sampled than any NIR and radial-velocity survey, and usually precede them in time.

However, the $griz$ LCTs that we provide in this paper are meant to be used in a different way: they will not be applied on poorly-sampled light curves of known variables for which a well-sampled $V$-band light curve is already available. Instead, they will be applied on candidate RRLs for which the early LSST time series will provide only an estimate of the pulsation period and a <$mag$> to be improved with the LCT itself. This means that, for the LCTs that we provide, the best phase anchors are $T_{ris}^{g}$/$T_{ris}^{r}$/$T_{ris}^{i}$/$T_{ris}^{z}$ for the $griz$ LCTs respectively.

To estimate $T_{ris}$ for the ZTF $gri_Z$ light curves, we followed a two-step method: 

Step 1) ---- After folding the light curves with the proper period and shifting them to a preliminary and arbitrary reference epoch ($T_0$=0), we fitted the light curves with Fourier series. We adopted a dynamical selection of the Fourier degree, starting from a base of 3/6 for RRc/RRab stars and increasing by one, up to a maximum degree of 12. At each step, $\chi^2_n$ (the $\chi^2$ for the $n^{th}$-degree fit) and  $\chi^2_{n+1}$, for the $n$-th and $n+1$-th degree fits are derived and their ratio ($\frac{\chi^2_{n+1}}{\chi^2_{n}}$) is derived. The larger is $\frac{\chi^2_{n+1}}{\chi^2_{n}}$, the less significant is the improvement of the fit when increasing the degree. We have set a 0.99 threshold ($th_{\chi^2}$) so that, when $\frac{\chi^2_{n+1}}{\chi^2_{n}}$ is smaller than that, we increase $n$. When $\frac{\chi^2_{n+1}}{\chi^2_{n}} \geq 0.99$, we stop the iteration and adopt $n$ as the degree of our best-fitting model. We point out that we did not apply any sigma clipping of the phase points because this would invalidate the $\chi^2$ comparison, since the residuals would be calculated on different datasets. The $\chi^2$ comparison might seem biased with a very large value of the threshold fraction and favoring too high degrees. However, to properly calculate \tris, it is important that the rising branch is very well fitted and, especially for RRab this usually requires high Fourier degrees. This is why we had set a quite high cap on the Fourier series degree (12) and selected carefully the training sample of the Neural Network (NN, see~\ref{sect_selection}). This was done in order to be sure that the estimate of \tris was not affected by features on the rising branch, e.g., the hump before the maximum. We point out that this iterative method for the degree selection, was selected over others (e.g., the Bayesian Information Criterion) for at least two reasons: 1) the direct control over the $\chi^2$ ratio; 2) its similarity with a reliable method like that adopted within the \gaia Collaboration (S. Leccia, private communication).

Step 2) ---- By adopting the model fit, we derived the <$mag$> of the star and, by searching its intersection with the rising branch fit \citep[see Appendix C.1 in][]{braga2021}, we derived \tris. Having an estimate of \tris for each light curve, the time series were re-phased by adopting this new reference epoch and the folded light curves were re-fitted, obtaining the final Fourier series model. The latter provided $Ampl$ and <$mag$>, integrated over an arbitrary flux scale.

For the $z_D$-band light curves, we followed a qualitatively identical algorithm to phase the light curves. The only difference is that we performed fits adopting both Fourier series and PLOESS models \citep[][and references therein]{braga16}, and selected the best-looking one, with a visual inspection. This approach was necessary because the DECam light curves are not as well sampled as the ZTF ones and the Fourier fit often failed, especially when large gaps are present.


\subsection{Selection of the light curves}\label{sect_selection}

Not all the light curves in our database are suitable to build the LCTs, therefore we added a selection algorithm to keep only the best light curves.

The ZTF dataset is so large that it was not possible to visually inspect all the light curves, therefore we adopted a NN to perform the selection. First, we selected a training sample of 681 $gri$ light curves, we visually inspected all the light curves and manually flagged the light curves that should be kept/rejected for the template building. We tested several combinations of predictors and NN architectures and found that the best solution is to adopt, as predictors, the parameters in Table~\ref{tab:predictors} and a NN architecture with two hidden layers, each with 14 perceptrons.

\begin{table}
\caption{Predictors used for the Neural Network}
\label{tab:predictors}
\footnotesize
\begin{tabular}{r p{7.3cm}}
Predictor & Description \\
\hline\\
$P$ & Pulsation period of the variable \\
$mag$ & Mean magnitude, intensity-averaged over the model fit \\
$Ampl$ & Light amplitude (maximum minus minimum of the model fit) \\
$\chi^2$ & Reduced $\chi^2$ of the model fit with respect to the data \\
$\chi_A^2$ & Reduced $\chi^2$ of the model fit with respect to the data, divided by $Ampl$ \\
$U$ & Uniformity parameter\footnote{Modified version of \texttt{UniformityMetric} by Peter
Yoachim (\url{https://rubin-sim.lsst.io/api/rubin_sim.maf.metrics.UniformityMetric.html\#rubin_sim.maf.metrics.UniformityMetric}), to compute how uniformly the observations are spaced in phase} \\
$U_{bin}$ & Same as $U$, but calculated on phase points grouped in phase bins of 0.05 \\
$n$ & Number of phase points \\
$\Delta\phi_{max}$ & Largest phase gap between two consecutive phase points \\
$Ku$ & Kurtosis of the light curve \\
$Sk$ & Skewness of the light curve \\
$r$ & Unweighted sum of the residuals divided by the degrees of freedom \\
$r_A$ & Same as $r$, divided by $Ampl$ \\
$out_1$ & Phase points at more than$\pm$1$\sigma$ distance from the model fit  \\
$out_3$ & Phase points at more than$\pm$3$\sigma$ distance from the model fit  \\
$out_5$ & Phase points at more than$\pm$5$\sigma$ distance from the model fit  \\
$out_{10}$ & Phase points at more than$\pm$10$\sigma$ distance from the model fit  \\
$A_1$ & Fourier-fit coefficient: 1$^{st}$-order amplitude \\
$A_2$ & Fourier-fit coefficient: 2$^{nd}$-order amplitude \\
$A_3$ & Fourier-fit coefficient: 3$^{rd}$-order amplitude \\
\hline\\
\end{tabular}
\end{table}


The NN was built by using the Python package \texttt{sklearn} and, more precisely, the \texttt{MLPClassifier}, with a maximum iteration number of 40000, an optimization \texttt{tolerance} of 0.0001, a \texttt{'tanh'} function for the \texttt{activation} and the default \texttt{'adam'} value for the \texttt{solver} parameter. We tested the accuracy of the NN and obtained a train/test converging ratio of 0.91/0.88.

The NN provided us a predictor that we have applied to the entire dataset of ZTF light curves, obtaining a True/False rejection flag for each light curve. After this task, the sample of light curves to be kept to build the LCTs was 14829/12239/1243 for the $g_Z/r_Z/i_Z$ bands, respectively. We have visually inspected a small sample of the light curves flagged by the NN, and found that a very high fraction of the selections were performed correctly. Therefore, we were satisfied with the adopted NN, and did not build a NN for each filter.


                        
We did not adopt a machine learning algorithm for the DECam data because the sample is too small, and because the visual inspection---that would be required anyway to set the training sample---does not take too much time on such a limited sample. After the fitting, we rejected all the light curves that are not suitable to build the LCTs, meaning those with less than 14 phase points, those for which a reliable measure of $Ampl$ is not available and those with a large dispersion around the fit of the light curve. We end up with 217 $z_D$-band light curves of RRab with 14-to-86 phase points, covering a period range between 0.356 and 0.822 days (see Fig.~\ref{fig:histoper}).

Tables~\ref{tab:ztf} and ~\ref{tab:decam} report, for the ZTF and DECam databases, the coordinates and pulsation properties; Table~\ref{tab:ztf} also reports the light curve fitting statistics needed to run the selection algorithm described above. Figure~\ref{fig:histoper} displays the distribution in period (left panels) and Bailey diagram (right panels) of the final sample of stars. One can see that all these datasets are made mostly by Oosterhoff I (OoI) RRLs and host a smaller fraction of Oosterhoff II (OoII) RRLs, where OoI and OoII RRLs are typically associated to metal-rich/metal-poor Globular Clusters of the Milky Way \citep{oosterhoff39}. Note that, for our $z_D$-band dataset, OoII RRLs seem to make up an even smaller fraction of the total number of RRLs---as expected due to the higher average metallicity of Bulge RRLs with respect to Halo RRLs---but we do not have enough statistics to assert this on a quantitative basis. 

We conclude this section by pointing out that, for each period bin (see Section~\ref{sect_periodbin}), we kept out of the selection, three RRLs that would be suitable to build the LCTs. These will be used as independent data for the validation of the LCTs themselves (see Section~\ref{sect_validation}).


\begin{table}
\caption{Pulsation and fitting properties of the DECam $z$ light curves of RRLs.}
\label{tab:decam}
\scriptsize
\begin{tabular}{rrrrrrrrrrr}
\hline\\
 S19 ID &  RA &     Dec &    $P$ &   $mag$ &  $Ampl$ &       \trisz &   n & Fit & rej \\
  &  [deg] &     [deg] &    [days] &   [mag] &  [mag] &       [days] &    &  &  \\
\hline\\
1131110 & 271.1231 & --29.5742 & 0.537080 &        15.978 &      0.463 & 2456424.8524 &    55 &     F & 0 \\
1131438 & 271.1378 & --29.5703 & 0.586080 &        15.851 &      0.381 & 2456453.6499 &    26 &     P & 0 \\
1132394 & 271.1834 & --29.6777 & 0.495820 &        15.670 &      0.728 & 2456423.6250 &    51 &     F & 0 \\
1132647 & 271.1952 & --29.6191 & 0.561640 &        16.003 &      0.425 & 2456423.4108 &    55 &     F & 0 \\
1133227 & 271.2210 & --29.6234 & 0.711310 &        15.593 &      0.633 & 2456423.2454 &    73 &     F & 0 \\
1289124 & 271.3183 & --29.6729 & 0.403450 &        16.326 &      0.797 & 2456423.4856 &    67 &     P & 0 \\
1290186 & 271.3692 & --29.5505 & 0.588420 &        16.983 &      0.406 & 2456424.3231 &    57 &     P & 0 \\
1290223 & 271.3707 & --29.6014 & 0.561410 &        15.892 &      0.385 & 2456423.1651 &    39 &     F & 0 \\
1290809 & 271.3985 & --29.6636 & 0.494860 &        16.016 &      0.720 & 2456423.4253 &    67 &     P & 0 \\
1291217 & 271.4183 & --29.5910 & 0.526810 &        15.883 &      0.663 & 2456452.4864 &    36 &     P & 0 \\
\hline\\
\end{tabular}
\tablefoot{Fit: ``F'' indicates that the best fit is a Fourier series; ``P'' indicates that the best fit is a PLOESS. Only the first ten lines are displayed. The full table is available on the CDS.}
\end{table}

\subsection{Period binning}\label{sect_periodbin}

The basic assumption for the LCTs is that different stars of the same variability class, same pulsation mode and with similar physical properties should have similar light-curve shapes. Recent studies demonstrated that stars with different pulsation properties might have very similar light curves \citep{jurcsik2022}, but these are a minor fraction and the quoted assumption is still valid when dealing with large datasets, which is the purpose of LCTs in this work.

Since the physical properties are not easily measurable, one should adopt one (or more) easily measurable parameters to group RRLs in template bins inside which all the variables have a similar shape of the light curve. We adopt two parameters: 1) the pulsation mode; 2) the pulsation period. The former is quite obvious and provides a dichothomic separation of light curves (almost sinusoidal for RRc and saw-tooth-like for RRab). The second should be better justified and we do this in the following.

In principle, the shape of the light curves of RRab and RRc depends on their physical properties (mass, temperature, chemical composition...) but these properties are not easy to measure. Luckily, the pulsation period ($P$) is an easily measurable quantity and is connected to the main physical properties of the stars by the pulsation relations \citep{vanalbada71,marconi15}, of the form 

\begin{equation}
\begin{aligned}
  \log P_F = &\alpha_F + \beta_F \log L/L_{\odot} + \gamma_F \log M/M_{\odot} + \\ & \delta_F \log T_{eff} + \eta_F \log Z  
\end{aligned}
\end{equation}

and

\begin{equation}
\begin{aligned}
  \log P_{FO} = &\alpha_{FO} + \beta_{FO} \log L/L_{\odot} + \gamma_{FO} \log M/M_{\odot} + \\ & \delta_{FO} \log T_{eff} + \eta_{FO} \log Z  
\end{aligned}
\end{equation}

As a consequence, the properties of RRLs scale with $P$. This is quite straightforward for <$mag$>, as it appears from the Period-Luminosity relations \citep[PLs,][]{longmore86,bono03c,madore13}; this is also true (although less clear due to non-linear effects) for $Ampl$, as one can see in a Bailey diagram \citep{bailey1902}. We point out that $P$ has several advantages with respect, e.g., the amplitude, which was adopted for RRLs by \citet{jones1996}. 1) It is the easiest parameter to measure: with LSST multi-band data, period measurements will be precise up to 1e-5 days with only six months of data for half of the stars with $g\sim$23.0 and for up to 70\% of the stars with $g\sim$21.0 \citep[][Di Criscienzo, private communication]{vanderplasivezic2015,dicriscienzo2023}; 2) also RRLs showing secondary modulations (Blazhko and double-mode) have a precise measure of the period, while the same is not true for the amplitude; 3) $P$, as $Ampl$, is independent of distance and reddening, that are not always easy to estimate.

This is the reason why a crucial operation is the selection of the thresholds of period bins: all the normalized light curves (NLCVs) of RRLs with a period inside a given bin should be merged, providing one CNLCV per bin, per band. Thus, the selection of the thresholds of the period bins is the key to separate properly the different light curve shapes. 


\begin{figure*}[!htbp]
\centering
\includegraphics[width=9cm]{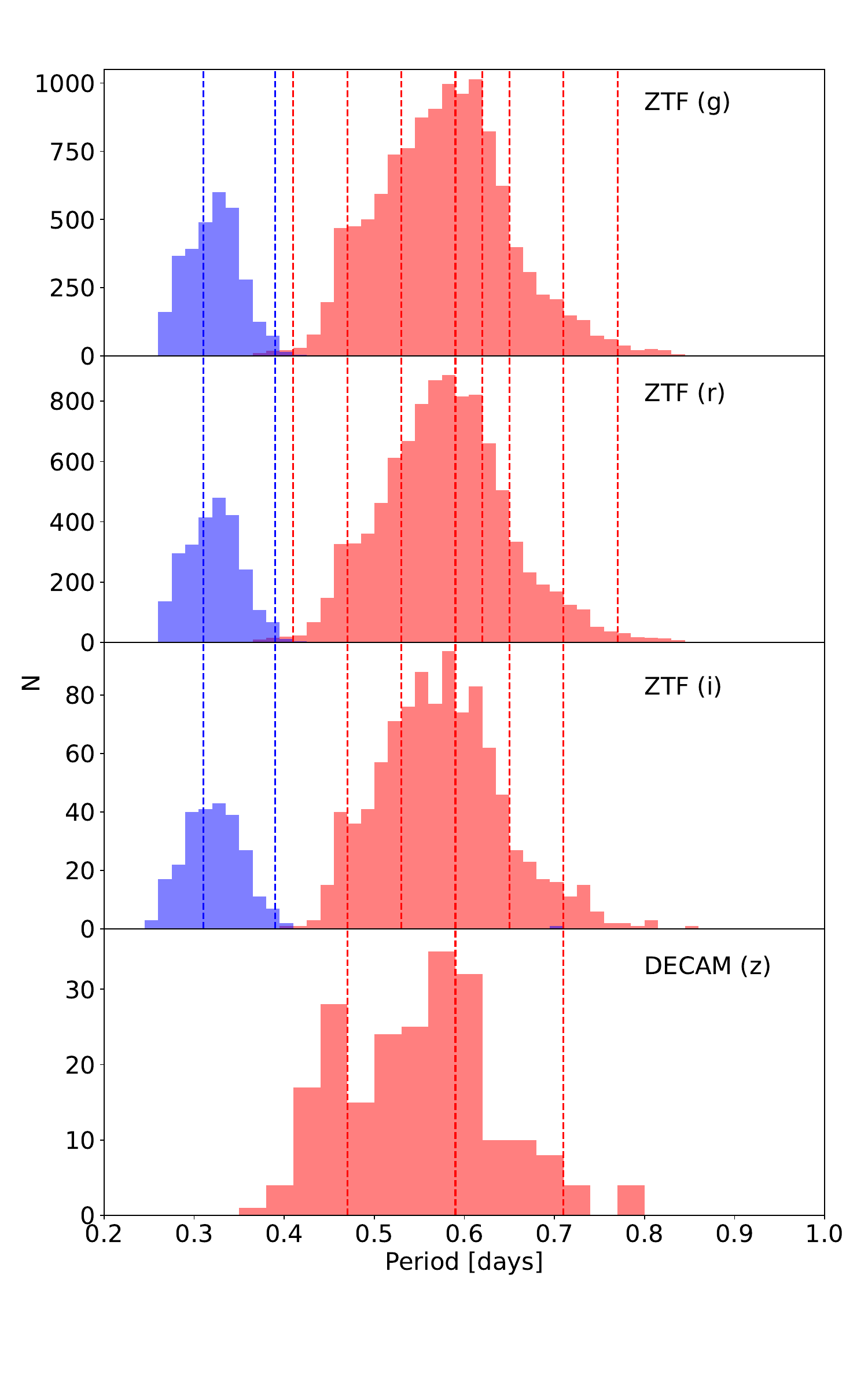}
\includegraphics[width=9cm]{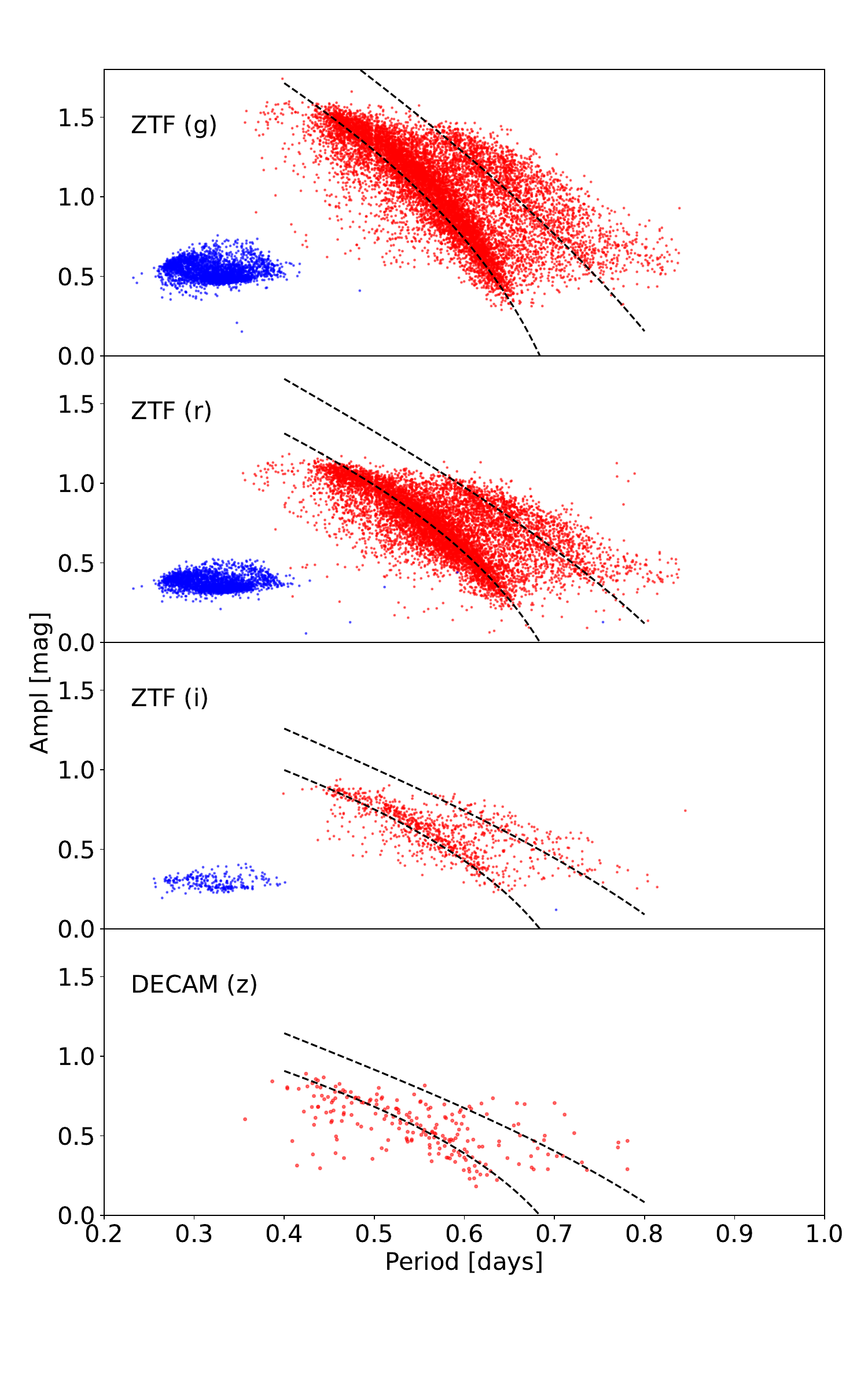}
\caption{Left panels: from top to bottom, period distribution of the RRLs adopted to build the templates in the $g_Zr_Zi_Zz_D$ bands, from top to bottom. RRab are displayed in red and RRc in blue. The red and blue dashed lines display the period thresholds of the LCTs derived in this work, for RRab and RRc, respectively.
Right panels: same as left, but for the Bailey diagram (period versus Ampl). Dashed black lines display the loci of the Oosterhoff tracks derived by \citet{fabrizio2019} and transformed into the $g_Zr_Zi_Zz_D$ bands using the ratios derived in \ref{sect:amplratio_appendix}.}
\label{fig:histoper}
\end{figure*}

Fig.~\ref{fig:histoper} displays the period distribution of the RRab and RRc that we will use to build the LCTs. Due to the different number of RRLs within the datasets in different passbands, we had to adopt different criteria to split the bins.

We started by selecting the bins for the RRab stars. The period range covered by RRLs for all the bands is 0.35-0.83. There are a few RRLs with period $>$0.83 days, but their number is small and we do not build any template if there are less than three RRLs in a bin. 

After several tests and deriving LCTs for different bin sizes, we decided to divide this period range into bins of 0.06 days for the ZTF ($gri_Z$) database, with only two exceptions: 1) due to the very large number of variables between 0.59 and 0.65 days, we splitted this bin in two for the $gr_Z$ data; 2) since, for the $i_Z$ band, we have fewer data, we merged the two bins 0.71-0.77 and 0.77-0.83 into one larger bin, to have sufficient statistics and derive a reliable LCT.

The bins for the $z$ band LCTs, however, must be wider, since the DECam dataset is considerably smaller (around 12,000 phase points in total, which is a factor of 400 smaller than the ZTF $g_Z$ and $r_Z$ samples). Therefore, in this case, we selected a bin width of 0.12, which is twice as large as that of the $gri_Z$ bands.

For the RRc stars, that cover around 0.17 days in pulsation periods, we have more than 1/1.5 million phase points for the $g_Z/r_Z$ bands, respectively, and almost 40,000 for the $i_Z$ band. We have tried several different bin sizes but the optimal solution is to adopt two bins of 0.08 days and a longer-period bin of 0.04 days for each band. Although smaller bins would be possible thanks to the large number of phase points, these would be useless since LCTs would be almost indistinguishable between these smaller bins. Indeed, the LCTs for the mid- and long-period RRc are very similar, apart from the hump at phase 0.05-0.10 in the $i_Z$-band, therefore we performed some tests by shifting the bins by steps of 0.01 days, to check how the LCTs changed by moving the thresholds. The tests revealed that, despite the small sample size of the longest bin, the difference in shape is solid and real. In fact, decreasing the low-period threshold of the long-period bin, the hump is less and less evident. Therefore we provide the three LCTs as described.

\subsection{Normalization and cumulation of the light curves}\label{sect_normalization}

Since the LCTs are provided as normalized curves, with mean 0 and amplitude 1, we have normalized the folded time series of all the RRLs in all the bins, by subtracting <$mag$> from all the measured magnitudes of each phase point, and then dividing them by the light amplitude. This process provided us the normalized light curves (NLCVs).

After this operation we collected, for each passband and each period bin, the phase points of all the NLCVs within that bin. The result is a single light curve, that is the cumulation of all the NLCVs in the same band, with similar periods and, therefore, with similar shapes. This is what we call the cumulated and normalized light curve (CNLCV). Table~\ref{tab:templates} displays the number of phase points and of RRLs for each CNLCV.

Note that we did not generate any CNLCV (meaning that we did not derive any template) for bins with less than three light curves, because such a limited statistics cannot really guarantee that there is no true difference in the shape for a period range with such a few data.

\subsection{Fitting}\label{sect_fitting}

Finally, we derived the analytical form of the LCTs by fitting, with a Fourier Series, the CNLCVs. However, we noted that a simple fit on all the phase points is not the best solution because it often generates models that are too smooth and for which the typical features of RRL light curves at that period are lost. This happens for any degree that is adopted for the Fourier series. Therefore, we have built 2D histograms in the phase-$Nmag$ (normalized magnitude) plane, on a $n_x \times n_y$ grid, from a minimum of 90$\times$90 to a maximum of 500$\times$500 cells, slicing the [0,1] phase range and the [1,-1] $Nmag$ range. $n_x$ and $n_y$ were individually selected for each of the CNLCVs. We point out that $n_x$ may be different from $n_y$. We counted the number of phase points in each cell ($N_{ij}$, with $i$ and $j$ running over $n_x$ and $n_y$) and generated a density plot. At first, we tried to simply take the densest cell for each vertical slice (that is, at each phase) and then perform a fit on these points. However, this technique provides inaccurate models, with non-realistic ripples.

\begin{figure}[!htbp]
\centering
\includegraphics[width=9cm]{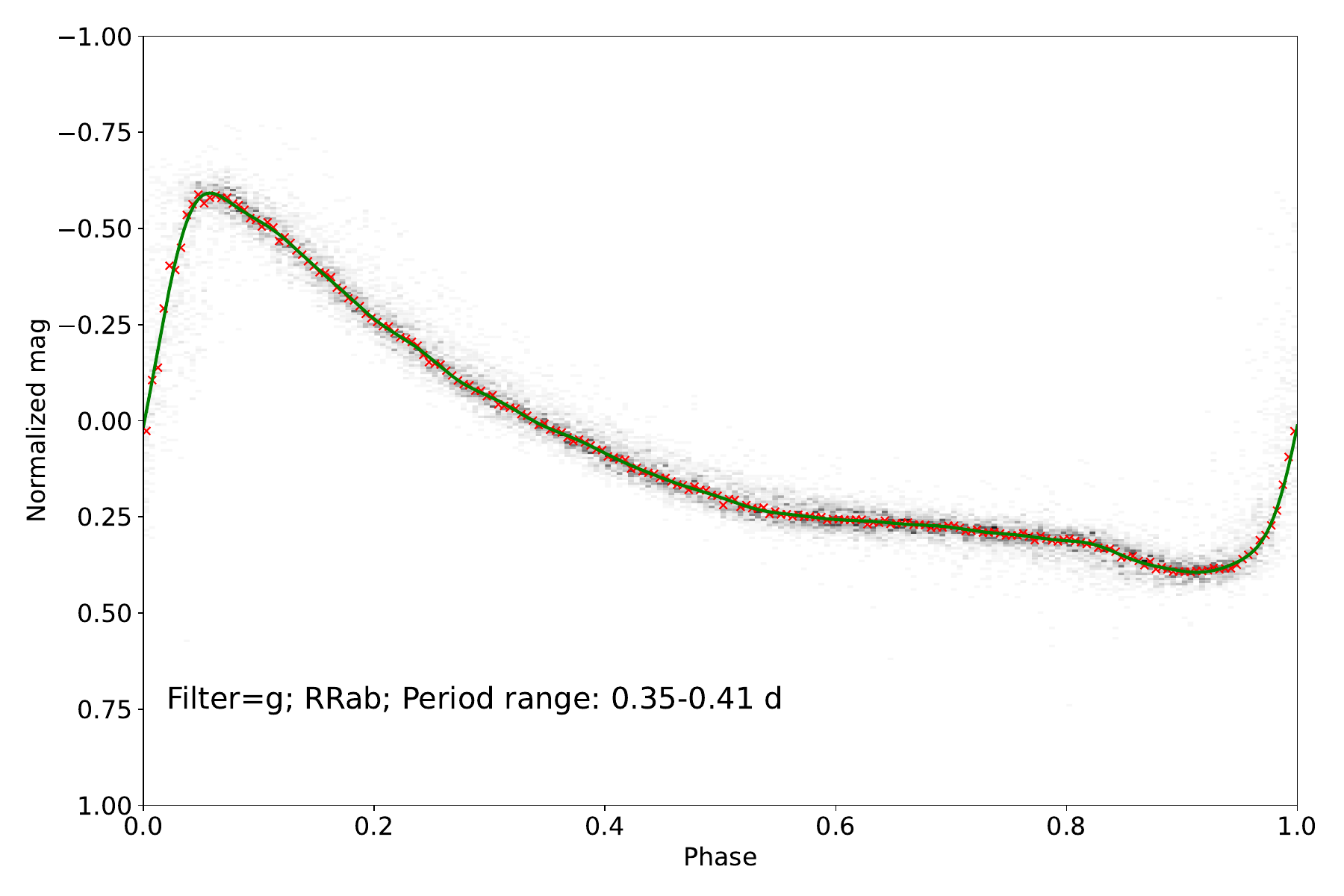}\\
\caption{Density plot of a CNLCV ($g$ band, RRab, period bin [0.35-0.41 days]). Darker cells are the most dense. The red crosses display the average of the four highest-density cells at each phase. The green, solid line shows the Fourier fit of these average points.}
\label{fig:cnlcv}
\end{figure}

Therefore, we decided to take, for each of the $n_x$ phases, an arbitrary number (between four and ten, selected individually for each CNLCV) of the densest cells ($n_{cell}$) of the vertical slice, and calculated the weighted average of their $Nmag$. As weights for each of the $n_{cell}$, we adopted $N_{ij}$, so that the densest cells have more weight in the average. In this way, we obtain a number $n_x$ of average magnitudes ($<Nmag_i>$), each associated to a specific phase. Finally, we fitted these points with a Fourier series and obtained the definitive, analytical form of the LCTs. The density plot and the fit of a sample LCT is displayed in Fig.~\ref{fig:cnlcv}. Note that, for all the fits, the degree of the Fourier series is dynamically selected by setting a $th_{\chi^2}$ of 0.995-to-0.999, and a maximum Fourier-series degree of 40.

Table~\ref{tab:templates} lists the Fourier series coefficients representing the analytical form of the LCTs and the statistics behind the template building. The figures showing the analytic forms of the templates are displayed in Appendix~\ref{sect:templates_appenidx}.

\subsection{Applicability to LSST photometric system}\label{sect_conversion}

To build the LCTs, we have used data from two different photometric systems (DECam and ZTF), neither of which is that of LSST. As shown by Table~\ref{tab:filters}, the difference of both the effective wavelength ($\lambda_{eff}$) and the effective width ($W_{eff}$) of the same passband in two different photometric systems can range from a few tens of  ~to a few hundreds of \AA, especially in the $z$ band. Therefore, we have to check whether and how the reliability of our LCTs is affected when used on LSST data.

\begin{table}
\caption{Characteristic wavelengths of the DECam, ZTF and LSST passbands}
\scriptsize
\begin{tabular}{lrrrr}\\
\hline\\            
FilterID &  $\lambda_{eff}$ &  $\lambda_{min}$ &  $\lambda_{max}$ &    $W_{eff}$ \\
 & \multicolumn{4}{c}{[\AA]} \\
\hline\\
CTIO/DECam.g &     4769.90 &     3850.48 &     5574.42 & 1003.72 \\
CTIO/DECam.r &     6370.44 &     5404.75 &     7295.86 & 1276.22 \\    
CTIO/DECam.i &     7774.30 &     6793.73 &     8694.71 & 1281.01 \\ 
CTIO/DECam.z &     9154.88 &     8252.42 &    10145.17 & 1289.35 \\    
Palomar/ZTF.g &     4746.48 &     3676.00 &     5613.82 & 1196.25 \\   
Palomar/ZTF.r &     6366.38 &     5497.60 &     7394.40 & 1417.53 \\   
Palomar/ZTF.i &     7829.03 &     6870.97 &     8964.61 & 1320.30 \\   
LSST/LSST.g &     4740.66 &     3876.02 &     5665.33 & 1253.26 \\  
LSST/LSST.r &     6172.34 &     5377.19 &     7055.16 & 1206.92 \\         
LSST/LSST.i &     7500.97 &     6765.77 &     8325.05 & 1174.77 \\  
LSST/LSST.z &     8678.90 &     8035.39 &     9375.47 &  997.51 \\  
\hline\\
\end{tabular}\\
\tablefoot{In this paper, we adopted filter properties from SVO\footnote{\url{http://svo2.cab.inta-csic.es/svo/theory/fps/index.php?mode=browse}}.}
\label{tab:filters}
\end{table}

For variable stars, there are three main properties that can be affected when one considers the differences between two photometric systems: 1) <$mag$>; 2) $Ampl$; 3) the shape of the light curve.

\subsubsection{Mean magnitude}
Luckily, when working with LCTs, one does not need to know <$mag$> in advance because the template can either be fitted leaving a free magnitude offset, or it can be anchored to a phase point, if the ephemerides are known. In both cases, <$mag$> is an output and not an input of the process. Therefore, we can ignore the effect of <$mag$> caused by adopting different photometric systems. Nonetheless, we performed a simple test, by fitting DECam light curves with $gri$ LCTs, derived from ZTF. The offset in <$mag$> is generally smaller than 0.01 mag, also for RRab with high amplitude, that have the sharpest and most asymmetrical light curves, being therefore the least easy to fit, at least with respect to RRab with smaller amplitudes and, in general, to RRc.

\subsubsection{Light amplitudes}
When using the LCTs, one has to rescale the normalized amplitude of the LCT to the real (or the expected) $Ampl$ of the observed variable. There are several methods to set the value of $Ampl$: 1) to leave $Ampl$ as a free parameter in the fit; 2) to set a fixed value of $Ampl$ based, e.g., on the Oosterhoff sequence relation ($logP$ versus $Ampl$) in the Bailey diagram \citep{cacciari05,fabrizio2021}, taking advantage of the knowledge of the pulsation period; 3) in the case that the LCT is applied to a RRL with known pulsation properties (including $Ampl(X)$ in a given $X$ passband) from a former survey, one can adopt $Ampl(X)$. Concerning case 1), $Ampl$ is an output of the process and not an input, meaning that one can use the LCTs on LSST data without any concern. Note well that, when the LCT is applied to \textit{Rubin} Observatory data (which would be the main use of LCTs provided in this paper) one will have to adopt a value of $Ampl$ in the LSST photometric system for a proper rescaling. In cases 2) and 3), $Ampl$ is an input and it should be converted into $Ampl(x_L)$, where $x_L$ is the LSST passband to which the template must be applied. The calculation and discussion on the amplitude ratios is extensive and detailed and would break the current discussion on how to adapt the LCTs to the LSST photometric system. Therefore, we present it in Appendix~\ref{sect:amplratio_appendix}.

\subsubsection{Light curve shape} 
Unfortunately, concerning the shape of the light curves, we cannot provide any quantitative estimate of differences between the LSST, DECam and ZTF passbands. In fact, the DECam light curves have too few phase point to derive the fit coefficients. Furthermore, we cannot even compare the synthetic light curves that we used for the amplitude ratios. In fact, while these can provide accurate estimates of <$mag$> and $Ampl$, they display some features (e.g., humps before the maximum) that are not observed in real-life photometric data. This means that any comparison with, e.g., Fourier series coefficients or principal components, would not be reliable when applied to empirical data. Even more importantly, since LCTs are meant to be used on a small number of observed phase points and the final goals are 1) to help with the separation between RRc and RRab and 2) to get a <$mag$> estimate, tiny differences in the shape lead to errors that are smaller than the intrinsic uncertainty of the LCTs and the propagation due to the amplitude rescaling. Also the idea of converting directly the observed light curves between the different datasets, although feasible in principle, has to be discarded: for a computationally heavier approach, we could introduce uncertainties due to the intrinsic spread of the conversion equations and to their limitations in color range and evolutionary stage. Moreover, even in the case that the shape of the light curves is actually improved, the change would be tiny and, as already  explained, the improvement would be smaller than the intrinsic uncertainty of the LCTs.

\section{Validation}\label{sect_validation}

After deriving the analytical form of the LCTs, we have tested their performance. As anticipated in Section~\ref{sect_selection}, we left a few variables (three for each period bin in every passband) out of the database that was used to build the LCTs. We used these RRLs to validate the LCTs with independent data. To perform our validation, we considered realistic future scenarios in which one will use RRLs for the investigation of Galactic structure and formation. In these cases, LCTs could be used to obtain accurate <$mag$> estimates and, in turn, distances, of RRLs during the commissioning or the first couple of years of observations with LSST. This is, as already mentioned in Section~\ref{sect_intro}, our aim in delivering these LCTs: to enhance early science with LSST, having more precise distances for these important old population tracers.

Using LSST data, our LCTs can be used to improve the estimate of mean magnitudes, adopting two different techniques: 1) LCT anchoring on a single (or more) phase points, when the period, reference epoch and light amplitude in a given passband are known; 2) LCT fitting of the light curve, in the case that the reference epoch is not known and a few phase points are available in a given passband. Case 2) splits in two: 2a) fixed amplitude in the fitting process (minimum three phase points required, plus the pulsation period and $Ampl$); 2b) amplitude as a free parameter in the fitting process (minimum four phase points required, plus the pulsation period). In the following sections, we discuss our tests using these three techniques.

\subsection{Single point anchoring validation}\label{sect_validation_singlepoint}

Applying LCTs on single phase points of RRLs with well-known pulsation properties is the original purpose of LCTs of RRLs \citep{jones1996} and variable stars, in general. The LCTs provided in this work are aimed to extract as much information as possible from LSST early data, including that narrow window (1 or 2 months from the beginning of the survey) when only one phase point per band per target will be available. In this case, one can apply the LCTs only to variables for which both the period, reference epoch and $Ampl$ are known, otherwise there would be not enough information to anchor the LCT on one point. A practical example of relevant scientific interest is that of RRLs found by OGLE in the Bulge \citep{soszynski2019}. In fact, for these OGLE RRLs, we have detailed information but only in the $I$ and $V$ bands. However, obtaining <$mag$> in more passbands, means to be able to use different PLs and PW. This approach allows both to validate the results obtained by OGLE itself, and to improve them thanks to, e.g., the higher slope and smaller dispersion of the PL($z$), or the insensitiveness of PW($r$,$g-r$) on metallicity \citep{marconi2022}. Moreover, having <$mag$> in three-four passbands, allows to follow the extinction law on a wider range of wavelengths, if one is interested in that.

A conservative estimate based 1) on the assumption of a saturation limit of 16 mag in the $i_{LSST}$ band \citep{lsst_main}; 2) a maximum $i_{DECam}-I$ color of 0.3 mag, verified on the subsample of OGLE RRLs matching those in \citep{saha2019}, assuming $i_{DECam} \approx i_{LSST}$; 3) $\sim$0.45 mag as half of the maximum $Ampl(i)$ (see Fig.~\ref{fig:histoper}), tells us that at least 41,000 RRLs with known pulsation properties from OGLE ($\sim$60\% of the total sample of $\sim$68,000) will not saturate in the LSST images. Unfortunately, the overlap between the OGLE RRLs and ZTF RRLs is poor and there are not enough RRLs to validate the $gri$ LCTs with the quoted sample. On the other hand, OGLE and DECam Bulge RRLs overlap almost completely, meaning that, for the $z$-band LCTs, we can simulate the quoted scientific case.

In the following, we describe the procedure that we adopted to validate the LCTs by using the single point anchoring method. For each variable, we randomly extracted a single phase and calculated the value of the fitting model (Fourier series) at that phase. Each extraction was replicated 100 times for each star in the validation sample, to randomly extract different phases. This operation was performed five times for each extraction, adopting a different noise level ($\sigma$ = 0.005, 0.01, 0.02, 0.05 and 0.10 mag\footnote{These correspond to the expected photometric error on the single visit for a star of magnitude $r\sim$20.2, 21.0, 22.0, 23.2 and 24.0 within the LSST \citep{lsst_main}}) and then adding the noise as $\sigma\cdot\mathcal{N}(0,\sigma^2)$, where $\mathcal{N}(0,\sigma^2)$ is a random value extracted from a normal distribution of mean 0 and variance $\sigma^2$. Since this procedure is applied to three test RRLs for each LCT, we obtain a grand total of 37 LCT $\times$ 3 RRLs per LCT $\times$ 5 levels of noise $\times$ 100 simulations per RRL = 55,500 re-sampled phase points. On each of these, we anchored the appropriate LCT based on the pulsation mode, period and passband. Note that, for the $gri$ bands, we re-scaled the LCTs by adopting the true $Ampl(g,r,i)$ value given in Table~\ref{tab:ztf} since we could not find enough OGLE matches. For the $z$ band, we adopted $Ampl(I)$ from OGLE and rescaled it to $Ampl(z)$ using the values in Table~\ref{tab:amplratio}.

Once we applied the rescaled LCT to the re-sampled phase point, we obtained an estimate of <$mag$>. To evaluate the improvement on the <$mag$> estimate introduced by the use of LCTs, with respect to the single-point measurement, we calculated the difference between the true <$mag$> (<$mag$>$_{true}$) and both the <$mag$> derived using the LCT (<$mag$>$_{LCT(n,\sigma)}$) and the magnitude of the re-sampled phase point (<$mag$>$_{single(n,\sigma)}$), where $n$ and $\sigma$ indicate the number of re-sampled points and the noise level; in this case, $n$ is always 1. We obtained $\delta_{L(n,\sigma)}$ = <$mag$>$_{LCT(n,\sigma)}$ - <$mag$>$_{true}$ and $\delta_{S(n,\sigma)}$ = <$mag$>$_{single(n,\sigma)}$ - <$mag$>$_{true}$. We calculated the mean values and the standard deviations of $\delta_{S(1,\sigma)}$ and $\delta_{L(1,\sigma)}$ on each set of 100 resampled phase points. The results are displayed in Table~\ref{tab:deltamag_singlepoint} and Table~\ref{tab:deltamag_singlepoint_simplemean}. As expected, all the averages of both $\delta_{S(1,\sigma)}$ and $\delta_{L(1,\sigma)}$ are zero within the standard deviations. We also found that, for all the 185 combinations of LCT shape and $\sigma$, $\delta_{S(n,\sigma)}$ is always larger than $\delta_{L(n,\sigma)}$, meaning that the improvement is real in all cases.

However, the fundamental result is that all the standard deviations of the $\delta_{L(1,\sigma)}$ are smaller than the $\delta_{S(1,\sigma)}$ by a factor 2-15. We also point out that the means of $\delta_{L(1,\sigma)}$ are significantly smaller than those of $\delta_{S(1,\sigma)}$ (by a factor $\sim$5-65). These ratios increase with decreasing $\sigma$, meaning that, in real life, the improvement introduced by using the templates will be more evident (by a factor $\sim$5) for bright stars.

\begin{table}
\caption{Averages and standard deviations of the $\delta_{L(1; \sigma)}$}
\scriptsize
\begin{tabular}{rrrrrr}
\hline
\hline
LCT ID & <$\delta$>$_{L(1; 0.005)}$ & <$\delta$>$_{L(1; 0.010)}$ & <$\delta$>$_{L(1; 0.020)}$ & <$\delta$>$_{L(1; 0.050)}$ & <$\delta$>$_{L(1; 0.100)}$ \\
\hline
0 &  0.001$\pm$0.016 & -0.000$\pm$0.019 & -0.000$\pm$0.025 & -0.002$\pm$0.052 &  0.011$\pm$0.103 \\
1 &  0.000$\pm$0.014 &  0.000$\pm$0.019 &  0.001$\pm$0.022 & -0.003$\pm$0.053 & -0.004$\pm$0.099 \\
2 &  0.001$\pm$0.014 &  0.003$\pm$0.016 & -0.002$\pm$0.023 &  0.004$\pm$0.056 & -0.011$\pm$0.093 \\
3 &  0.001$\pm$0.012 & -0.000$\pm$0.015 &  0.001$\pm$0.021 &  0.001$\pm$0.052 &  0.004$\pm$0.099 \\
4 & -0.000$\pm$0.010 &  0.000$\pm$0.013 &  0.002$\pm$0.021 &  0.003$\pm$0.048 & -0.002$\pm$0.096 \\
5 &  0.000$\pm$0.015 & -0.002$\pm$0.019 & -0.001$\pm$0.027 & -0.003$\pm$0.054 &  0.002$\pm$0.094 \\
6 &  0.000$\pm$0.010 &  0.001$\pm$0.014 &  0.001$\pm$0.021 &  0.000$\pm$0.051 &  0.005$\pm$0.102 \\
7 &  0.000$\pm$0.014 & -0.001$\pm$0.018 & -0.001$\pm$0.021 & -0.003$\pm$0.053 &  0.007$\pm$0.103 \\
8 &  0.001$\pm$0.008 &  0.000$\pm$0.012 & -0.000$\pm$0.022 & -0.002$\pm$0.052 & -0.004$\pm$0.097 \\
9 &  0.001$\pm$0.027 &  0.003$\pm$0.028 & -0.002$\pm$0.033 &  0.001$\pm$0.062 & -0.009$\pm$0.106 \\
\hline
\end{tabular}
\tablefoot{Averages derived from each set of 100 resampled phase points, with <$mag$> derived anchoring the LCT to a single point. Only the first ten lines are displayed. The full version of this table is available in electronic format.}
\label{tab:deltamag_singlepoint}
\end{table}

\begin{table}
\caption{Averages and standard deviations of the $\delta_{S(1; \sigma)}$}
\scriptsize
\begin{tabular}{rrrrrr}
\hline
\hline
LCT ID & <$\delta$>$_{S(1; 0.005)}$ & <$\delta$>$_{S(1; 0.010)}$ & <$\delta$>$_{S(1; 0.020)}$ & <$\delta$>$_{S(1; 0.050)}$ & <$\delta$>$_{S(1; 0.100)}$ \\
\hline
0 & 0.005$\pm$0.184 &  0.024$\pm$0.195 &  0.018$\pm$0.193 &  0.016$\pm$0.200 &  0.041$\pm$0.222 \\
1 & 0.015$\pm$0.183 &  0.014$\pm$0.173 &  0.012$\pm$0.179 &  0.015$\pm$0.182 &  0.002$\pm$0.191 \\
2 & 0.010$\pm$0.202 &  0.009$\pm$0.209 &  0.003$\pm$0.209 &  0.012$\pm$0.214 &  0.008$\pm$0.227 \\
3 & 0.005$\pm$0.139 &  0.018$\pm$0.133 &  0.000$\pm$0.134 &  0.017$\pm$0.152 &  0.015$\pm$0.169 \\
4 & 0.001$\pm$0.127 &  0.016$\pm$0.131 &  0.020$\pm$0.131 &  0.006$\pm$0.139 & -0.004$\pm$0.152 \\
5 & 0.009$\pm$0.128 & -0.001$\pm$0.130 &  0.015$\pm$0.132 &  0.008$\pm$0.145 &  0.014$\pm$0.156 \\
6 & 0.002$\pm$0.107 & -0.008$\pm$0.108 &  0.009$\pm$0.110 &  0.005$\pm$0.120 &  0.007$\pm$0.144 \\
7 & 0.012$\pm$0.110 & -0.003$\pm$0.108 & -0.001$\pm$0.112 & -0.004$\pm$0.124 &  0.012$\pm$0.157 \\
8 & 0.017$\pm$0.107 &  0.009$\pm$0.104 & -0.003$\pm$0.110 &  0.006$\pm$0.112 &  0.005$\pm$0.142 \\
9 & 0.111$\pm$0.447 &  0.112$\pm$0.452 &  0.127$\pm$0.452 &  0.150$\pm$0.467 &  0.138$\pm$0.455 \\
\hline
\end{tabular}
\tablefoot{Averages derived from each set of 100 resampled phase points, assuming, as <$mag$>, the magnitude of the single phase point. The full version of this table is available in electronic format.}
\label{tab:deltamag_singlepoint_simplemean}
\end{table}


\subsection{Template fit}\label{sect_validation_templatefit}

The LCTs can be used as fitting functions by minimizing the $\chi^2$ on two/three free parameters: 1 [mandatory]) an offset in phase ($\Delta\phi$); 2 [mandatory]) an offset in magnitude ($\Delta mag$); 3 [optional]) the light amplitude ($Ampl$). Therefore, at least three/four phase points in a given passband should be available to perform a least-square LCT fitting, depending on whether $Ampl$ is fixed or not in the fitting procedure.

Starting from the fourth/fifth month from the start of the LSST survey, at least three/four phase points per band per star will be available, according to the predictions, and around ten observation per target per band per year \citep{bianco2022}. Moreover, already from the sixth month, period estimates for candidate variables will be accurate within 2$\cdot$10$^{-5}$ days, (a 0.005\% relative error). This means that, already a few months into the survey, one with access to early data, will already have thousands (or tens of thousands) of candidate RRLs with accurate-enough periods and enough phase points to adopt the LCT fitting method, either by fixing the amplitude or leaving it free (cases 2a and 2b described in Section~\ref{sect_validation}).

\subsubsection{Fixed-amplitude template fit}\label{sect_validation_templatefit_fixedampl}

If the LSST source is associated to a variable with known $Ampl$ from another survey, in principle, one can fix the value of $Ampl$ in the LCT fitting procedure, using the ratios derived in Appendix~\ref{sect:amplratio_appendix}. The advantage of this approach is that one can apply it also when only three phase points are available. This means that, with respect to the free-amplitude fit for which four phase points are available, this method can be used one-two months earlier. The trade-off is that this method can only be adopted on already-known RRLs. However, one can also fix the value of $Ampl$, based on the estimated period of the RRL and on the Oosterhoff tracks in the Bailey diagram \citep{kunder13,fabrizio2019}. This second method to fix $Ampl$ can be applied also to new candidate RRLs, without any knowledge from previous surveys, but the value of $Ampl$ can be over/underestimated up to a factor of 2, especially for RRc stars and short-period RRab. Therefore we strongly advice, when using this method for new candidate RRLs, to always plan a follow-up check by leaving the amplitude as a free parameter in the following months, with 2-3 additional phase points per band available.

The data resampling for this test is similar to that adopted in Sect.~\ref{sect_validation_singlepoint}, with the difference that we did not resample a single phase point 100 times but, for each test variable, we extracted a set of four, eight and twelve random phase points 100 times. Also in this case, we simulated five levels of noise ($\sigma$ = 0.005, 0.01, 0.02, 0.05 and 0.10 mag) for each set of phase points. Note that we performed a different random extraction of the $\mathcal{N}(0,\sigma^2)$ factor for each phase point. This procedure is applied to three test RRLs for each LCT, therefore the grand total is 37 LCT $\times$ 3 RRLs per LCT $\times$ 5 levels of noise $\times$ 3 sets of four, eight and twelve phase points $\times$ 100 simulations per RRL = 166,500 re-sampled light curves on which we anchored the LCTs.

On each set of re-sampled light curves, we calculated the <$mag$> with two techniques: 1) by simply averaging the resampled points (<$mag$>$_{avg(n,\sigma)}$); 2) by performing a LCT fit and calculating the <$mag$> on the fitted LCT (<$mag$>$_{LCT(n,\sigma)}$). In both cases, we converted magnitudes to fluxes, averaged the fluxes, and reconverted the mean flux to <$mag$>. Throughout the rest of the paper, we will refer to <$mag$> as magnitudes averaged in flux. The aim is to compare the <$mag$> derived by a simple mean (which is the solution adopted for candidate variables when neither templates nor light curve fits are not available) and those derived by adopting the LCTs. As a next step, we calculated the difference between <$mag$>$_{true}$ and the ones obtained with both methods ($\delta_{L(n,\sigma)}$ = <$mag$>$_{LCT(n,\sigma)}$ - <$mag$>$_{true}$ ; $\delta_{A(n,\sigma)}$ = <$mag$>$_{avg(n,\sigma)}$ - <$mag$>$_{true}$). We calculated the mean values and the standard deviations of $\delta_{A(n,\sigma)}$ and $\delta_{L(n,\sigma)}$ on each set of 100 resampled light curves. The results are displayed in Table~\ref{tab:deltamag_freeamplitude} and Table~\ref{tab:deltamag_freeamplitude_simplemean}. 

All the averages are zero within the errors, both for $\delta_{L(n,\sigma)}$ and $\delta_{A(n,\sigma)}$, but the relevant comparison is that of the standard deviations, both by deriving the difference ($\Delta(\sigma\delta)_{S(n,\sigma)}$ = $\sigma\delta_{A(n,\sigma)}$ - $\sigma\delta_{L(n,\sigma)}$) and the ratio ($R(\sigma\delta)_{S(n,\sigma)}$ = $\dfrac{\sigma\delta_{A(n,\sigma)}}{\sigma\delta_{L(n,\sigma)}})$ for each combination of LCT shape, $n$ and $\sigma$. $R(\sigma\delta)_{S(n,\sigma)}$ gives a quantitative evidence of how the LCT fitting improves the <$mag$> estimate, ranging from $\sim$1.0 to $\sim$3.5. Note that the ratio is $\sim$1.0 only for $n$=4 and $\sigma \geq$ 0.05 mag, that is the poorest light curve sampling for faint stars. We also found that, in 445 over 555 cases, $\Delta(\sigma\delta)_{S(n,\sigma)}$ is positive, meaning that the LCT fit improves the robustness of the <$mag$> estimate. The improvement is even more clear when only considering the resampled light curves with 8 or 12 phase points (322 over 370 cases) or light curves with photometric errors smaller than 0.100 mag (368 over 444 cases).


\subsubsection{Free-amplitude template fit}\label{sect_validation_templatefit_freeampl}

The fitting technique that will be adopted the most is likely the LCT fit by leaving the amplitude as a free parameter. In fact, in this case, it is not required any previous knowledge of $Ampl$. Moreover, four phase points per target per band will already be available around five months from the start of the survey.

To test the improvement introduced by the LCTs with respect to the simple average of the phase points, we follow the same approach described in Section~\ref{sect_validation_templatefit_fixedampl}. Also in this case, we obtain 166,500 resampled time series and calculated the mean values and the standard deviations of $\delta_{A(n,\sigma)}$ and $\delta_{L(n,\sigma)}$ on each set of 100 resampled light curves. The results are displayed in Table~\ref{tab:deltamag_freeamplitude} and Table~\ref{tab:deltamag_freeamplitude_simplemean}. 

As in all the cases discussed before, the averages of both $\delta_{L(n,\sigma)}$ and $\delta_{S(n,\sigma)}$ are 0 within their standard deviations, in all cases. As for the fixed amplitude LCT fit technique, we derived $\Delta(\sigma\delta)_{S(n,\sigma)}$ = $\sigma\delta_{A(n,\sigma)}$ - $\sigma\delta_{L(n,\sigma)}$ and $R(\sigma\delta)_{S(n,\sigma)}$ = $\dfrac{\sigma\delta_{A(n,\sigma)}}{\sigma\delta_{L(n,\sigma)}}$ for each combination of LCT shape, $n$ and $\sigma$. $R(\sigma\delta)_{S(n,\sigma)}$ ranges between $\sim$1 and 12. We point out that the smaller ratios (around 1 and sometimes even smaller) are only found for very large $\sigma$ and $n$=4. We tested this also finding that, in 438 over 555 cases, ($\Delta(\sigma\delta)_{S(n,\sigma)}$) is positive, meaning that the LCT fit improves the robustness of the mean magnitude estimate. The improvement is even more clear if we limit ourselves to the resampled light curves with 8 or 12 phase points (362 over 370 cases) or to light curves with photometric errors smaller than 0.100 mag (370 over 444 cases). 

\subsubsection{Free amplitude template fit (pulsation mode test)}\label{sect_validation_templatefit_freeampl_ab_or_c}

The ranges of typical pulsation periods of RRab and RRc stars overlap between $\sim$0.35 d and $\sim$0.55 d. This means that, for newly discovered RRL candidates with a pulsation period within this range and few phase points (which will be the most typical case for the use of our LCTs) one cannot know in advance the pulsation mode of the RRL candidate. Unfortunately, the periods of our RRc and RRab LCTs do not overlap over such a large range, but we do provide LCTs for both RRc and RRab between 0.37 and 0.43 d. Therefore, on all candidate RRLs within this period range, one can apply both an RRc and an RRab LCT. 

We performed a test to check whether our LCTs can provide a solid classification in these cases. For this test, we followed the same process used in Section~\ref{sect_validation_templatefit_freeampl}: we fitted 36,000 resampled light curves of both RRab and RRc by adopting, for each star, the LCT of both RRc and RRab. Since we know in advance the true pulsation mode of the star, we can calculate the ratio between the $\chi^2$ of the correct LCT and that of the wrong LCT. Our assumption is that the ratio $\dfrac{\chi^2_{correct}}{\chi^2_{wrong}}$ is larger than 1. This assumption is satisfied for 35,999 tests: the only exception is for the resampled light curve of the star 387113300010133 (ZTF ID) with photometric error 0.10 mag and number of resampled phases equal to 4 (that is, the worst possible conditions). In this case, the ratio $\dfrac{\chi^2_{correct}}{\chi^2_{wrong}}$ is 0.92 while, in all other cases, it is---as expected---larger than 1 and can be as high as $\sim$350, generally increasing with decreasing photometric error. This means that our LCTs are not only useful to achieve more accurate mean magnitudes, but are also solid pulsation mode indicators.

\subsection{Improvement on distance estimates}\label{sect_distance}

We have checked that, under a variety of conditions, using the LCTs instead of simply averaging the magnitudes, improves the precision on the mean magnitude estimate. The main scientific outcome is, in turn, an improvement on the distance estimates when adopting the relations to derive them. In this section, we quantify the improvement on the precision of distances obtained from both PLs and PWs.

For our test, we adopted the PLs($i$), PL($z$) and the PW($r$,$g-r$). We selected the coefficients provided by \citet{marconi2022}, based on pulsation models. Note that we have selected the separated FU and FO PLs/PWs \citet[see Table 4 in][]{marconi2022} for our RRab and RRc, respectively. In this way, we could test the effect on distance estimates using all our LCTs. Note that, in this section, we are not interested in providing accurate, absolute distances for our validation stars. The crucial information that we want to extract from our simulation is the  difference between the distances obtained applying the PLs/PWs to <$mag$>$_{true}$ and those obtained applying the same relations to the simple mean magnitude (<$mag$>$_{avg(n,\sigma)}$) and to the mean magnitude from the LCT (<$mag$>$_{LCT(n,\sigma)}$). We name the three distances obtained in this way as $d_{true}$, $d_{avg(n,\sigma)}$ and $d_{LCT(n,\sigma)}$, respectively. In the case of distance estimate with a single phase point, $d_{avg(n,\sigma)}$ is obtained from the magnitude of the phase point itself without any averaging operation, as instead is done when more phase points are simulated.

Following the same method that we adopted to validate the mean magnitudes, we obtained, for each simulation of each star, the differences $\delta d_{L(n,\sigma)}$ = $d_{LCT(n,\sigma)}$ - $d_{true}$ and $\delta d_{S(n,\sigma)}$ = $d_{avg(n,\sigma)}$ - $d_{true}$. We derived their averages (<$\delta d_{L(n,\sigma)}$> and <$\delta d_{S(n,\sigma)}$>) and standard deviations ($\sigma\delta d_{L(n,\sigma)}$ and $\sigma\delta d_{S(n,\sigma)}$) for each case and, finally, obtained the relative averages and standard deviations, by dividing by $d_{true}$: <$\delta d_{rel[L/S](n,\sigma)}$> = $\dfrac{<\delta d_{[L/S](n,\sigma)}>}{d_{true}}$ and $\sigma\delta d_{rel[L/S](n,\sigma)}$ = $\dfrac{\sigma{\delta d_{[L/S](n,\sigma)}}}{d_{true}}$.

In the following paragraphs, we do not report all the cases but only a few examples, to demonstrate how, and in which cases, the LCTs allow to improve the distance estimate of a RRL. We point out that these are only examples that do not represent all the possible situations in which one might want to apply the LCTs and then derive the distances, therefore we will give a qualitative discussions and only a few quantitative data, because the latter are clearly affected by the specific time series available.

\subsubsection{Distances from PL(i) and PL(z); single point anchoring}

\begin{figure}[!htbp]
\centering
\includegraphics[width=9cm]{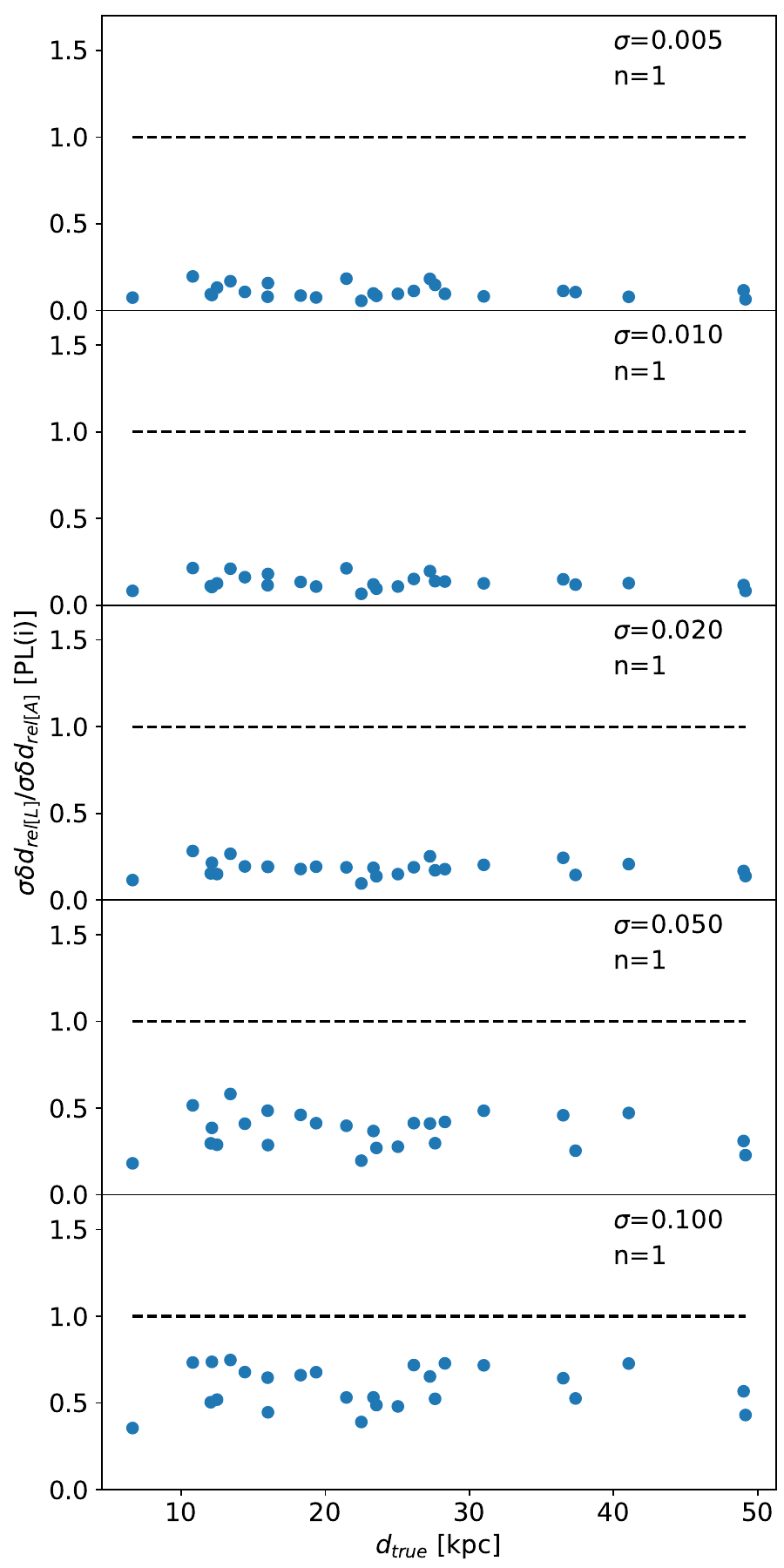}
\caption{Relative standard deviations of the distance offsets obtained with the LCTs over those obtained with the simple average. The ratios are plotted as a function of $d_{true}$. From top to bottom, the photometric error adopted for the simulation increases. The distances were obtained with PL($i$) relations and mean magnitudes from single-point LCT anchoring.}
\label{fig:distance_single_pli}
\end{figure}

\begin{figure}[!htbp]
\centering
\includegraphics[width=9cm]{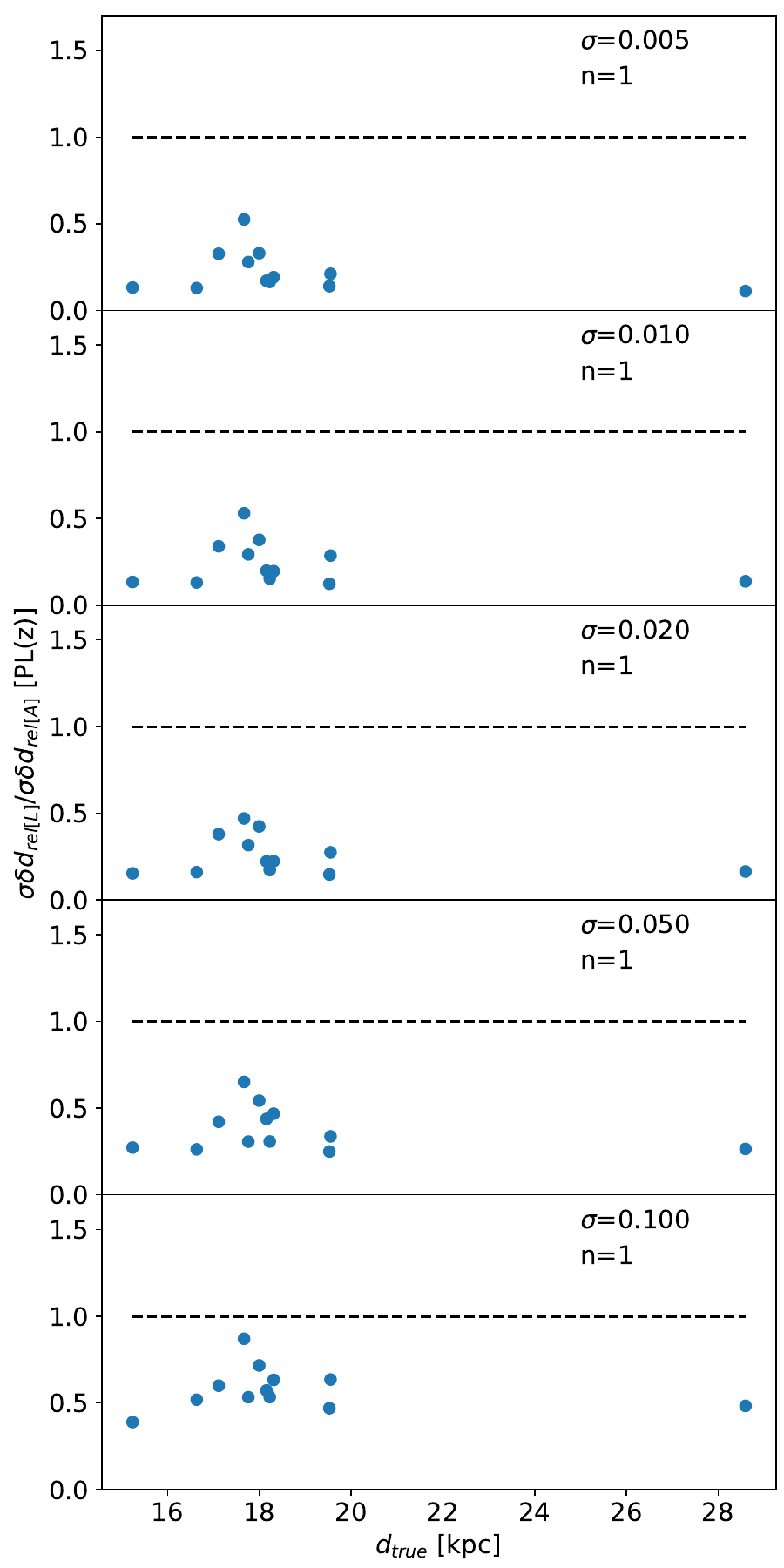}
\caption{Same as Fig.~\ref{fig:distance_single_pli} but the distances were obtained with PL($z$) relations.}
\label{fig:distance_single_plz}
\end{figure}

We tested the case for distances derived both with the PL($i$) and PL($z$), using the single point anchor approach. We found that the $\sigma$ of the offsets are a factor $\sim$1.5-to-7 smaller when using the LCTs. Figures~\ref{fig:distance_single_pli} and ~\ref{fig:distance_single_plz} show a decreasing trend of the ratio with decreasing photometric error. We also note that the <$\delta d_{[L/s](n,\sigma)}$> are closer to zero when using LCTs with respect to the simple mean. By looking at the absolute values, using the LCTs, $\sigma$ drops below 1 kpc in all cases, and below $\sim$0.5 kpc when the photometric error is smaller than 0.050 mag. This means relative uncertainties smaller than $\sim$5-10\% in the two cases. As expected, when the period, $Ampl$ and reference epoch of the variable are well known, and one can assume that there was no phase shift or significant period change between the observations and the time series adopted to derive the pulsation properties, applying the LCT is always an advantage.

\subsubsection{Distances from PL(i) and PL(z); LCT fitting at fixed amplitude}

We have checked the behavior of distances obtained from PL($i$) and PL($z$), using also the mean magnitudes derived with the LCT fitting at fixed $Ampl$. Fig.~\ref{fig:distance_amplfixed_pli} and ~\ref{fig:distance_amplfixed_plz} display the ratio of the standard deviations of the distance offsets in the two cases of LCT fitting and simple average. The trend with photometric error is the same as for the single-point anchoring case: with increasing photometric error, the LCTs are less convenient with respect to the simple average. Nonetheless, LCTs still provide a better precision in the majority of cases, at least if the number of phase points is larger than four, or the photometric error is smaller than 0.020 mag. We note that, for distances obtained with the PL($z$), the LCTs are not as effective as for PL($i$). This might be due 1) to the smaller number of LCTs for the $z$ band, meaning that the variability of light curve morphology cannot be as well sampled as for the $g$, $r$ and $i$ bands; 2) to the fact that we do not have $z$-band RRc LCTs: RRc are easier to fit due to their almost sinusoidal light curves and not having them might worsen the average ratios of the $z$ band.

\subsubsection{Distances from PW(r,g-r); LCT fitting at free amplitude}

Fig.~\ref{fig:distance_amplfree} displays the ratios of offsets obtained by estimating the distances with the PW($r$,$g-r$) for Halo stars. In this case, the improvement in the standard deviations can be as high as a factor 12, with the most significant improvements obtained when the photometric error is small. There are a few cases in which the precision on the distance estimate is generally worse when using the template, more specifically when only four points are available and the photometric error is larger than 0.010 mag. This is consistent with what we found for <$mag$> offsets. We therefore advice to adopt the LCT fitting with free $Ampl$ when more than four points are available, or when the photometric error is 0.01 mag or lower.

\section{Summary and final remarks}\label{sect_discussion}

We have built LCTs of RRLs in the $gri$ LSST passbands for both RRab and RRc type stars, plus LCTs in the $z$ LSST passband for RRab. These were conceived to enhance the science performed with early data from the \textit{Rubin} Observatory, improving the mean magnitude and distance estimate or RRLs, that are valuable Galactic structure and formation probes, since they trace the very old population \citep{layden94,fabrizio2019,iorio2021}.

In total, we delivered a set of 37 LCTs in analytical form, providing the coefficients of the Fourier series (up to the 40$^{th}$ order) representing them. The LCTs of RRab highlight, as expected, the known behavior of this pulsation type, with light curves that are more and more asymmetrical and with steeper rising branches, at bluer bands and shorter periods. On the other hand, RRc LCTs display the well-known almost sinusoidal morphology, with either a hump or a flattening before the maximum of pulsation.

We have tested our LCTs using three different methods: 1) the classical LCT anchoring to a single observed phase point \citep{jones1996}; 2) the LCT fitting, where, if at least three observed phase points are available, one can use the LCT as a fitting function, minimizing the $\chi^2$ on two free parameters (phase offset and magnitude offset) \citep{braga2019}; 3) the LCT fitting with free amplitude, that is a brand new approach, proposed in this paper: if at least four observed phase points are available, one can use the LCT as a fitting function, minimizing the $\chi^2$ on three free parameters (phase offset, magnitude offset and the amplitude of the LCT). The latter will be particularly helpful for the main applications of these LCTs, namely, early science of newly-discovered RRLs by LSST. In the first one-two years, for hundreds of thousands previously unknown RRLs, LSST will provide reliable periods but, since these will be new candidates, one will have no information on their pulsation properties, including the amplitude. Therefore, being able to apply LCTs without the need of knowing the amplitude, will mean to have a sample of RRLs with precise distances at least a factor of 2-3 larger, on which to base any Galactic archaeology investigation.

We checked that, in almost all cases, the LCTs do provide better mean magnitudes and distances, but there are a few exceptions. More specifically, when either the number of observations is very low, meaning around four, or the LSST photometric uncertainty is large ($\gtrsim$ 0.05 mag), one should pay attention and consider just calculating the mean magnitudes as simple flux averages. This means that, in the very first months of the survey, when only 3-5 observations per target will be available, we recommend to apply the templates only on objects brighter than $r\sim$23 mag. However, starting with DR1, we recommend to adopt LCTs on all available targets, based on our quantitative validation.

We also report an interesting finding. The LCTs in this work were derived for the LSST photometric system, but using data in the ZTF and DECam photometric systems. Thus, an important part of our work was the comparison of RRL light curves in these three systems, especially concerning the amplitude ratios (See Appendix~\ref{sect:amplratio_appendix}). These are crucial for two reasons. First of all, our investigation of the amplitude ratios allowed us to quantitatively validate the usability of our LCTs with \textit{Rubin} Observatory data. Secondly, once certified that ZTF, DECam and LSST amplitudes are the same within 1$\sigma$, we compared the $griz_Z$ amplitudes also with \gaia $G$ and OGLE ($VI$) amplitudes, to provide amplitude ratios to be used for the conversion between the amplitudes of known RRLs in these large surveys, into $griz_L$ amplitudes. This work has unveiled some interesting results. On one side, the ratios $\dfrac{Ampl(x_Z)}{Ampl(I)}$ display a behavior similar to that of the RRLs of $\omega$ Cen \citep{braga2018}, with RRc stars having a higher ratio than RRab in bluer bands ($g_Zr_Z$), and a similar one at redder bands ($i_Zz_D$). However, we also found an unexpected and completely new behavior concerning the ratios $\dfrac{Ampl(I)}{Ampl(V)}$ of RRab. We found that, for Bulge RRLs, this ratio follows a positive trend with the period, while in the LMC and SMC the ratio is constant. Both blending and metallicity might be playing a role but it is not within the aim of this paper to investigate this feature.

Finally, let us mention that we are aware that another group is submitting a similar work (Baeza-Villagra, in prep.). However, our works are completely independent, based on different datasets and our LCTs have different purposes.

\begin{acknowledgements}
Based on observations obtained with the \textit{Samuel Oschin} 48-inch Telescope at the Palomar Observatory as part of the Zwicky Transient Facility project. ZTF is supported by the National Science Foundation under Grant No. AST-1440341 and a collaboration including Caltech, IPAC, the Weizmann Institute for Science, the Oskar Klein Center at Stockholm University, the University of Maryland, the University of Washington, Deutsches Elektronen-Synchrotron and Humboldt University, Los Alamos National Laboratories, the TANGO Consortium of Taiwan, the University of Wisconsin at Milwaukee, and Lawrence Berkeley National Laboratories. Operations are conducted by COO, IPAC, and UW.

V.F.B. acknowledges the INAF project ``Participation in LSST - Large Synoptic Survey Telescope'' (LSST inkind contribution ITA-INA-S22, PI: G. Fiorentino), OB.FU. 1.05.03.06.

M.Mo. and V.F.B acknowledge financial support from the ACIISI, Consejer\'ia de Econom\'ia, Conocimiento y Empleo del Gobierno de
Canarias and the European Regional Development Fund (ERDF) under the grant with reference ProID2021010075.

M.Mo. acknowledges support from Spanish Ministry of Science, Innovation and Universities (MICIU) through the Spanish State Research Agency under the grants "RR Lyrae stars, a lighthouse to distant galaxies and early galaxy evolution" and the European Regional Development Fun (ERDF) with reference PID2021-127042OB-I00 and from the Severo Ochoa Programe 2020-2023 (CEX2019-000920-S).

M.D.O. achknowledges the INAF GO Project: ``The GAlactic bulGE with pleiadi'' (PI: M. Dall'Ora), OB.FU.: 1.05.23.05.24.

M.Ma. acknowledges Project PRIN MUR 2022 (code 2022ARWP9C) ``Early Formation and Evolution of Bulge and HalO (EFEBHO)'', PI: Marconi, M., funded by European Union – Next Generation EU and Large grant INAF 2023 MOVIE (PI: M. Marconi).

C.G. acknowledges support from AEI-MCINN under grant ``At the forefront of Galactic Archaeology: evolution of the luminous and dark matter components of the Milky Way and Local Group dwarf galaxies in the \gaia era'' with reference PID2020-118778GB-I00/10.13039/501100011033 and from the Spanish Ministry of Science, Innovation and University (MICIN) through the Spanish State Research Agency, under Severo Ochoa Centres of Excellence Programme 2020-2023 (CEX2019-000920-S).

R.S. acknowledges SNN-147362 and the KKP-137523 ‘Seismo-Lab’ \'Elvonal grants of the Hungarian Research, Development and Innovation Office (NKFIH).

This work was also supported by the NKFIH excellence grant TKP2021-NKTA-64

\end{acknowledgements}

\bibliographystyle{aa}
\bibliography{ms}

\begin{appendix}

\section{Additional material}\label{sect:addditional}

Hereby, we display the tables and figures that are too large to be put within the text and would severely affect the readability of the paper.

\begin{sidewaystable}
\caption{Pulsation, fitting and light curve properties of the ZTF $gri$ light curves of RRLs}
\scriptsize
\begin{tabular}{rrlrrlrrrrrrrrrrrrrrrr}
\hline\\
 ZTF ID &  \gaia DR3 Source ID & Mode &       RA &     Dec & band &   $P$ &   $mag$ &  $Ampl$ &       \trisv &   n &  $\chi^2$ &  $\chi_A^2$ &  $\sigma\delta$ &    $r$ &  $r_A$ &  $U$ &  $U_{bin}$ &  $\Delta\phi_{max}$ &  $Sk$ &  $Ku$ &  $N_F$ \\  
 &   &  &       [deg] &     [deg] &  &   [days] &   [mag] &  [mag] &       [days] &    &  [mag] &   &  [mag$^2$] &    [mag$^2$] &  [mag] &   &   &   &   &   &   \\  
 &&&&&&&&&&& $A_0$ &  $\phi_0$ & $A_1$ &  $\phi_1$ &   $A_2$ &  $\phi_2$ &  $out_1$ &  $out_3$ &  $out_5$ &  $out_{10}$ &  rej \\
 &&&&&&&&&&& [mag] &  & [mag] &  &   [mag] &  &  &  &  &  &  \\
 \hline\\
1596106200001713 & 2847261780479880704 &  RRc &   1.4031 & 22.1126 &   zr & 0.326170 & 16.378 &     0.304 & 2458334.8832 &  84 &    1.438 &      4.729 &          0.0132 & 0.0002 &  0.0007 & 0.102 &     0.022 &            0.080 &    0.054 &    --1.621 &       4 \\
&&&&&&&&&&&  0.153 &    1.574 &  0.027 &    1.331 &  0.016 &    1.170 &     0.5000 &     0.2381 &     0.1548 &      0.0595 &       0 \\
1596115400001888 & 2850150816001358976 & RRab &   0.9086 & 25.0989 &   zg & 0.629568 & 16.761 &     0.999 & 2458337.9152 &  80 &   48.661 &     48.692 &          0.0799 & 0.0079 &  0.0079 & 0.112 &     0.057 &            0.085 &    --0.782 &    --0.643 &       6 \\
&&&&&&&&&&&  0.354 &    --4.408 &  0.158 &    1.308 &  0.122 &    1.227 &     0.5000 &     0.3250 &     0.2375 &      0.1500 &       1 \\
1596115400001888 & 2850150816001358976 & RRab &   0.9086 & 25.0989 &   zr & 0.629585 & 16.542 &     0.743 & 2458335.4038 &  82 &    8.235 &     11.080 &          0.0335 & 0.0013 &  0.0018 & 0.077 &     0.025 &            0.084 &    --0.682 &    --0.708 &       5 \\
&&&&&&&&&&&  0.234 &    --4.460 &  0.116 &    1.498 &  0.093 &    1.463 &     0.4938 &     0.2840 &     0.1235 &      0.0247 &       0 \\
1642116100000085 & 1881392148928780672 & RRab & 336.1911 & 26.2265 &   zr & 0.632368 & 16.489 &     0.338 & 2458311.4142 & 153 &    2.285 &      6.766 &          0.0185 & 0.0004 &  0.0011 & 0.063 &     0.020 &            0.042 &    --0.045 &    --1.186 &       3 \\
&&&&&&&&&&&  0.139 &    1.451 &  0.052 &    0.978 &  0.025 &    0.767 &     0.4967 &     0.2026 &     0.0980 &      0.0261 &       0 \\
1644110400001540 & 2827873164235514368 & RRab & 355.1114 & 23.2773 &   zg & 0.523223 & 17.583 &     1.318 & 2458314.2309 &  82 &    1.966 &      1.491 &          0.0235 & 0.0007 &  0.0006 & 0.116 &     0.050 &            0.072 &    --0.633 &    --1.299 &       9 \\
&&&&&&&&&&& ---0.443 &    --1.094 &  0.209 &    1.576 &  0.158 &    1.394 &     0.4938 &     0.2593 &     0.1728 &      0.0741 &       0 \\
1644111400003784 & 2828004727673530880 & RRab & 353.4258 & 23.0716 &   zr & 0.617585 & 15.808 &     0.921 & 2458326.9733 &  82 &  169.112 &    183.558 &          0.1127 & 0.0161 &  0.0175 & 0.081 &     0.044 &            0.073 &    --0.901 &    --0.298 &       7 \\
&&&&&&&&&&&  0.285 &    --4.449 &  0.145 &    1.450 &  0.104 &    1.575 &     0.5000 &     0.2317 &     0.1707 &      0.0854 &       1 \\
1644113200000180 & 2852854167201887616 & RRab & 356.1166 & 26.2224 &   zg & 0.653005 & 17.433 &     0.758 & 2458314.0938 & 154 &    7.160 &      9.448 &          0.0507 & 0.0030 &  0.0040 & 0.073 &     0.022 &            0.056 &    --0.631 &    --0.953 &       9 \\
&&&&&&&&&&&  0.265 &    --4.376 &  0.126 &    1.373 &  0.071 &    1.344 &     0.4967 &     0.3179 &     0.2384 &      0.1060 &       0 \\
1644113200000180 & 2852854167201887616 & RRab & 356.1166 & 26.2224 &   zr & 0.653124 & 17.110 &     0.534 & 2458334.9412 & 161 &    1.822 &      3.414 &          0.0210 & 0.0005 &  0.0009 & 0.065 &     0.025 &            0.049 &    --0.466 &    --1.066 &       5 \\
&&&&&&&&&&& ---0.192 &    --1.133 & ---0.088 &    --1.223 & ---0.056 &    --0.901 &     0.4969 &     0.3043 &     0.1801 &      0.0745 &       0 \\
1645110200000706 & 2861854086487076352 & RRab &   1.7327 & 31.4703 &   zg & 0.457301 & 16.687 &     1.181 & 2458315.1761 &  81 &   52.432 &     44.385 &          0.0800 & 0.0087 &  0.0073 & 0.109 &     0.076 &            0.077 &    --1.020 &    --0.405 &       9 \\
&&&&&&&&&&&  0.456 &    --4.293 & ---0.178 &    4.580 &  0.119 &    1.407 &     0.4938 &     0.2346 &     0.1358 &      0.1111 &       1 \\
1645110200000706 & 2861854086487076352 & RRab &   1.7327 & 31.4703 &   zr & 0.457299 & 16.546 &     0.813 & 2458327.0632 &  86 &   26.564 &     32.675 &          0.0509 & 0.0032 &  0.0040 & 0.090 &     0.047 &            0.107 &    --0.587 &    --1.085 &       7 \\
&&&&&&&&&&&  0.327 &    --4.359 &  0.139 &    1.418 &  0.078 &    1.147 &     0.5000 &     0.2442 &     0.1977 &      0.1047 &       1 \\
\hline\\
\end{tabular}
\tablefoot{The parameters are those already presented in Table~\ref{tab:predictors} plus a few more: $N_F$: the degree of the Fourier series fitting the data; $\sigma\delta$: the standard deviation of the residuals from the fit; $N_F$: the degree of the Fourier series fitting the data. Only the first ten lines are displayed. The full table is available on the CDS.}
\label{tab:ztf}
\end{sidewaystable}

\begin{sidewaystable*}
\caption{Coefficients of the LCTs and parameters used to derive them.}
\footnotesize
\begin{tabular}{rllrrrrrrrrrrrrrrrrrrrlr}\\
\hline\\
LCT ID & Mode & Filter & $P_{lo}$ & $P_{hi}$ &  $d_F$ &  $n_{RRLs}$ &  $n_{\phi p}$ &  $n_{pix(x)}$ &  $n_{pix(y)}$ &  $n_{bins}$ &    $c$ &  $A_{0}$ &  $\phi_{0}$ &  $A_{1}$ &  $\phi_{1}$ &  $A_{2}$ &  $\phi_{2}$ &  $A_{3}$ &  $\phi_{3}$ & $A_{4}$ &  $\phi_{4}$ &  ... &  $\sigma$ \\
\hline\\    0 &  RRc &      g &     0.23 &     0.31 &      11 &        1065 &        353886 &           200 &           300 &          10 & 0.0325 &  0.5076 &  7.9305 &  --0.0877 &      4.8006 &   0.0389 &     14.1241 &   0.0297 &     14.4088 &  --0.0202 &     11.0263 & \ldots &    0.0002 \\     1 &  RRc &      g &     0.31 &     0.39 &       9 &        1966 &        705748 &           200 &           300 &          10 & 0.0301 &  0.5093 &  7.8771 &   0.0629 &      1.6244 &  --0.0342 &     --1.3052 &  --0.0170 &     --0.9824 &  --0.0099 &     --1.0794 & \ldots &    0.0001 \\     2 &  RRc &      g &     0.39 &     0.43 &      10 &          29 &         11401 &           100 &           150 &          10 & 0.0335 &  0.5061 &  7.8882 &   0.0758 &     14.1954 &  --0.0395 &     --1.5560 &  --0.0139 &     --0.8323 &  --0.0074 &     --0.7143 & \ldots &    0.0008 \\     3 &  RRc &      r &     0.23 &     0.31 &      10 &         884 &        510279 &           200 &           300 &          10 & 0.0217 &  0.5017 &  7.8923 &   0.0881 &      7.9003 &  --0.0395 &     --1.5970 &  --0.0259 &     --1.3297 &   0.0181 &     14.2151 & \ldots &    0.0002 \\     4 &  RRc &      r &     0.31 &     0.39 &       9 &        1598 &        918102 &           200 &           300 &          10 & 0.0209 &  0.5035 &  1.5676 &   0.0669 &      1.6424 &  --0.0355 &     --1.2898 &  --0.0177 &     --1.0535 &  --0.0102 &     --1.0592 & \ldots &    0.0002 \\     5 &  RRc &      r &     0.39 &     0.43 &       8 &          28 &         16408 &           100 &           150 &          10 & 0.0179 &  0.4913 &  1.5993 &   0.0718 &      1.6391 &  --0.0390 &     --1.3667 &  --0.0142 &     --1.1682 &  --0.0075 &     --1.0144 & \ldots &    0.0017 \\     6 &  RRc &      i &     0.23 &     0.31 &      11 &          97 &         16553 &           100 &           100 &          10 & 0.0162 &  0.4894 &  7.8795 &   0.0918 &      7.9331 &  --0.0374 &      4.6312 &   0.0212 &     --4.4111 &  --0.0164 &     11.0853 & \ldots &    0.0006 \\     7 &  RRc &      i &     0.31 &     0.39 &       9 &         150 &         21592 &           100 &           100 &          10 & 0.0157 &  0.4966 &  7.8272 &   0.0763 &      1.6413 &   0.0340 &      1.8021 &  --0.0200 &     --1.2124 &  --0.0103 &     --0.9667 & \ldots &    0.0009 \\     8 &  RRc &      i &     0.39 &     0.43 &       5 &           5 &           755 &           100 &           150 &          10 & 0.0178 &  0.5003 &  1.5006 &   0.0680 &      1.7357 &  --0.0351 &     --1.2689 &  --0.0091 &     --1.2503 &  --0.0092 &      0.7470 & \ldots &    0.0031 \\     9 & RRab &      g &     0.35 &     0.41 &      15 &          52 &         15374 &           200 &           300 &          10 & 0.0616 & --0.3389 & --1.1482 &   0.1821 &      7.8568 &  --0.1187 &      4.6701 &   0.0739 &      1.4132 &   0.0505 &      1.3897 & \ldots &    0.0006 \\    10 & RRab &      g &     0.41 &     0.47 &      24 &         775 &        240306 &           300 &           500 &           8 & 0.0669 & --0.3546 & --1.0877 &   0.1625 &      7.8625 &  --0.1206 &      4.5749 &  --0.0765 &      4.5281 &   0.0549 &      1.2917 & \ldots &    0.0003 \\    11 & RRab &      g &     0.47 &     0.53 &      13 &        2307 &        745647 &           300 &           500 &          10 & 0.0587 & --0.3520 & --1.1051 &   0.1610 &      1.5699 &   0.1203 &      1.4363 &   0.0774 &      1.3586 &   0.0552 &      1.2981 & \ldots &    0.0003 \\    12 & RRab &      g &     0.53 &     0.59 &      39 &        3541 &       1178347 &           500 &           500 &           8 & 0.0472 & --0.3506 & --1.1752 &   0.1675 &      1.5006 &   0.1192 &      1.4052 &   0.0764 &      1.3379 &   0.0476 &      1.3296 & \ldots &    0.0002 \\    13 & RRab &      g &     0.59 &     0.62 &      29 &        1973 &        680613 &           200 &           300 &          10 & 0.0407 &  0.3641 & --4.3909 &  --0.1697 &      4.5764 &   0.1120 &      1.3334 &   0.0641 &      1.3059 &  --0.0311 &      4.4693 & \ldots &    0.0004 \\    14 & RRab &      g &     0.62 &     0.65 &      19 &        1447 &        509742 &           200 &           300 &          10 & 0.0351 &  0.3819 & --4.4409 &  --0.1649 &      4.5346 &   0.0989 &      1.2961 &  --0.0492 &      4.4970 &  --0.0228 &      4.6438 & \ldots &    0.0003 \\    15 & RRab &      g &     0.65 &     0.71 &      24 &        1140 &        361019 &           200 &           300 &          10 & 0.0382 &  0.3697 & --4.4380 &  --0.1710 &      4.5887 &  --0.1072 &      4.4831 &  --0.0563 &      4.5137 &  --0.0277 &      4.5069 & \ldots &    0.0002 \\    16 & RRab &      g &     0.71 &     0.77 &      15 &         415 &        142495 &           200 &           300 &          10 & 0.0324 &  0.3811 &  1.7914 &   0.1711 &      1.4100 &   0.1002 &      1.3277 &   0.0453 &      1.3965 &   0.0206 &      1.4844 & \ldots &    0.0004 \\    17 & RRab &      g &     0.77 &     0.83 &      14 &         108 &         35247 &           100 &           150 &           5 & 0.0285 &  0.3988 &  1.7084 &   0.1691 &      1.3799 &   0.0875 &      1.3584 &   0.0320 &      1.5004 &  --0.0162 &     --1.3002 & \ldots &    0.0009 \\    18 & RRab &      r &     0.35 &     0.41 &      15 &          44 &         29142 &           200 &           300 &          10 & 0.0380 &  0.3274 & --4.3660 &  --0.1764 &      4.7058 &  --0.1167 &      4.7007 &   0.0754 &      1.4829 &   0.0516 &      1.4573 & \ldots &    0.0005 \\    19 & RRab &      r &     0.41 &     0.47 &      17 &         566 &        301516 &           200 &           300 &           6 & 0.0414 &  0.3376 & --4.3137 &   0.1618 &      1.5795 &   0.1209 &      1.4609 &   0.0774 &      1.4057 &   0.0548 &      1.3229 & \ldots &    0.0003 \\    20 & RRab &      r &     0.47 &     0.53 &      15 &        1763 &        905847 &           300 &           500 &           6 & 0.0370 &  0.3350 & --4.3438 &   0.1587 &      1.5582 &   0.1203 &      1.4570 &   0.0792 &      1.3892 &   0.0549 &      1.3525 & \ldots &    0.0003 \\    21 & RRab &      r &     0.53 &     0.59 &      23 &        3214 &       1611518 &           300 &           500 &           7 & 0.0302 &  0.3388 & --4.4381 &   0.1653 &      1.4862 &   0.1177 &      1.4334 &   0.0749 &      1.4047 &   0.0451 &      1.4067 & \ldots &    0.0004 \\    22 & RRab &      r &     0.59 &     0.62 &      24 &        1637 &        857507 &           300 &           500 &          10 & 0.0258 &  0.3534 & --4.5183 &   0.1668 &      1.4401 &   0.1093 &      1.4047 &   0.0613 &      1.4409 &   0.0309 &      1.5122 & \ldots &    0.0002 \\    23 & RRab &      r &     0.62 &     0.65 &      24 &        1165 &        610347 &           300 &           500 &          10 & 0.0246 &  0.3670 &  1.7271 &   0.1635 &      1.4276 &   0.1000 &      1.3981 &   0.0517 &      1.4859 &  --0.0242 &     --1.5177 & \ldots &    0.0002 \\    24 & RRab &      r &     0.65 &     0.71 &      19 &         927 &        471496 &           300 &           800 &          10 & 0.0264 &  0.3540 &  1.7284 &   0.1692 &      1.4635 &   0.1080 &      1.4105 &   0.0578 &      1.4763 &   0.0278 &      1.5097 & \ldots &    0.0003 \\    25 & RRab &      r &     0.71 &     0.77 &      15 &         322 &        171344 &           200 &           300 &          10 & 0.0217 &  0.3705 &  1.6710 &   0.1685 &      1.4270 &   0.0998 &      1.4328 &   0.0451 &      1.5489 &  --0.0224 &     --1.4270 & \ldots &    0.0004 \\    26 & RRab &      r &     0.77 &     0.83 &      15 &          76 &         43816 &           150 &           200 &           5 & 0.0214 &  0.3922 &  7.8551 &   0.1654 &      1.4073 &  --0.0854 &     --1.6414 &   0.0330 &      7.9369 &  --0.0179 &     --1.0887 & \ldots &    0.0006 \\    27 & RRab &      i &     0.41 &     0.47 &      13 &          59 &         11410 &           200 &           300 &          10 & 0.0318 &  0.3309 & --4.3774 &  --0.1568 &      4.7323 &  --0.1188 &      4.6112 &   0.0777 &      1.4201 &   0.0524 &      1.3882 & \ldots &    0.0007 \\    28 & RRab &      i &     0.47 &     0.53 &      15 &         205 &         39002 &           200 &           300 &          10 & 0.0276 &  0.3254 & --4.4197 &  --0.1529 &      4.7035 &  --0.1193 &     --1.6880 &   0.0801 &      1.4150 &   0.0546 &      1.3742 & \ldots &    0.0005 \\    29 & RRab &      i &     0.53 &     0.59 &      24 &         336 &         57637 &           200 &           300 &          10 & 0.0215 &  0.3291 & --4.5260 &   0.1602 &      1.4887 &   0.1171 &      1.4645 &   0.0758 &      1.4787 &   0.0467 &      1.5479 & \ldots &    0.0004 \\    30 & RRab &      i &     0.59 &     0.65 &      21 &         265 &         47278 &           150 &           200 &           8 & 0.0212 &  0.3432 &  1.6794 &   0.1629 &      1.4660 &   0.1111 &      1.4796 &   0.0664 &      1.5379 &  --0.0335 &     --1.4866 & \ldots &    0.0004 \\    31 & RRab &      i &     0.65 &     0.71 &      22 &          83 &         13454 &            80 &           100 &           8 & 0.0199 &  0.3478 &  7.9238 &  --0.1679 &      4.6042 &   0.1097 &      1.4898 &   0.0620 &      1.5955 &  --0.0311 &     --1.4483 & \ldots &    0.0006 \\    32 & RRab &      i &     0.71 &     0.83 &      11 &          40 &          6821 &           180 &           250 &           6 & 0.0203 &  0.3644 &  1.5620 &   0.1644 &      1.4323 &   0.0989 &      1.5269 &  --0.0503 &     --1.3911 &  --0.0228 &     --1.1315 & \ldots &    0.0014 \\    33 & RRab &      z &     0.35 &     0.47 &       9 &          50 &          2828 &           100 &           100 &           6 & 0.0262 &  0.3343 & --4.4958 &   0.1634 &      1.5069 &   0.1062 &      1.4856 &   0.0646 &      1.3999 &   0.0425 &      1.3598 & \ldots &    0.0018 \\    34 & RRab &      z &     0.47 &     0.59 &       9 &          99 &          5925 &           100 &           100 &           6 & 0.0208 &  0.3320 & --4.5818 &   0.1781 &      1.4676 &   0.1110 &      1.5030 &   0.0688 &      1.5040 &  --0.0358 &      4.6322 & \ldots &    0.0016 \\    35 & RRab &      z &     0.59 &     0.71 &       9 &          60 &          3338 &           100 &           100 &           6 & 0.0182 &  0.3534 &  1.5611 &   0.1721 &      1.5094 &   0.0994 &      1.5917 &   0.0488 &      1.6774 &  --0.0258 &     --1.4412 & \ldots &    0.0022 \\    36 & RRab &      z &     0.71 &     0.83 &       5 &           8 &           451 &            55 &            55 &           6 & 0.0221 &  0.3720 &  1.5138 &   0.1812 &      1.6073 &  --0.0902 &     --1.2868 &  --0.0304 &     --1.4031 &  --0.0186 &     --0.8477 & \ldots &    0.0050 \\
\hline\\
\end{tabular}
\tablefoot{$P_{lo}$ and $P_{hi}$ indicate the period thresholds of the bins, $d_F$ is the degree of the Fourier series, $n_{RRLs}$ is the number of RRLs used to derive. For visualization purposes, only the coefficients of the first four orders are displayed. The full table is available in machine-readable format.}
\label{tab:templates}
\end{sidewaystable*}

\begin{sidewaystable}
\caption{Averages and standard deviations of the $\delta_{L(n,\sigma)}$ with fixed amplitudes.}
\scriptsize
\begin{tabular}{rrrrrrrrrrrrrrrr}
\hline
\hline
LCT ID & <$\delta$>$_{L(4; 0.005)}$ & <$\delta$>$_{L(4; 0.010)}$ & <$\delta$>$_{L(4; 0.020)}$ & <$\delta$>$_{L(4; 0.050)}$ & <$\delta$>$_{L(4; 0.100)}$ & <$\delta$>$_{L(8; 0.005)}$ & <$\delta$>$_{L(8; 0.010)}$ & <$\delta$>$_{L(8; 0.020)}$ & <$\delta$>$_{L(8; 0.050)}$ & <$\delta$>$_{L(8; 0.100)}$ & <$\delta$>$_{L(12; 0.005)}$ & <$\delta$>$_{L(12; 0.010)}$ & <$\delta$>$_{L(12; 0.020)}$ & <$\delta$>$_{L(12; 0.050)}$ & <$\delta$>$_{L(12; 0.100)}$ \\
\hline
0 &  0.001$\pm$0.021 &  0.002$\pm$0.021 &  0.000$\pm$0.053 & -0.004$\pm$0.063 &  0.007$\pm$0.094 &  0.002$\pm$0.011 &  0.002$\pm$0.012 &  0.003$\pm$0.013 &  0.003$\pm$0.021 &  0.000$\pm$0.037 &  0.002$\pm$0.007 &  0.002$\pm$0.007 &  0.001$\pm$0.009 &  0.002$\pm$0.018 &  0.002$\pm$0.031 \\
1 &  0.000$\pm$0.011 &  0.003$\pm$0.023 &  0.002$\pm$0.032 & -0.001$\pm$0.057 & -0.008$\pm$0.093 &  0.000$\pm$0.007 &  0.000$\pm$0.008 &  0.001$\pm$0.010 &  0.001$\pm$0.020 & -0.004$\pm$0.042 &  0.000$\pm$0.004 & -0.001$\pm$0.005 &  0.000$\pm$0.007 &  0.001$\pm$0.016 &  0.001$\pm$0.030 \\
2 &  0.005$\pm$0.032 & -0.002$\pm$0.032 &  0.005$\pm$0.047 &  0.000$\pm$0.072 &  0.009$\pm$0.096 &  0.005$\pm$0.012 &  0.004$\pm$0.013 &  0.005$\pm$0.014 &  0.003$\pm$0.022 &  0.000$\pm$0.043 &  0.004$\pm$0.008 &  0.004$\pm$0.009 &  0.005$\pm$0.011 &  0.003$\pm$0.017 &  0.004$\pm$0.031 \\
3 &  0.000$\pm$0.024 &  0.001$\pm$0.020 & -0.005$\pm$0.035 & -0.002$\pm$0.063 &  0.000$\pm$0.083 &  0.001$\pm$0.005 &  0.000$\pm$0.006 &  0.000$\pm$0.010 &  0.002$\pm$0.022 &  0.001$\pm$0.039 &  0.000$\pm$0.004 &  0.000$\pm$0.004 &  0.000$\pm$0.007 &  0.002$\pm$0.015 & -0.002$\pm$0.035 \\
4 & -0.002$\pm$0.017 & -0.001$\pm$0.012 & -0.001$\pm$0.034 & -0.001$\pm$0.059 &  0.002$\pm$0.086 & -0.001$\pm$0.005 & -0.001$\pm$0.006 &  0.000$\pm$0.008 & -0.001$\pm$0.021 & -0.006$\pm$0.046 & -0.001$\pm$0.004 & -0.001$\pm$0.004 & -0.001$\pm$0.007 & -0.001$\pm$0.015 &  0.002$\pm$0.032 \\
5 &  0.001$\pm$0.010 &  0.000$\pm$0.020 &  0.003$\pm$0.034 &  0.001$\pm$0.058 &  0.004$\pm$0.092 &  0.000$\pm$0.004 &  0.001$\pm$0.006 &  0.000$\pm$0.009 &  0.000$\pm$0.020 &  0.001$\pm$0.042 &  0.000$\pm$0.003 &  0.000$\pm$0.004 &  0.001$\pm$0.007 &  0.000$\pm$0.017 &  0.004$\pm$0.032 \\
6 &  0.001$\pm$0.014 &  0.000$\pm$0.017 &  0.006$\pm$0.029 &  0.001$\pm$0.045 & -0.002$\pm$0.066 &  0.001$\pm$0.006 &  0.001$\pm$0.007 &  0.000$\pm$0.010 &  0.002$\pm$0.022 &  0.003$\pm$0.041 &  0.001$\pm$0.005 &  0.001$\pm$0.005 &  0.002$\pm$0.007 &  0.001$\pm$0.016 &  0.002$\pm$0.029 \\
7 & -0.001$\pm$0.017 &  0.000$\pm$0.028 &  0.002$\pm$0.038 &  0.002$\pm$0.052 & -0.001$\pm$0.068 &  0.001$\pm$0.008 &  0.001$\pm$0.010 &  0.001$\pm$0.012 &  0.000$\pm$0.022 &  0.003$\pm$0.040 &  0.001$\pm$0.007 &  0.001$\pm$0.007 &  0.001$\pm$0.010 &  0.000$\pm$0.017 &  0.000$\pm$0.030 \\
8 &  0.000$\pm$0.010 &  0.001$\pm$0.012 & -0.001$\pm$0.032 &  0.006$\pm$0.052 & -0.003$\pm$0.078 &  0.000$\pm$0.005 &  0.001$\pm$0.005 & -0.001$\pm$0.010 & -0.002$\pm$0.022 &  0.002$\pm$0.038 &  0.001$\pm$0.004 &  0.000$\pm$0.004 &  0.000$\pm$0.007 &  0.002$\pm$0.016 &  0.002$\pm$0.030 \\
9 &  0.128$\pm$0.124 &  0.133$\pm$0.128 &  0.138$\pm$0.140 &  0.136$\pm$0.148 &  0.128$\pm$0.158 &  0.113$\pm$0.087 &  0.111$\pm$0.092 &  0.103$\pm$0.094 &  0.108$\pm$0.091 &  0.098$\pm$0.092 &  0.083$\pm$0.076 &  0.091$\pm$0.073 &  0.083$\pm$0.072 &  0.095$\pm$0.074 &  0.088$\pm$0.080 \\
\hline
\end{tabular}
\tablefoot{Averages derived from each set of 100 resampled light curves, with <$mag$> derived using the fixed-amplitude LCT fit. Only the first ten lines are displayed. The full version of this table is available in electronic format.}
\label{tab:deltamag_fixedamplitude}
\end{sidewaystable}

\begin{sidewaystable}
\caption{Averages and standard deviations of the $\delta_{A(n,\sigma)}$ with fixed amplitudes.}
\scriptsize
\begin{tabular}{rrrrrrrrrrrrrrrr}
\hline
\hline
LCT ID & <$\delta$>$_{A(4; 0.005)}$ & <$\delta$>$_{A(4; 0.010)}$ & <$\delta$>$_{A(4; 0.020)}$ & <$\delta$>$_{A(4; 0.050)}$ & <$\delta$>$_{A(4; 0.100)}$ & <$\delta$>$_{A(8; 0.005)}$ & <$\delta$>$_{A(8; 0.010)}$ & <$\delta$>$_{A(8; 0.020)}$ & <$\delta$>$_{A(8; 0.050)}$ & <$\delta$>$_{A(8; 0.100)}$ & <$\delta$>$_{A(12; 0.005)}$ & <$\delta$>$_{A(12; 0.010)}$ & <$\delta$>$_{A(12; 0.020)}$ & <$\delta$>$_{A(12; 0.050)}$ & <$\delta$>$_{A(12; 0.100)}$ \\
\hline
0 &  0.007$\pm$0.099 &  0.007$\pm$0.097 &  0.002$\pm$0.094 & -0.004$\pm$0.106 &  0.007$\pm$0.115 &  0.004$\pm$0.062 &  0.007$\pm$0.075 &  0.006$\pm$0.071 &  0.005$\pm$0.073 & -0.008$\pm$0.074 &  0.000$\pm$0.056 &  0.001$\pm$0.056 & -0.002$\pm$0.056 &  0.001$\pm$0.060 &  0.004$\pm$0.063 \\
1 &  0.010$\pm$0.089 &  0.003$\pm$0.093 &  0.007$\pm$0.084 &  0.002$\pm$0.087 & -0.002$\pm$0.094 &  0.005$\pm$0.059 &  0.000$\pm$0.060 & -0.003$\pm$0.064 &  0.000$\pm$0.064 & -0.009$\pm$0.068 &  0.005$\pm$0.046 &  0.004$\pm$0.051 &  0.003$\pm$0.050 &  0.001$\pm$0.054 &  0.000$\pm$0.056 \\
2 &  0.005$\pm$0.105 & -0.003$\pm$0.102 &  0.014$\pm$0.105 &  0.000$\pm$0.105 &  0.006$\pm$0.113 &  0.005$\pm$0.073 & -0.004$\pm$0.071 &  0.004$\pm$0.072 &  0.001$\pm$0.066 &  0.000$\pm$0.081 &  0.003$\pm$0.058 &  0.000$\pm$0.061 & -0.001$\pm$0.057 & -0.002$\pm$0.061 &  0.000$\pm$0.067 \\
3 &  0.004$\pm$0.069 & -0.002$\pm$0.066 & -0.009$\pm$0.069 &  0.001$\pm$0.070 &  0.001$\pm$0.083 & -0.002$\pm$0.046 &  0.000$\pm$0.050 &  0.001$\pm$0.045 &  0.004$\pm$0.048 &  0.000$\pm$0.055 &  0.001$\pm$0.037 &  0.002$\pm$0.038 &  0.000$\pm$0.039 &  0.000$\pm$0.042 & -0.005$\pm$0.050 \\
4 &  0.002$\pm$0.065 &  0.003$\pm$0.066 &  0.009$\pm$0.066 &  0.001$\pm$0.070 &  0.002$\pm$0.081 &  0.003$\pm$0.045 &  0.005$\pm$0.045 &  0.000$\pm$0.043 &  0.000$\pm$0.046 & -0.003$\pm$0.059 &  0.001$\pm$0.036 &  0.003$\pm$0.038 &  0.000$\pm$0.036 & -0.003$\pm$0.042 & -0.008$\pm$0.047 \\
5 &  0.005$\pm$0.065 &  0.009$\pm$0.061 &  0.005$\pm$0.063 &  0.000$\pm$0.066 &  0.000$\pm$0.087 &  0.003$\pm$0.047 & -0.002$\pm$0.048 &  0.002$\pm$0.045 &  0.000$\pm$0.051 & -0.004$\pm$0.060 &  0.000$\pm$0.038 &  0.000$\pm$0.036 &  0.002$\pm$0.039 &  0.000$\pm$0.041 & -0.001$\pm$0.049 \\
6 &  0.005$\pm$0.056 &  0.001$\pm$0.053 &  0.008$\pm$0.056 &  0.002$\pm$0.061 & -0.005$\pm$0.068 &  0.000$\pm$0.037 &  0.001$\pm$0.036 & -0.002$\pm$0.037 &  0.001$\pm$0.043 &  0.000$\pm$0.052 &  0.001$\pm$0.032 & -0.001$\pm$0.031 &  0.004$\pm$0.033 & -0.002$\pm$0.035 & -0.001$\pm$0.044 \\
7 & -0.002$\pm$0.052 &  0.003$\pm$0.057 &  0.005$\pm$0.057 &  0.003$\pm$0.061 & -0.006$\pm$0.071 &  0.002$\pm$0.038 &  0.001$\pm$0.040 & -0.002$\pm$0.040 &  0.002$\pm$0.044 & -0.001$\pm$0.050 &  0.002$\pm$0.036 &  0.001$\pm$0.031 &  0.000$\pm$0.035 & -0.001$\pm$0.036 & -0.007$\pm$0.041 \\
8 & -0.003$\pm$0.053 &  0.001$\pm$0.049 &  0.001$\pm$0.051 &  0.005$\pm$0.058 & -0.001$\pm$0.077 &  0.000$\pm$0.038 &  0.001$\pm$0.037 & -0.003$\pm$0.034 & -0.002$\pm$0.040 & -0.004$\pm$0.052 &  0.003$\pm$0.031 &  0.000$\pm$0.031 & -0.004$\pm$0.030 &  0.003$\pm$0.036 & -0.003$\pm$0.038 \\
9 &  0.011$\pm$0.265 &  0.024$\pm$0.265 &  0.033$\pm$0.241 &  0.023$\pm$0.264 &  0.021$\pm$0.261 &  0.029$\pm$0.180 &  0.026$\pm$0.192 &  0.014$\pm$0.203 &  0.020$\pm$0.181 &  0.006$\pm$0.192 & -0.006$\pm$0.156 &  0.009$\pm$0.149 & -0.003$\pm$0.156 &  0.019$\pm$0.154 &  0.005$\pm$0.160 \\
\hline
\end{tabular}
\tablefoot{Averages derived from each set of 100 resampled light curves, with <$mag$> derived using the fixed-amplitude LCT fit. Only the first ten lines are displayed. The full version of this table is available in electronic format.}
\label{tab:deltamag_fixedamplitude_simplemean}
\end{sidewaystable}

\begin{sidewaystable}
\caption{Averages and standard deviations of the $\delta_{L(n,\sigma)}$ with free amplitudes.}
\scriptsize
\begin{tabular}{rrrrrrrrrrrrrrrr}
\hline
\hline
LCT ID & <$\delta$>$_{L(4; 0.005)}$ & <$\delta$>$_{L(4; 0.010)}$ & <$\delta$>$_{L(4; 0.020)}$ & <$\delta$>$_{L(4; 0.050)}$ & <$\delta$>$_{L(4; 0.100)}$ & <$\delta$>$_{L(8; 0.005)}$ & <$\delta$>$_{L(8; 0.010)}$ & <$\delta$>$_{L(8; 0.020)}$ & <$\delta$>$_{L(8; 0.050)}$ & <$\delta$>$_{L(8; 0.100)}$ & <$\delta$>$_{L(12; 0.005)}$ & <$\delta$>$_{L(12; 0.010)}$ & <$\delta$>$_{L(12; 0.020)}$ & <$\delta$>$_{L(12; 0.050)}$ & <$\delta$>$_{L(12; 0.100)}$ \\
\hline
0  & -0.002$\pm$0.030 & 0.000$\pm$0.035 & -0.001$\pm$0.067 & -0.010$\pm$0.102 & -0.011$\pm$0.129 &  0.000$\pm$0.008 & -0.001$\pm$0.008 & 0.000$\pm$0.014 & -0.002$\pm$0.027 & 0.001$\pm$0.046 &  0.000$\pm$0.006 &  0.000$\pm$0.006 & -0.001$\pm$0.010 & 0.002$\pm$0.018 & -0.001$\pm$0.032 \\
1  & 0.001$\pm$0.027 & -0.005$\pm$0.047 & -0.005$\pm$0.061 & -0.001$\pm$0.087 & 0.001$\pm$0.126 & 0.000$\pm$0.007 &  0.000$\pm$0.009 & 0.001$\pm$0.011 &  0.000$\pm$0.028 &  0.000$\pm$0.048 & 0.001$\pm$0.005 & 0.001$\pm$0.006 & 0.001$\pm$0.008 & 0.001$\pm$0.016 & 0.000$\pm$0.033 \\
2  & 0.000$\pm$0.032 &  0.000$\pm$0.032 & -0.004$\pm$0.065 & -0.003$\pm$0.076 & -0.004$\pm$0.123 & -0.001$\pm$0.007 &  0.000$\pm$0.008 & 0.000$\pm$0.012 & -0.002$\pm$0.025 & -0.007$\pm$0.049 & 0.000$\pm$0.005 &  0.000$\pm$0.006 & 0.000$\pm$0.007 & -0.001$\pm$0.015 & -0.004$\pm$0.039 \\
3  & 0.001$\pm$0.028 & -0.003$\pm$0.036 & -0.004$\pm$0.060 & -0.010$\pm$0.071 & -0.015$\pm$0.105 & 0.000$\pm$0.006 & 0.000$\pm$0.007 & 0.000$\pm$0.011 & 0.001$\pm$0.022 & 0.004$\pm$0.049 & 0.000$\pm$0.004 & 0.000$\pm$0.005 &  0.000$\pm$0.008 & 0.001$\pm$0.017 &  0.000$\pm$0.032 \\
4  & 0.000$\pm$0.028 & 0.002$\pm$0.031 & 0.004$\pm$0.053 & 0.002$\pm$0.071 & -0.012$\pm$0.100 &  0.000$\pm$0.005 & 0.000$\pm$0.008 & -0.001$\pm$0.010 & -0.001$\pm$0.023 & 0.001$\pm$0.047 & 0.000$\pm$0.004 &  0.000$\pm$0.004 &  0.000$\pm$0.007 & 0.001$\pm$0.017 & -0.004$\pm$0.034 \\
5  & 0.003$\pm$0.035 & 0.001$\pm$0.033 & -0.001$\pm$0.041 & -0.006$\pm$0.078 & -0.011$\pm$0.115 & 0.000$\pm$0.004 & 0.000$\pm$0.007 & 0.001$\pm$0.012 &  0.000$\pm$0.024 & -0.001$\pm$0.047 & 0.000$\pm$0.003 & 0.000$\pm$0.004 &  0.000$\pm$0.008 & -0.001$\pm$0.017 & -0.003$\pm$0.032 \\
6  & -0.002$\pm$0.032 & 0.000$\pm$0.028 & -0.002$\pm$0.038 & 0.001$\pm$0.071 & -0.005$\pm$0.091 &  0.000$\pm$0.004 & 0.000$\pm$0.006 & -0.001$\pm$0.011 &  0.000$\pm$0.026 & -0.007$\pm$0.045 & 0.000$\pm$0.003 &  0.000$\pm$0.004 & -0.001$\pm$0.007 & -0.002$\pm$0.016 & -0.002$\pm$0.037 \\
7  & 0.000$\pm$0.028 & -0.002$\pm$0.036 &  0.000$\pm$0.049 & 0.003$\pm$0.071 & -0.018$\pm$0.098 &  0.000$\pm$0.007 &  0.000$\pm$0.009 & -0.001$\pm$0.013 & -0.001$\pm$0.026 & -0.007$\pm$0.048 &  0.000$\pm$0.004 &  0.000$\pm$0.006 & 0.000$\pm$0.008 & 0.000$\pm$0.018 & -0.002$\pm$0.036 \\
8  & 0.001$\pm$0.019 & -0.001$\pm$0.026 & 0.001$\pm$0.042 & -0.005$\pm$0.061 & -0.018$\pm$0.096 & 0.000$\pm$0.004 &  0.000$\pm$0.008 &  0.000$\pm$0.009 & -0.001$\pm$0.022 & 0.001$\pm$0.045 & 0.000$\pm$0.003 & 0.000$\pm$0.004 &  0.000$\pm$0.007 & -0.001$\pm$0.016 & -0.001$\pm$0.031 \\
9  & 0.057$\pm$0.145 & 0.064$\pm$0.149 & 0.056$\pm$0.150 & 0.071$\pm$0.175 & 0.079$\pm$0.192 & 0.011$\pm$0.061 & 0.008$\pm$0.055 & 0.009$\pm$0.063 & 0.020$\pm$0.090 & 0.039$\pm$0.122 & 0.007$\pm$0.033 & 0.004$\pm$0.038 & 0.008$\pm$0.042 & 0.014$\pm$0.061 & 0.012$\pm$0.074 \\
\hline
\end{tabular}
\tablefoot{Averages derived from each set of 100 resampled light curves, with <$mag$> derived using the free-amplitude LCT fit. Only the first ten lines are displayed. The full version of this table is available in electronic format.}
\label{tab:deltamag_freeamplitude}
\end{sidewaystable}

\begin{sidewaystable}
\caption{Averages and standard deviations of the $\delta_{A(n,\sigma)}$ with free amplitudes.}
\scriptsize
\begin{tabular}{rrrrrrrrrrrrrrrr}
\hline
\hline
LCT ID & <$\delta$>$_{A(4; 0.005)}$ & <$\delta$>$_{A(4; 0.010)}$ & <$\delta$>$_{A(4; 0.020)}$ & <$\delta$>$_{A(4; 0.050)}$ & <$\delta$>$_{A(4; 0.100)}$ & <$\delta$>$_{A(8; 0.005)}$ & <$\delta$>$_{A(8; 0.010)}$ & <$\delta$>$_{A(8; 0.020)}$ & <$\delta$>$_{A(8; 0.050)}$ & <$\delta$>$_{A(8; 0.100)}$ & <$\delta$>$_{A(12; 0.005)}$ & <$\delta$>$_{A(12; 0.010)}$ & <$\delta$>$_{A(12; 0.020)}$ & <$\delta$>$_{A(12; 0.050)}$ & <$\delta$>$_{A(12; 0.100)}$ \\
\hline
0 &  0.011$\pm$0.092 &  0.005$\pm$0.103 & -0.003$\pm$0.096 &  0.008$\pm$0.103 & -0.007$\pm$0.115 &  0.000$\pm$0.065 &  0.002$\pm$0.066 &  0.004$\pm$0.068 &  0.004$\pm$0.072 &  0.000$\pm$0.077 & -0.004$\pm$0.053 & -0.002$\pm$0.056 & -0.002$\pm$0.052 &  0.007$\pm$0.058 & -0.003$\pm$0.063 \\
1 & -0.003$\pm$0.086 & -0.001$\pm$0.091 &  0.001$\pm$0.094 &  0.002$\pm$0.090 & -0.008$\pm$0.100 &  0.003$\pm$0.059 &  0.005$\pm$0.062 & -0.005$\pm$0.059 &  0.002$\pm$0.065 & -0.003$\pm$0.070 &  0.003$\pm$0.051 &  0.001$\pm$0.048 &  0.005$\pm$0.052 &  0.006$\pm$0.054 & -0.004$\pm$0.055 \\
2 &  0.013$\pm$0.105 &  0.004$\pm$0.103 & -0.006$\pm$0.102 &  0.008$\pm$0.103 &  0.010$\pm$0.115 & -0.002$\pm$0.070 &  0.001$\pm$0.069 &  0.004$\pm$0.071 & -0.001$\pm$0.074 &  0.001$\pm$0.082 &  0.002$\pm$0.056 & -0.002$\pm$0.057 &  0.001$\pm$0.059 & -0.004$\pm$0.062 & -0.004$\pm$0.069 \\
3 &  0.005$\pm$0.064 &  0.004$\pm$0.068 & -0.002$\pm$0.070 &  0.003$\pm$0.072 &  0.002$\pm$0.090 &  0.003$\pm$0.044 &  0.002$\pm$0.049 & -0.004$\pm$0.049 & -0.002$\pm$0.048 &  0.001$\pm$0.061 & -0.001$\pm$0.036 & -0.001$\pm$0.040 &  0.005$\pm$0.040 &  0.000$\pm$0.042 & -0.002$\pm$0.052 \\
4 & -0.003$\pm$0.067 & -0.003$\pm$0.066 &  0.005$\pm$0.066 & -0.010$\pm$0.066 & -0.005$\pm$0.076 & -0.003$\pm$0.043 &  0.001$\pm$0.046 &  0.008$\pm$0.048 &  0.003$\pm$0.047 &  0.002$\pm$0.058 &  0.003$\pm$0.038 & -0.002$\pm$0.035 &  0.001$\pm$0.035 & -0.005$\pm$0.040 & -0.003$\pm$0.048 \\
5 &  0.002$\pm$0.066 & -0.007$\pm$0.068 & -0.008$\pm$0.061 &  0.003$\pm$0.067 & -0.005$\pm$0.090 &  0.002$\pm$0.043 &  0.005$\pm$0.046 &  0.002$\pm$0.048 &  0.003$\pm$0.046 & -0.001$\pm$0.058 &  0.003$\pm$0.038 & -0.002$\pm$0.036 &  0.004$\pm$0.037 & -0.001$\pm$0.040 & -0.004$\pm$0.044 \\
6 &  0.005$\pm$0.058 & -0.004$\pm$0.053 & -0.001$\pm$0.056 &  0.002$\pm$0.062 &  0.001$\pm$0.076 &  0.001$\pm$0.037 &  0.005$\pm$0.040 & -0.003$\pm$0.039 &  0.001$\pm$0.042 & -0.007$\pm$0.050 &  0.002$\pm$0.030 &  0.002$\pm$0.031 &  0.000$\pm$0.032 & -0.005$\pm$0.033 &  0.000$\pm$0.046 \\
7 &  0.006$\pm$0.053 &  0.000$\pm$0.057 &  0.003$\pm$0.060 &  0.004$\pm$0.061 & -0.007$\pm$0.075 & -0.001$\pm$0.037 &  0.003$\pm$0.040 & -0.002$\pm$0.040 &  0.001$\pm$0.042 & -0.009$\pm$0.051 & -0.001$\pm$0.031 &  0.002$\pm$0.033 &  0.000$\pm$0.032 & -0.001$\pm$0.034 & -0.007$\pm$0.044 \\
8 &  0.001$\pm$0.053 & -0.002$\pm$0.052 &  0.006$\pm$0.053 &  0.001$\pm$0.059 & -0.007$\pm$0.075 &  0.000$\pm$0.035 & -0.002$\pm$0.036 & -0.004$\pm$0.036 &  0.002$\pm$0.037 &  0.000$\pm$0.051 &  0.003$\pm$0.028 &  0.001$\pm$0.030 &  0.001$\pm$0.030 & -0.001$\pm$0.032 & -0.004$\pm$0.043 \\
9 &  0.041$\pm$0.259 &  0.049$\pm$0.256 &  0.024$\pm$0.263 &  0.022$\pm$0.249 &  0.039$\pm$0.257 &  0.022$\pm$0.188 &  0.014$\pm$0.180 &  0.020$\pm$0.183 &  0.015$\pm$0.190 &  0.032$\pm$0.196 &  0.016$\pm$0.151 &  0.012$\pm$0.149 &  0.017$\pm$0.161 &  0.020$\pm$0.147 & -0.003$\pm$0.158 \\
\hline
\end{tabular}
\tablefoot{Averages derived from each set of 100 resampled light curves, with <$mag$> derived using the free-amplitude LCT fit. Only the first ten lines are displayed. The full version of this table is available in electronic format.}
\label{tab:deltamag_freeamplitude_simplemean}
\end{sidewaystable}

\begin{figure*}[!htbp]
\centering
\includegraphics[width=15cm]{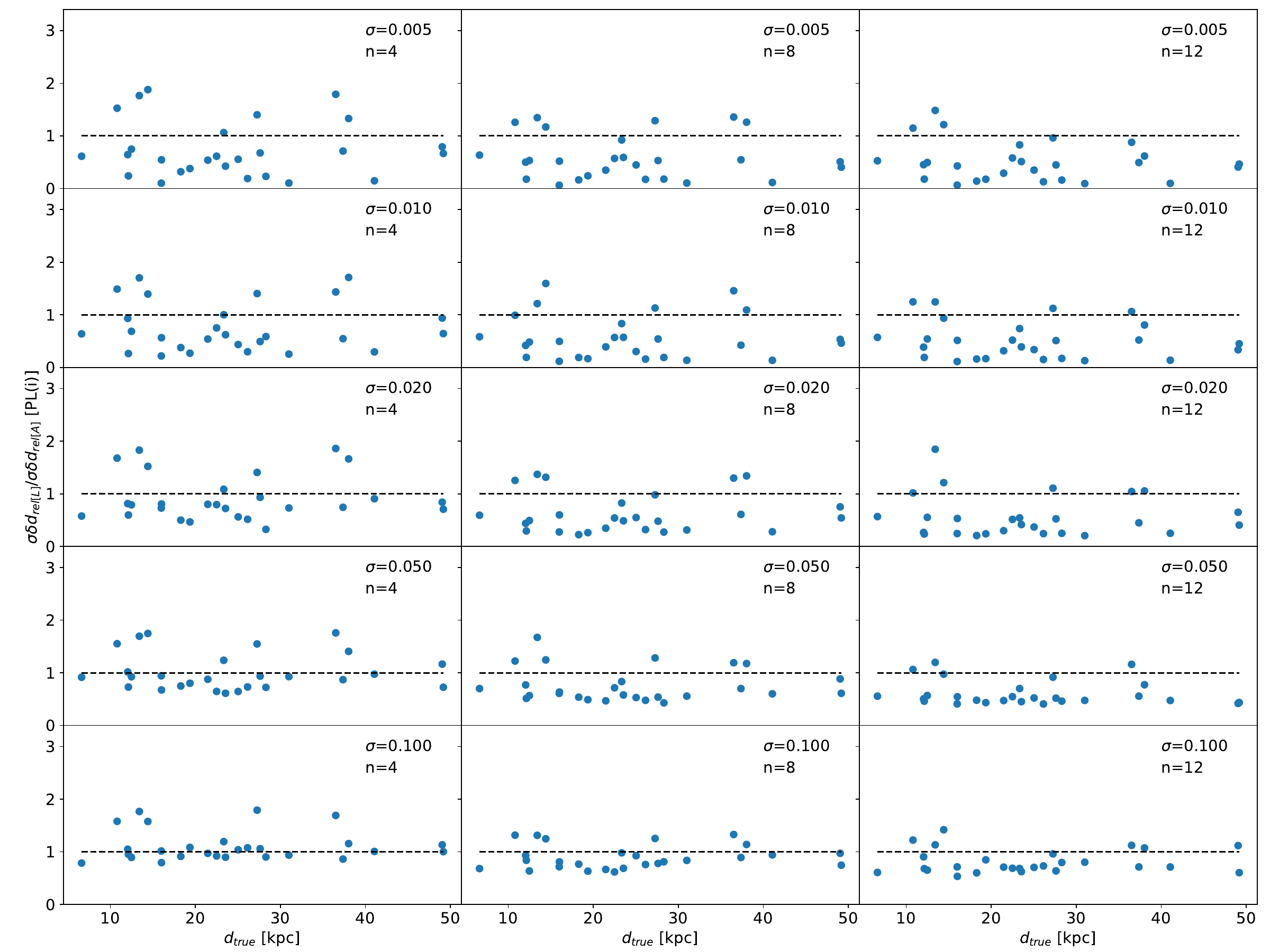}
\caption{Same as Fig.~\ref{fig:distance_single_pli} but the mean magnitudes were obtained by fitting the simulated time series with the LCT with fixed amplitude.}
\label{fig:distance_amplfixed_pli}
\end{figure*}

\begin{figure*}[!htbp]
\centering
\includegraphics[width=15cm]{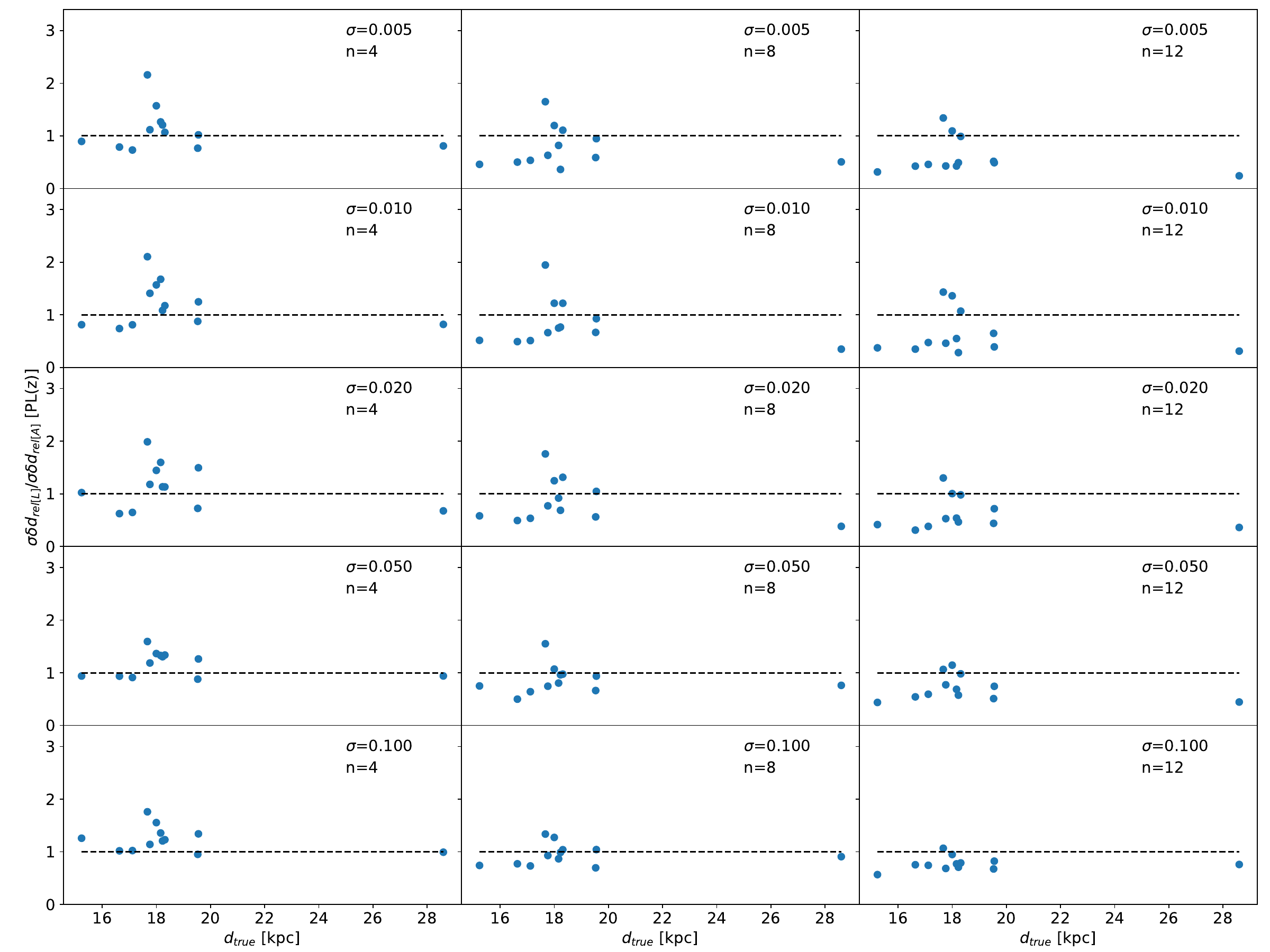}
\caption{Same as Fig.~\ref{fig:distance_single_pli} but the mean magnitudes were obtained by fitting the simulated time series with the LCT with fixed amplitude and the distances were obtained with PL($z$) relations.}
\label{fig:distance_amplfixed_plz}
\end{figure*}

\begin{figure*}[!htbp]
\centering
\includegraphics[width=18cm]{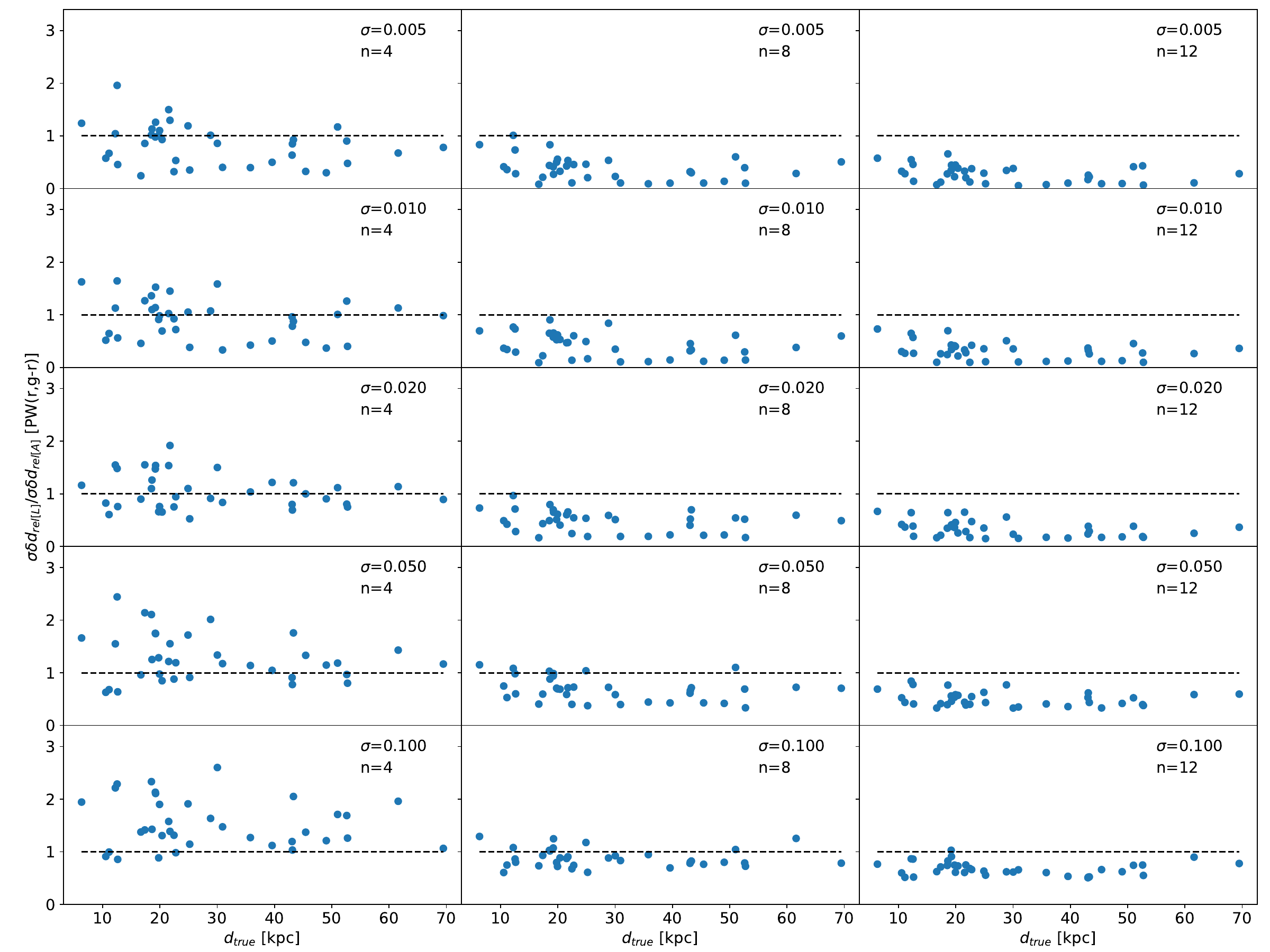}
\caption{Same as Fig.~\ref{fig:distance_single_pli} but the mean magnitudes were obtained by fitting the simulated time series with the LCT with free amplitude and the distances were obtained with PW($r$,$g-r$) relations.}
\label{fig:distance_amplfree}
\end{figure*}

\clearpage
\section{Figures of the light curve templates}\label{sect:templates_appenidx}

\begin{figure}[!htbp]
\centering
\includegraphics[width=9cm]{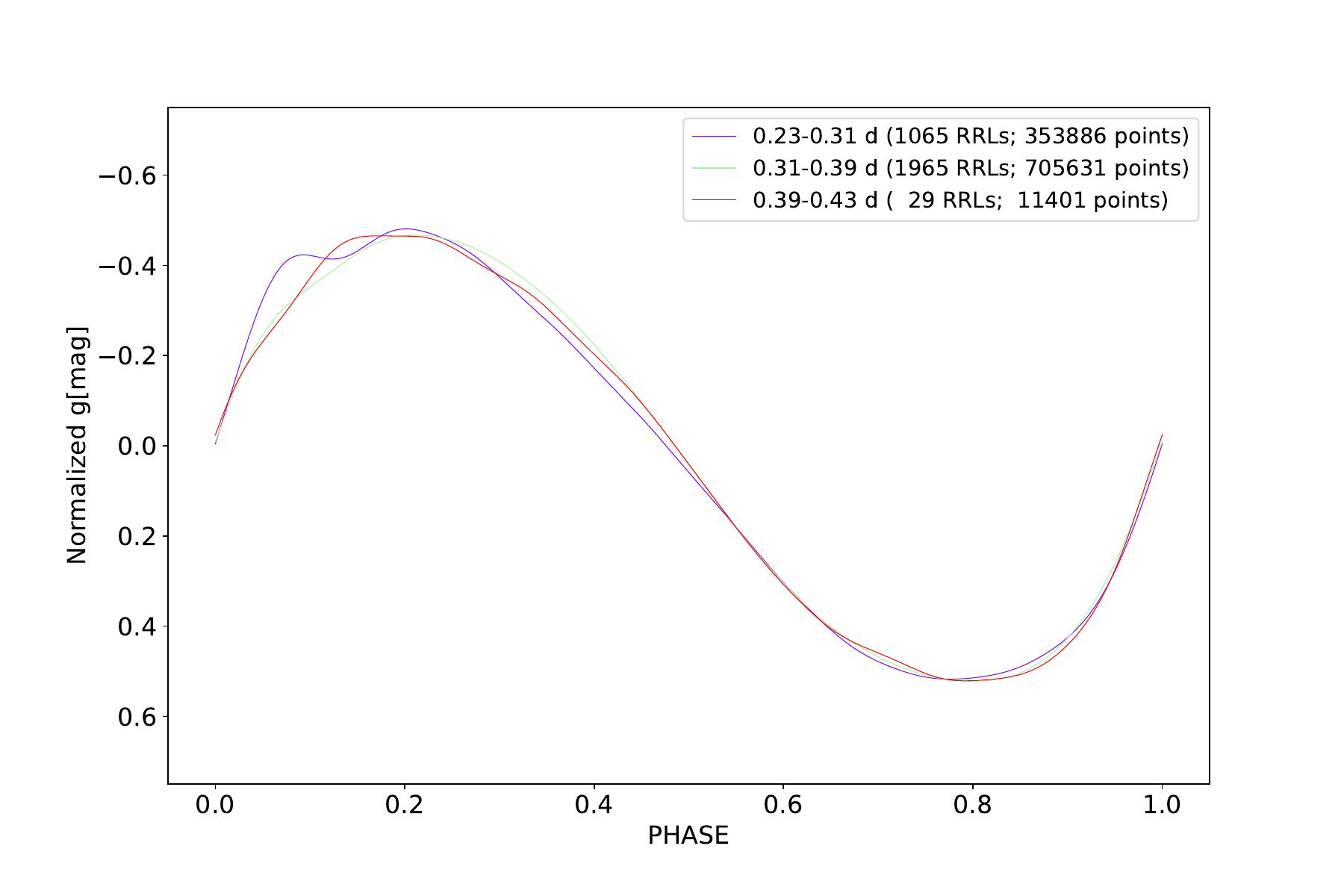}
\caption{LCTs of RRc in the passband $g$.}
\label{fig:template_rrc_g}
\end{figure}

\begin{figure}[!htbp]
\centering
\includegraphics[width=9cm]{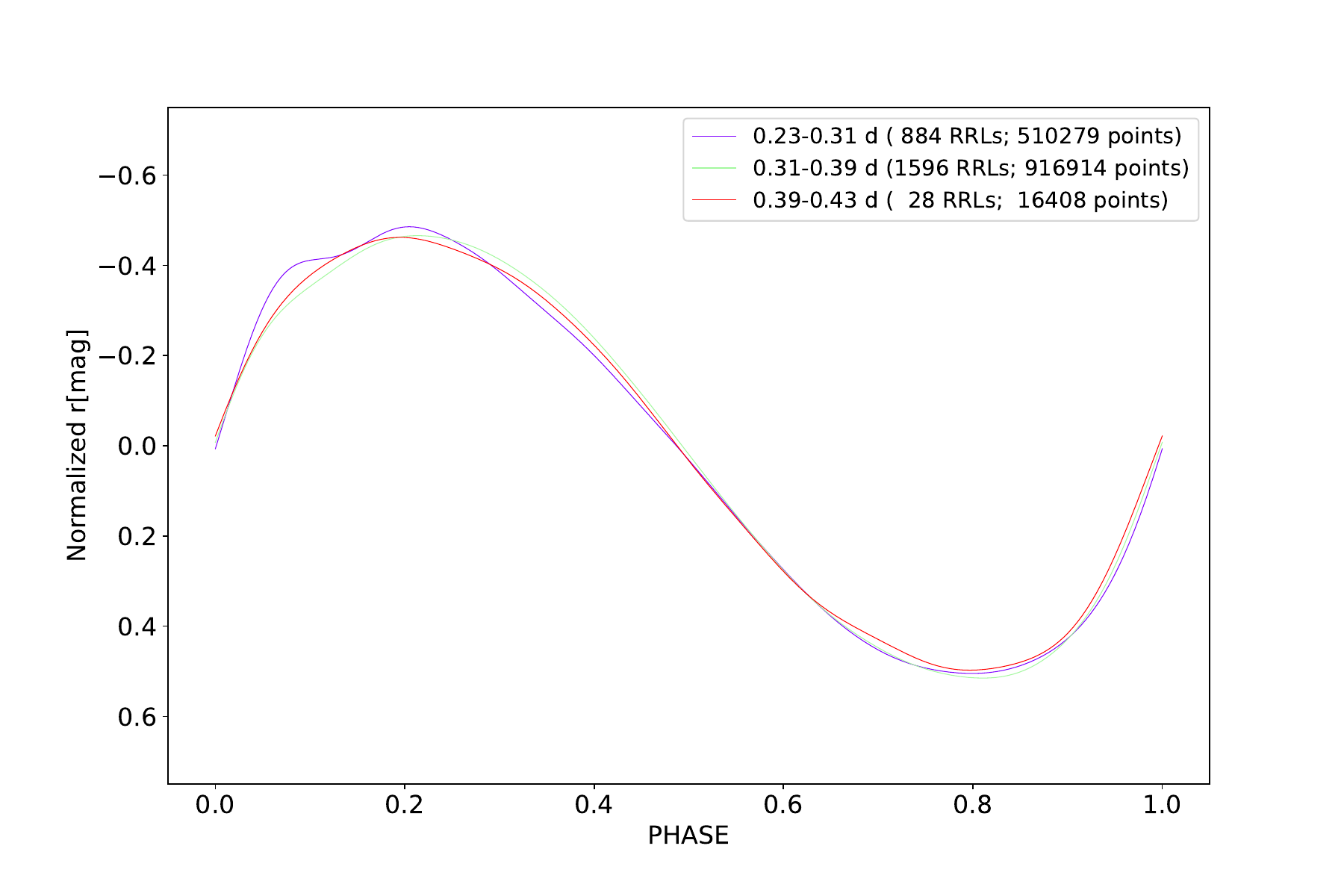}
\caption{LCTs of RRc in the passband $r$.}
\label{fig:template_rrc_r}
\end{figure}

\begin{figure}[!htbp]
\centering
\includegraphics[width=9cm]{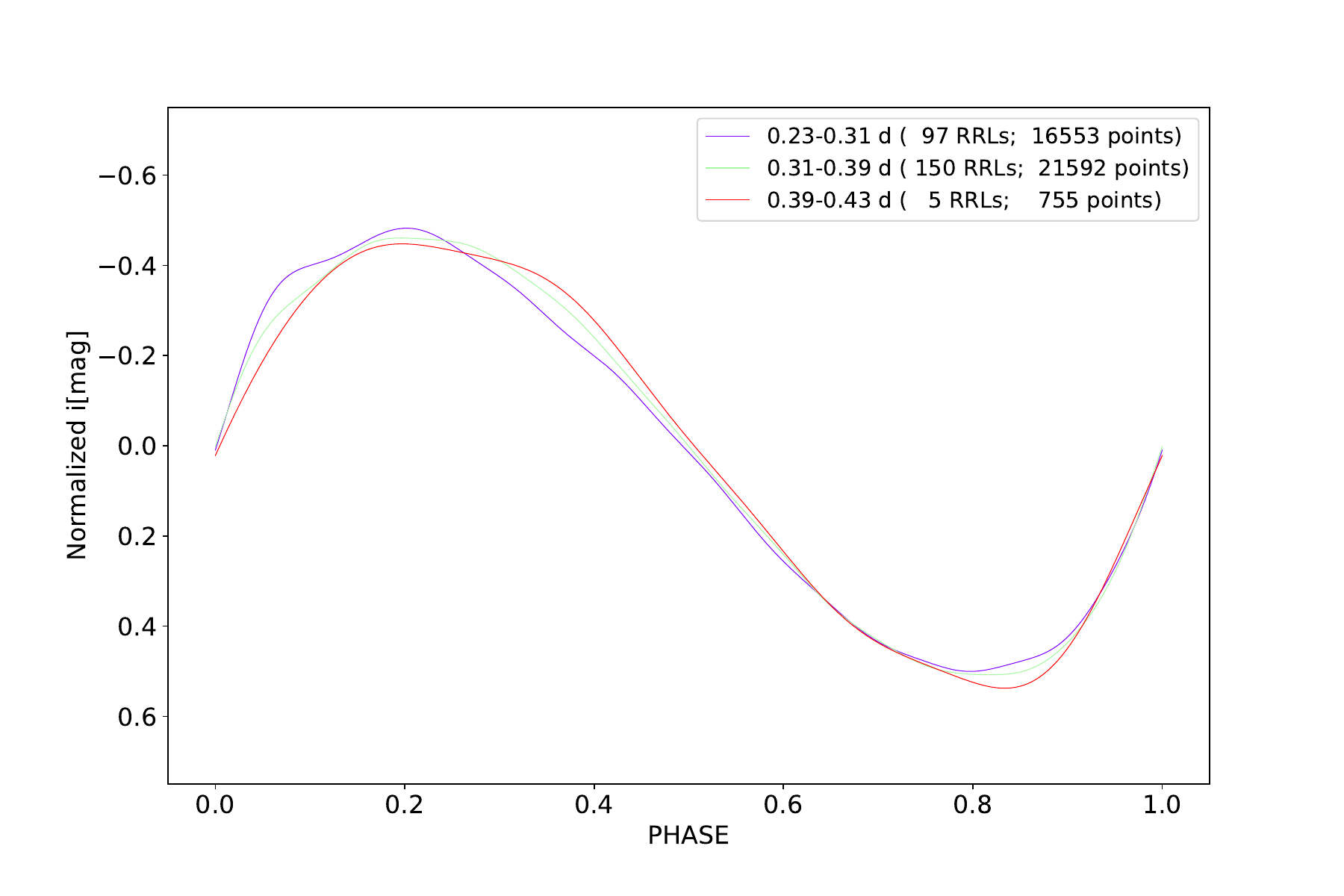}
\caption{LCTs of RRc in the passband $i$.}
\label{fig:template_rrc_i}
\end{figure}

Figs.~\ref{fig:template_rrc_g}, ~\ref{fig:template_rrc_r} and ~\ref{fig:template_rrc_i} make immediately clear that, for the RRc, the LCTs at shortest period are significantly different from other periods. In fact, they display a clear shoulder before the real pulsation maximum while, at longer periods, a flattening of the sinusoid appears. This is, in principle, in agreement with the suggestion of a hump progression with the period of RRc \citep{petersen1984}. However, we cannot draw firm conclusions from this feature, because: 1) metallicity plays a role and we do not have iron abundance estimates for a sizeable sample of our RRLs; 2) data for long-period RRc are a factor of 30-60 less than data for shorter-period RRc, meaning that a comparison might be misleading.

One can also notice that, in the $i$ band, the two bins with longer periods (0.31-0.39 days and 0.39-0.43 days) display a clear difference that spans almost half of the pulsation cycle. However, in the $g$ and $r$ bands, the LCTs are very similar, with small differences ($\ltsim$0.05 normalized mag) only around the maximum. Nonetheless, we decided to keep the same bin separation in all bands, including $g$ and $r$, to have a homogeneous binning of the periods for RRc.

\begin{figure}[!htbp]
\centering
\includegraphics[width=9cm]{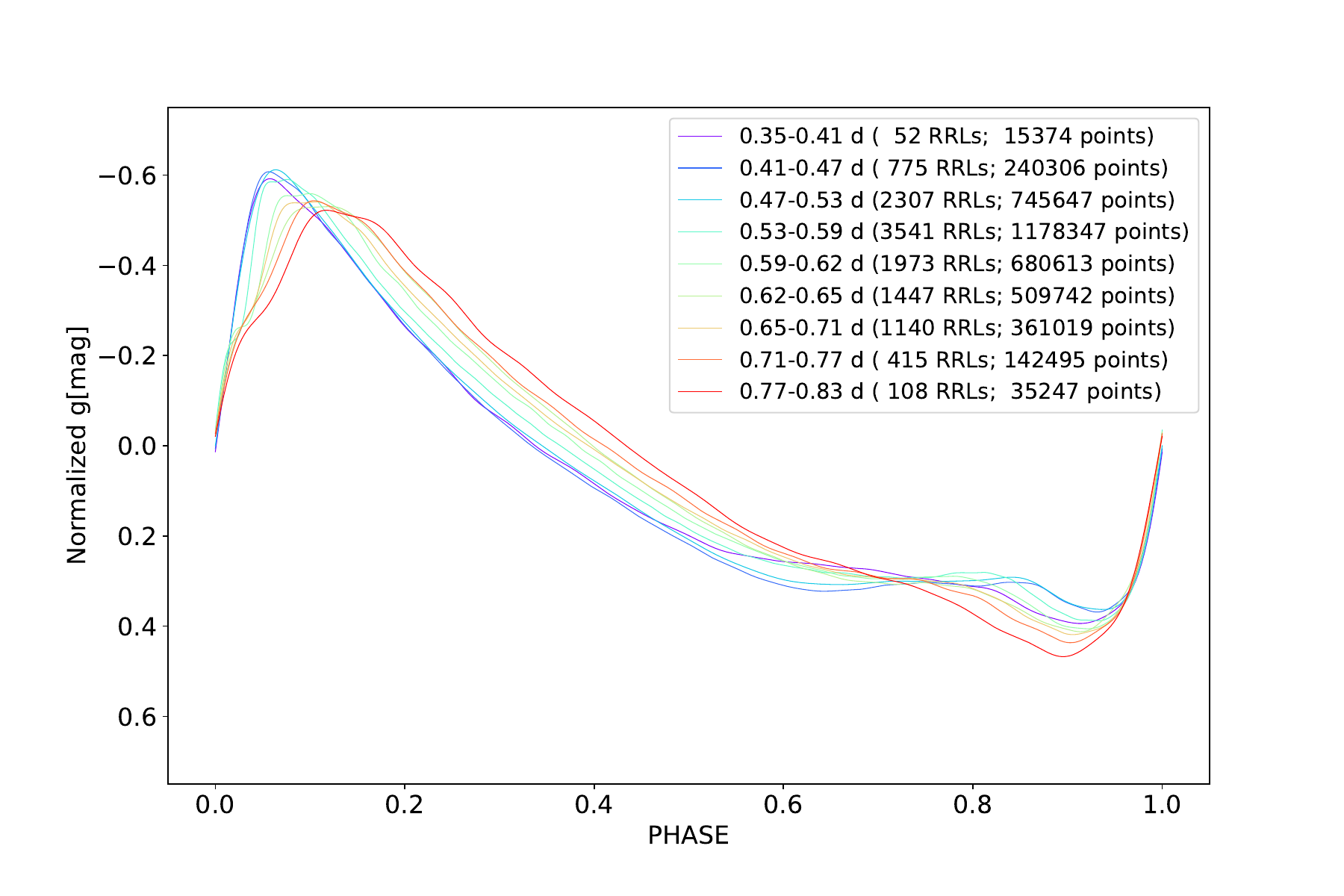}
\caption{LCTs of RRab in the passband $g$.}
\label{fig:template_rrab_g}
\end{figure}

\begin{figure}[!htbp]
\centering
\includegraphics[width=9cm]{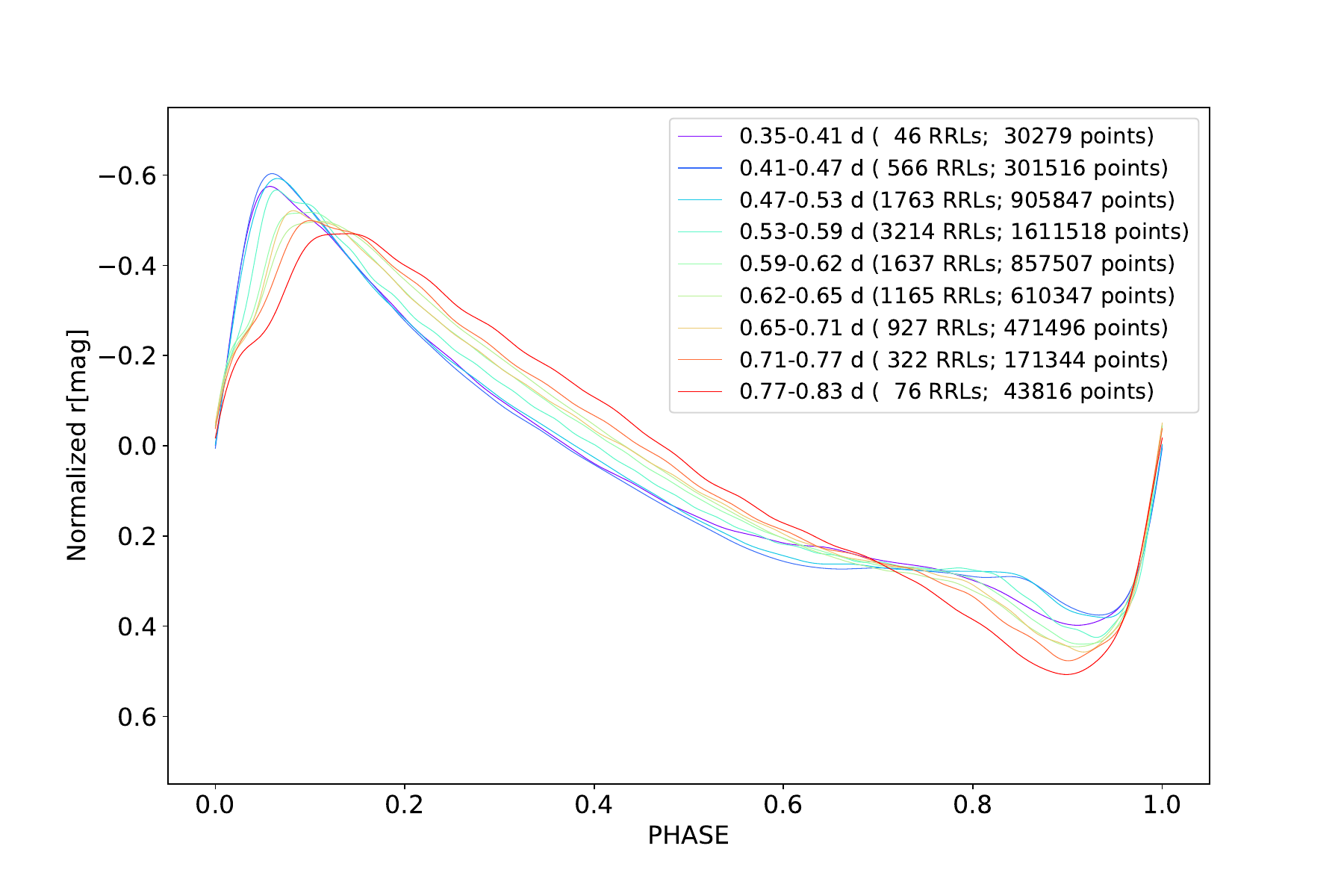}
\caption{LCTs of RRab in the passband $r$.}
\label{fig:template_rrab_r}
\end{figure}

\begin{figure}[!htbp]
\centering
\includegraphics[width=9cm]{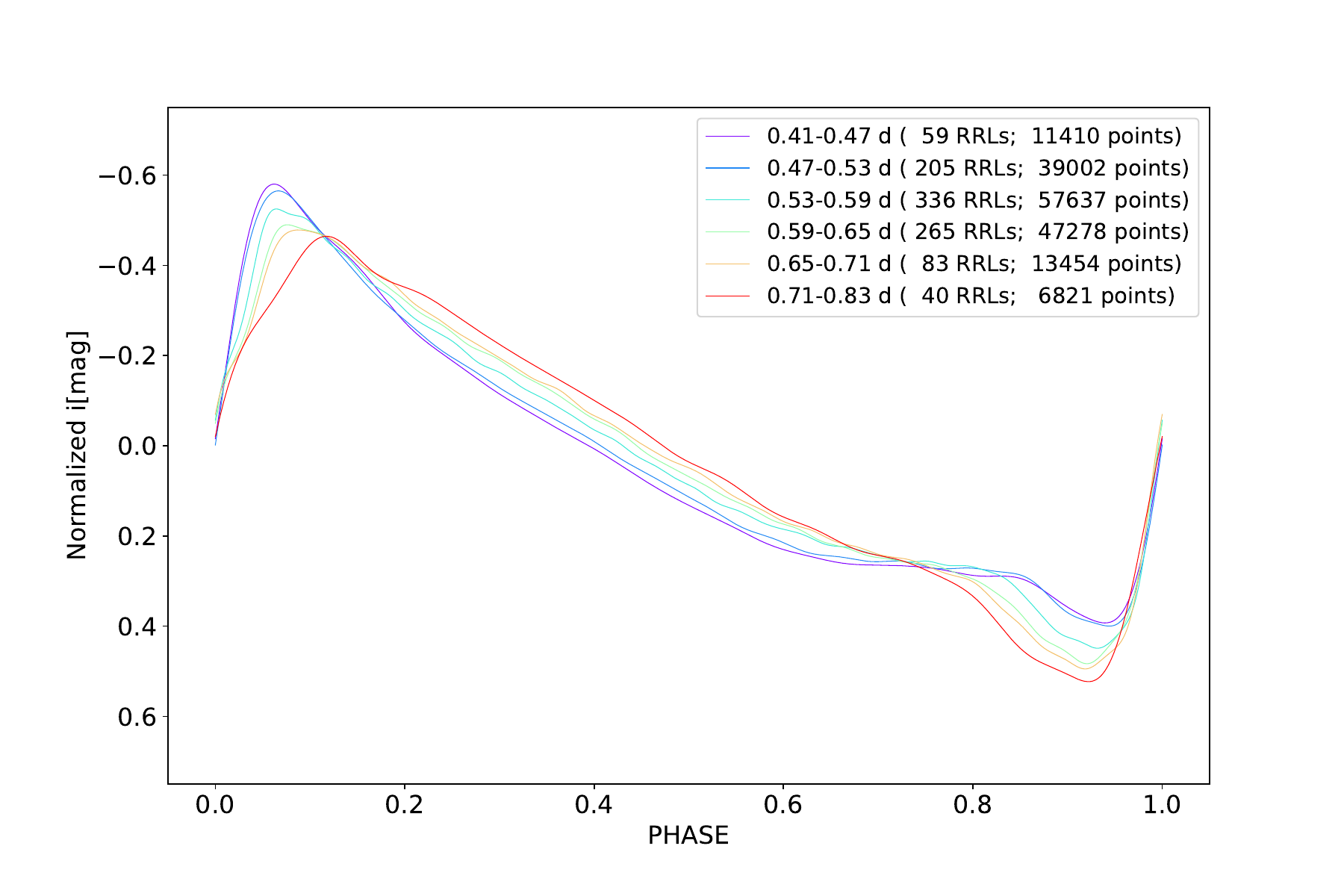}
\caption{LCTs of RRab in the passband $i$.}
\label{fig:template_rrab_i}
\end{figure}

\begin{figure}[!htbp]
\centering
\includegraphics[width=9cm]{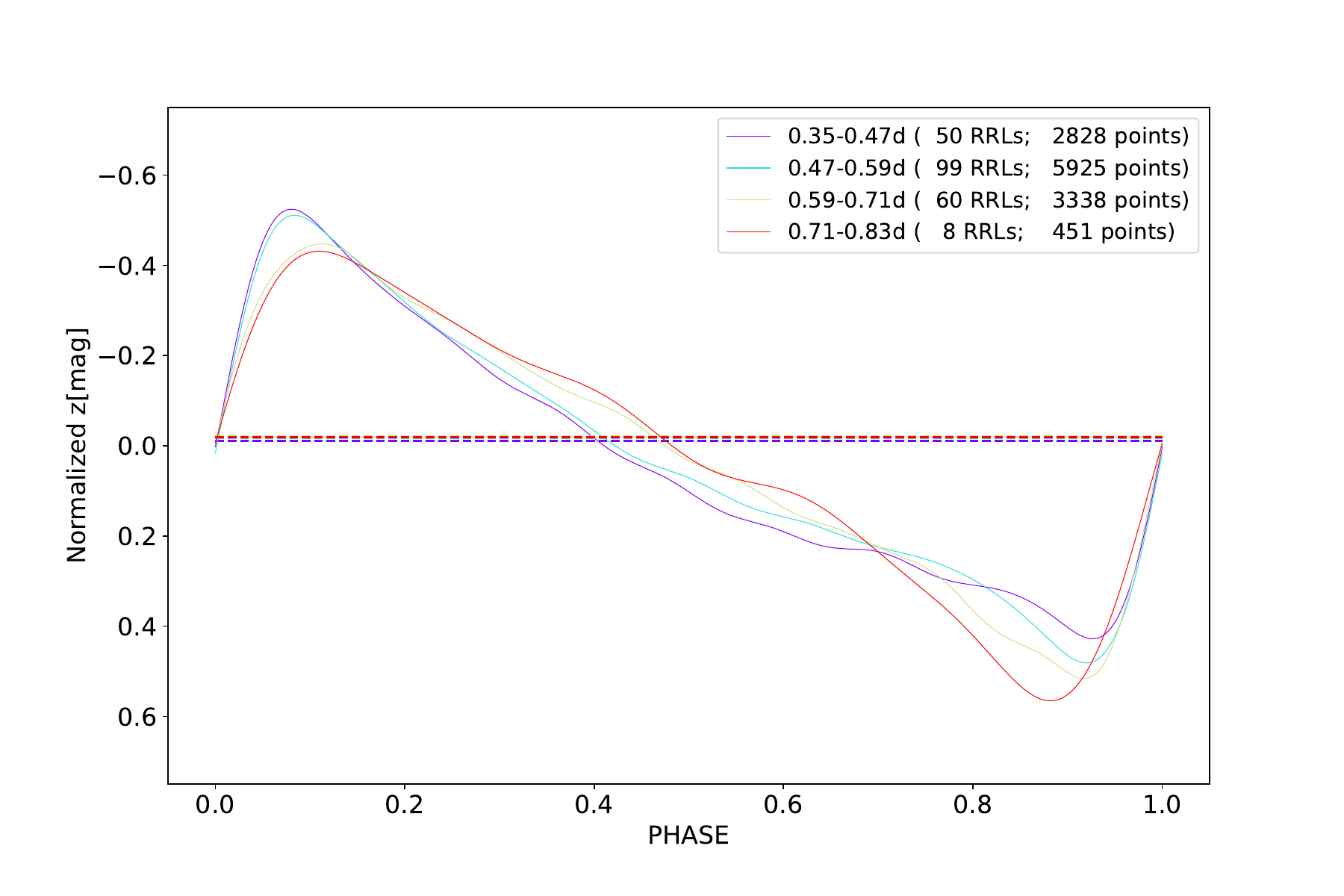}
\caption{LCTs of RRab in the passband $z$.}
\label{fig:template_rrab_z}
\end{figure}

Concerning RRab (Figs.~\ref{fig:template_rrab_g}, ~\ref{fig:template_rrab_r}, ~\ref{fig:template_rrab_i} and ~\ref{fig:template_rrab_z}), the change in shape of the LCTs is progressive for all period ranges and passbands. Unfortunately, for the long-period RRLs (0.71-0.83 days), we do not have enough $z$-band data to clearly reproduce the hump before maximum light. This feature is visible in all other bands, especially $g$ and $r$, where it starts to appear already in the 0.65-0.71 days bin.

\clearpage

\section{Comparison with Sesar 2010 templates}\label{sect:appendix_sesar}

Within a study of the spatial distribution of halo RRLs, S10 developed  LCTs of RRLs in the SDSS $ugriz$ passbands. Since we demonstrated, in sections~\ref{sect_conversion} and ~\ref{sect:amplratio_appendix} that there are no relevant differences between all the $ugriz$ photometric systems, one might think that to develop new LCTs is not necessary. However, by construction, S10 LCTs are not associated to any pulsation property, except the pulsation mode. This approach allowed S10 to deliver a set of LCTs covering all morphologies, and to be as general as possible, but this also means that, anyone wanting to use S10 LCTs for mean magnitude estimates, cannot know which specific LCT to adopt for a given target, since no pulsation property is associated to these LCTs.

A practical example for the $g$ passband, where 22 RRab LCTs are available in S10: even having all the pulsation properties available for a target and being able to correctly anchor the LCTs to any single observed phase point, one would have no way to decide which is the best LCT to use. Even when several phase points are available, one would still have to perform 22 fits instead of one, and only then it would be possible to select the best LCT. This means to increase the computing time by more than one order of magnitude for each target.

Aside from these difficulties in applying S10 LCTs, there are quantitative reasons for having developed our new LCTs. First of all, our database consists on more than 28000 light curves versus $\sim$2500 for S10). Secondly, our Halo sample (ZTF) covers the entire sky above declination --20 degrees, while S10 sample is limited to the Stripe 82 region of the SDSS survey, where substructures may more easily bias the properties of RRLs due, e.g., to an anomalous metallicity distribution.

\section{Amplitude ratios between different photometric systems}\label{sect:amplratio_appendix}

In the following, we provide light amplitude ratios between the most commonly used photometric systems. More precisely, we derived the following amplitude ratios:

\begin{itemize}
    \item the LSST $uriz$-over-$g$ ratios
    \item the LSST-over-DECam $ugrizy$ ratios
    \item the LSST-over-ZTF $gr$ ratios
    \item the LSST-over-SDSS $gr$ ratios
    \item the ZTF/DECam-over-$I$ ratios
    \item the ZTF/DECam-over-\gaia($G$) ratios
    \item the OGLE $I$-over-$V$ ratios
\end{itemize}

For the ratios between amplitudes in the same band but of different photometric systems, we adopt the following notation: $r(x)_{A/B}$ is the ratio of $Ampl$ between the $x$ bands of the A and B photometric systems, e.g., $r(g)_{L/D} = \frac{Ampl(g)_L}{Ampl(g)_D}$. 

Since we have no available empirical LSST light curves, the only way to estimate $r(g)_{L/D}$, $r(g)_{L/Z}$ and the LSST $\dfrac{Ampl(u_L/r_L/i_L/z_L)}{Ampl(g_L)}$ ratios was to adopt synthetic light curves obtained from stellar pulsation models \citep{marconi15,marconi2022}. For this purpose, we adopted models at Z=0.0003, Z=0.001 and Z=0.008, which are typical abundances for RRLs for the metal-poor end, the peak of the Galactic halo and the metal-rich tail, respectively \citep{fabrizio2021,mullen2022,cabreragarcia2024}.

Unfortunately, synthetic light curves are only available in the LSST and DECam passbands, but not in the ZTF. Therefore, we first derived $r(x)_{L/D}$ directly from pulsation models and then we obtained the $r(x)_{L/Z}$ ratios by multiplying $r(x)_{L/D}$ by $r(x)_{Z/D}$. The latter ratios ($r(x)_{Z/D}$) were obtained from observed light curves of objects in common between DECam and ZTF, namely, RRLs in the globular cluster M5. Note that the photometric reduction process of the M5 data in \citet{vivas2017} is the same as that adopted for Bulge data by \citet{saha2019}.

\subsection{LSST $uriz$-over-$g$ ratios}
\begin{figure*}[!htbp]\label{fig:amplratio_LSST}
\centering
\includegraphics[width=9cm]{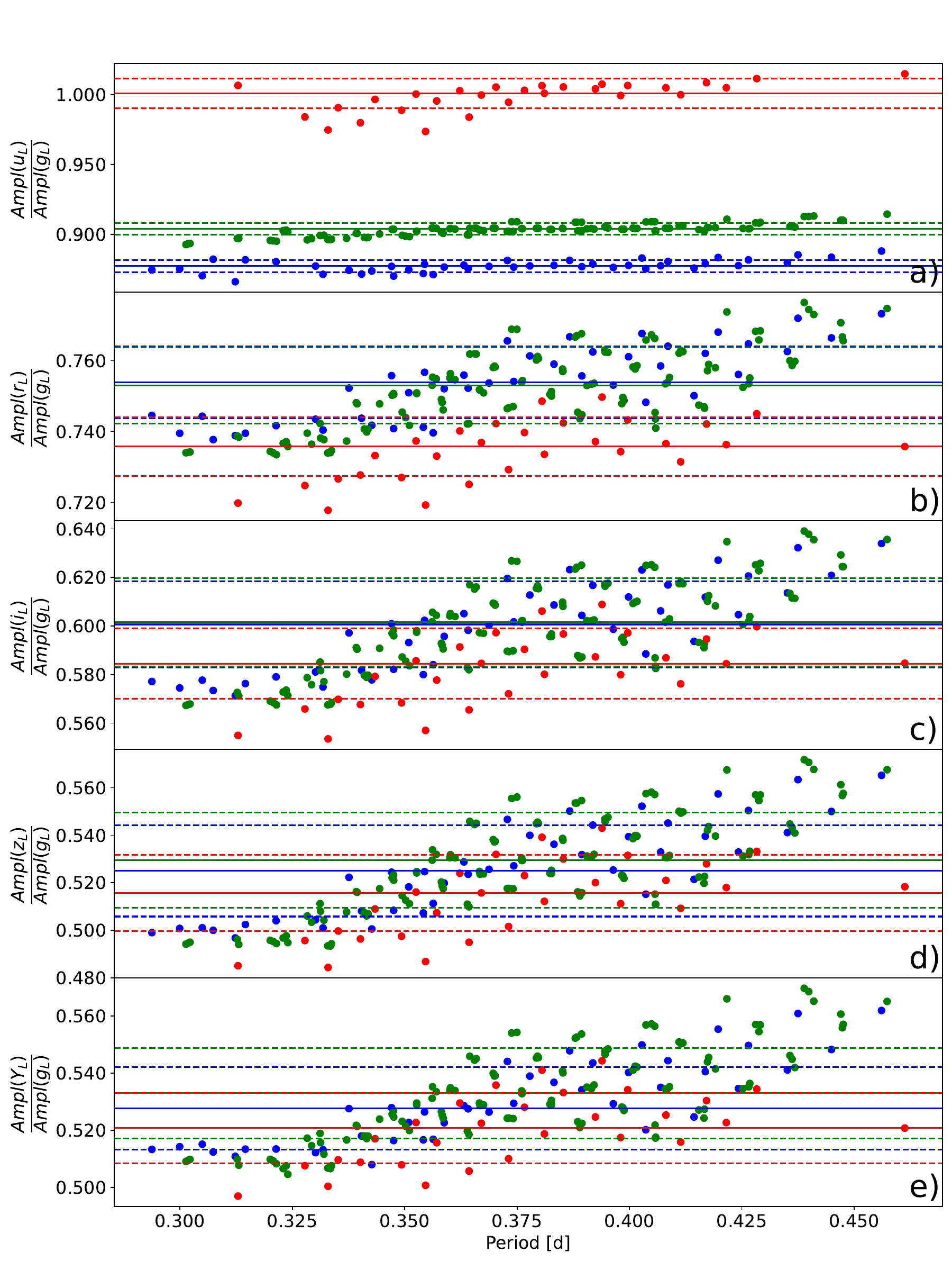}
\includegraphics[width=9cm]{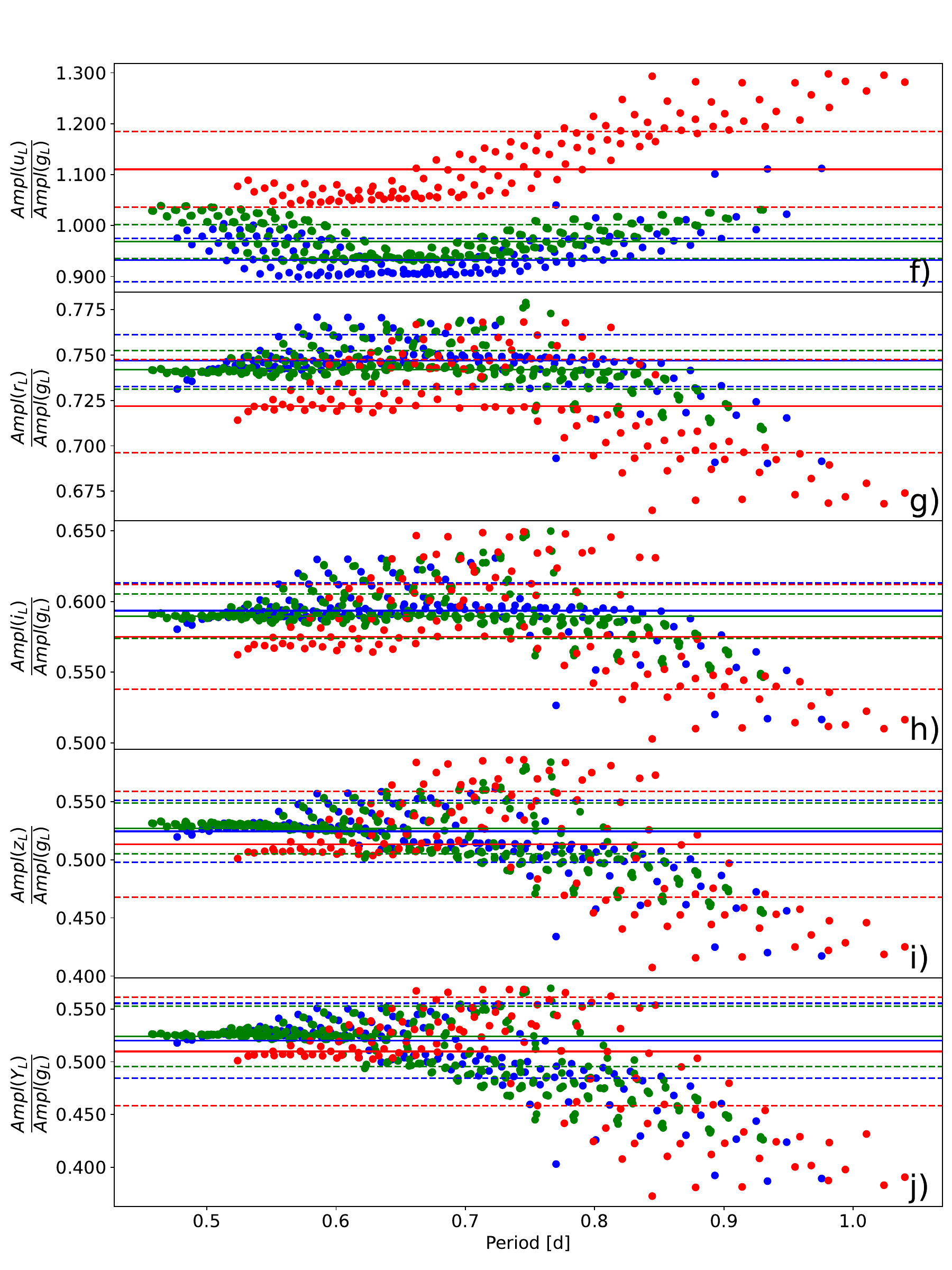}
\caption{Left panels (a-e): from top to bottom, $\dfrac{Ampl(u_L/r_L/i_L/z_L/Y_L)}{Ampl(g_L)}$ amplitude ratios versus pulsation period for Z=0.0003 (blue), Z=0.001 (green), Z=0.008 (red) RRc models. Solid/dashed lines display the average/standard deviations of each set of models. Right panels (f-j): Same as left, but for the RRab models.}
\label{fig:amplratio_lsst}
\end{figure*}

Figure~\ref{fig:amplratio_LSST} displays, in the left and right panels respectively, the ratios $\dfrac{Ampl(u_L/r_L/i_L/z_L/Y_L)}{Ampl(g_L)}$ versus periods of the RRc and RRab stars, color-coded by metallicity. The mean and standard deviations of the ratios are reported in Table~\ref{tab:amplratio} and displayed as solid and dashed lines in the figure, respectively. Four facts are immediately clear: 1) the ratios at fixed metallicity span a very narrow range; 2) almost all the mean ratios of RRab and RRc stars match within 1$\sigma$; 3) the ratio $\dfrac{Ampl(u_L)}{Ampl(g_L)}$ of RRab stars seems to follow a V shape; 4) The $\dfrac{Ampl(u_L)}{Ampl(g_L)}$ of both RRab and RRc displays a positive trend with metallicity. LSST will be the only photometric survey allowing to check empirically these ratios, thanks to its unprecedented multiband coverage.

\subsection{LSST-over-DECam $ugrizy$ ratios}
\begin{figure*}[!htbp]
\centering
\includegraphics[width=9cm]{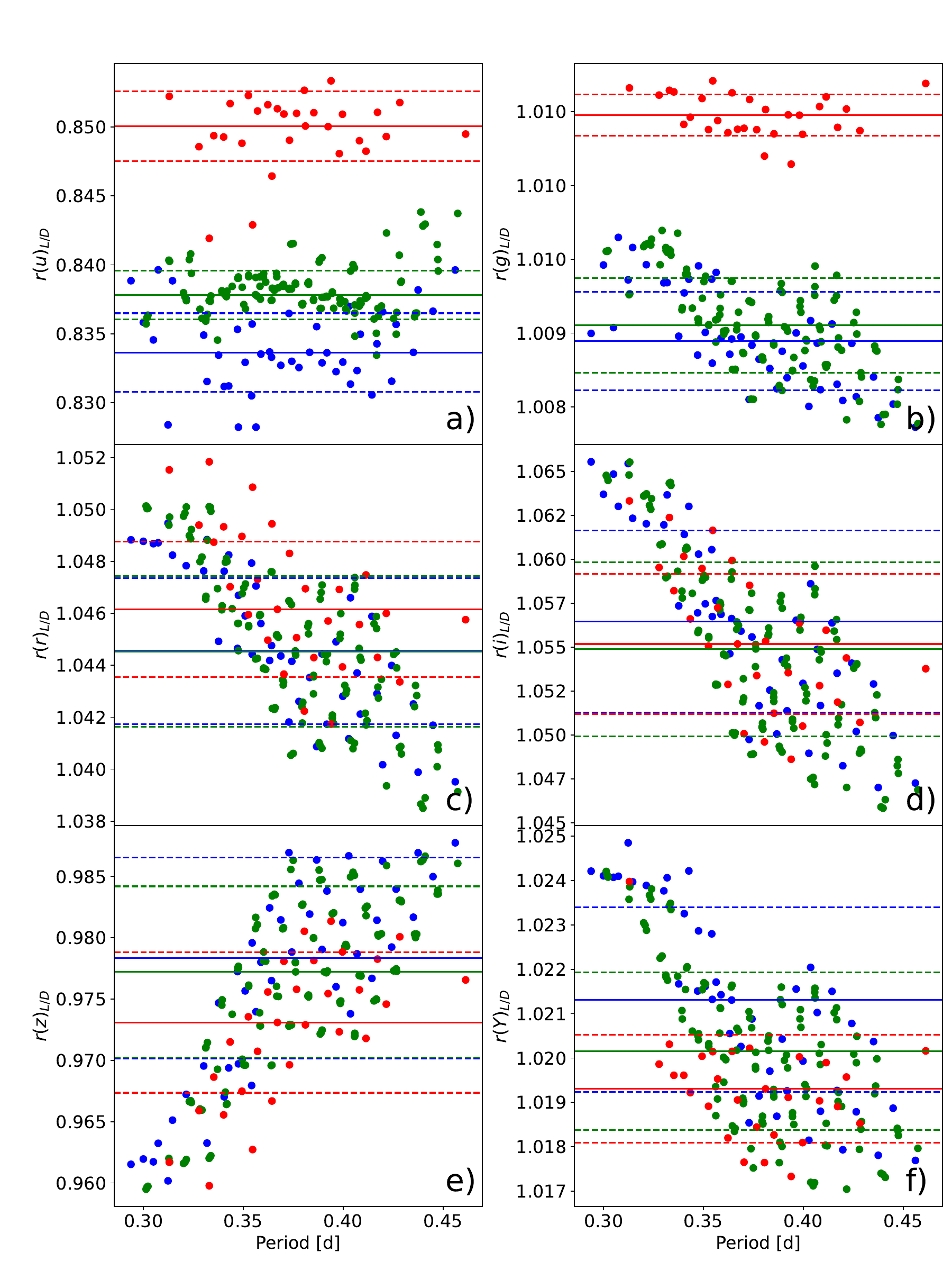}
\includegraphics[width=9cm]{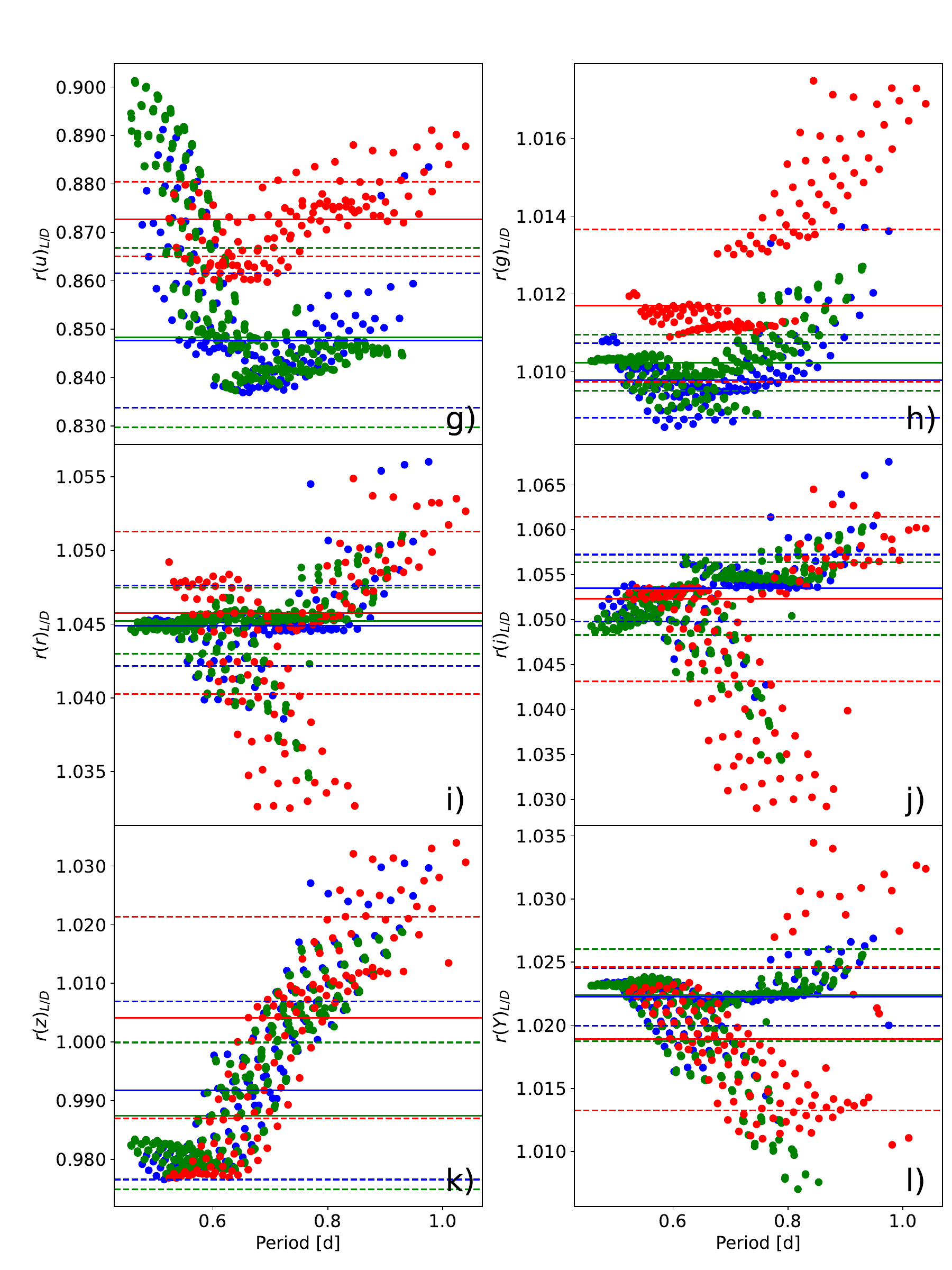}
\caption{Left panels (a-f): from top to bottom, left to right: $r(x)_{L/D}$ amplitude ratios versus pulsation period for Z=0.0003 (blue), Z=0.001 (green), Z=0.008 (red) RRc models. Solid/dashed lines display the average/standard deviations of each set of models. Right panels (g-l): Same as left, but for the RRab models.}\label{fig:decam_lsst_ratio}
\end{figure*}

Figure~\ref{fig:decam_lsst_ratio} displays, in the a)-f) and g)-l) right panels, the ratios $r(x)_{L/D}$ versus periods of the RRc and RRab stars, respectively. First of all we note that, within the dispersion, the ratios can be assumed to be constant with period. The only, marginal exceptions are $r(u)_{L/D}$ (that is useless for our templates) which displays a V-shape and $r(z)_{L/D}$ band for RRab. Note, however, that the range of variation of these ratios is, at most 0.05. This corresponds to a 1-5\% error on the estimate of $Ampl$, that is significantly less than typical errors on $Ampl$ \citep{braga2022}. Table~\ref{tab:amplratio} displays the average ratios and the standard deviations. To sum up, the $r(x)_{L/D}$ ratios are close to 1, with a maximum offset of $\sim$0.06 (if we exclude the $u$ band which is not useful for our templates). The standard deviations are also very small, being at most 0.01. 

\subsection{LSST-over-ZTF $gr$ ratios}
\begin{figure}[!htbp]
\centering
\includegraphics[width=9cm]{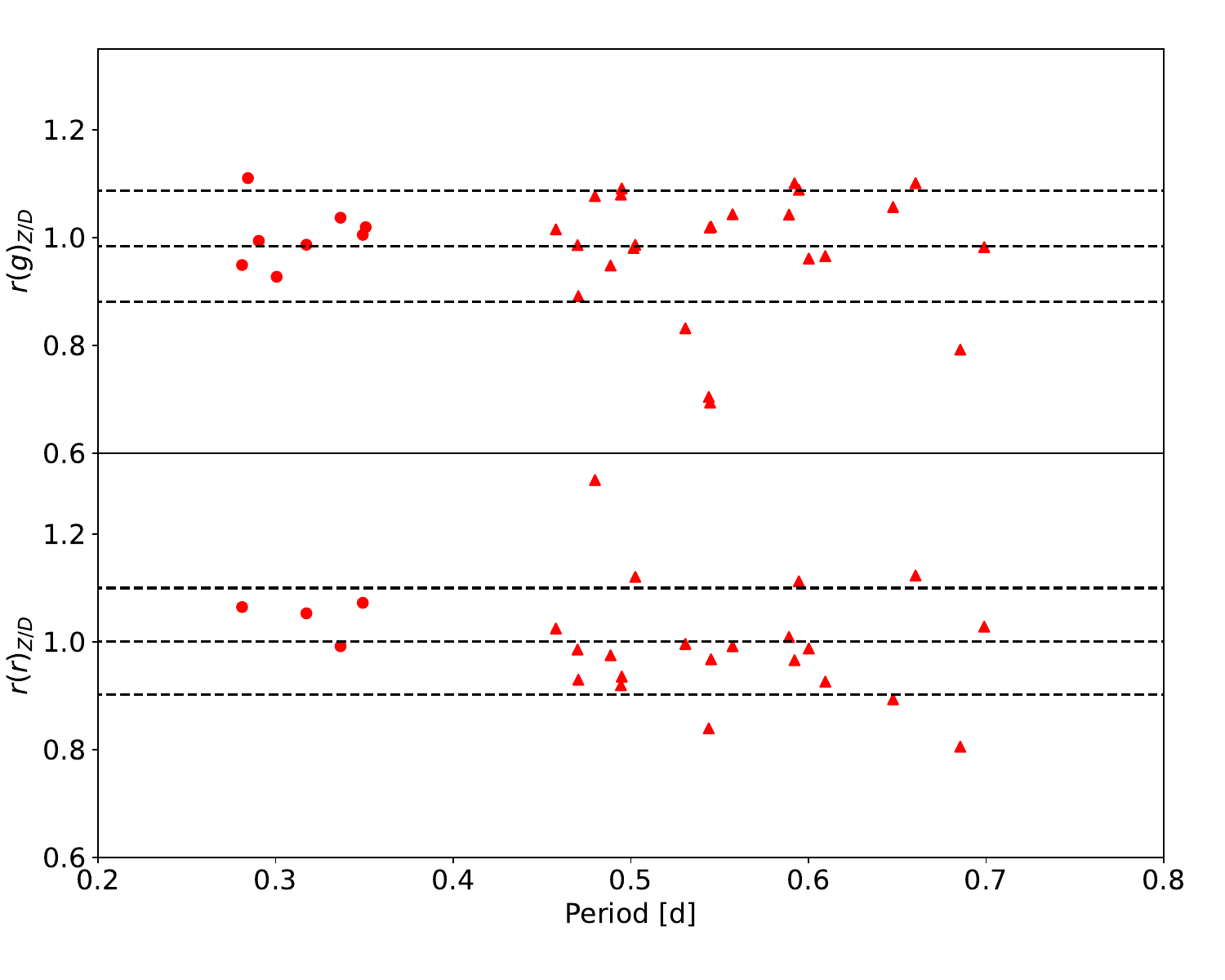}
\caption{Left: $r(g)_{Z/D}$ (top) and $r(r)_{Z/D}$ (bottom) versus $Ampl(g)_Z$ and $Ampl(r)_Z$, respectively, of the RRLs for which we have light curves from both DECam and ZTF. Right: same as left but for ZTF vs SDSS photometric systems.}
\label{fig:amplratio_ztf}
\end{figure}

As already mentioned, we have no way to directly compare LSST and ZTF data, neither empirical nor synthetic, therefore we have to follow a two-step process, first deriving $r(x)_{Z/D}$ and then obtaining $r(x)_{L/Z} = \dfrac{r(x)_{L/D}}{r(x)_{Z/D}}$. The $r(x)_{Z/D}$ ratios were obtained from empirical data and are displayed in Fig.~\ref{fig:amplratio_ztf}. It is immediately clear that both $r(g)_{Z/D}$ and $r(r)_{Z/D}$ are consistent, within the dispersion, with a ratio of 1, meaning that also $r(x)_{L/Z}$ is 1, within the errors. The values are listed in Table~\ref{tab:amplratio}. Unfortunately, we do not have any data in common for the $i$ band, meaning that we cannot quantitatively derive $r(i)_{Z/D}$. The fact that both $r(x)_{L/D}$ and $r(x)_{L/Z}$ are consistent with 1, well within one sigma means that, despite our LCTs were built based on ZTF and DECam data, they can be used on LSST data, which is our main aim.

\subsection{LSST-over-SDSS $gr$ ratios}
\begin{figure}[!htbp]
\centering
\includegraphics[width=9cm]{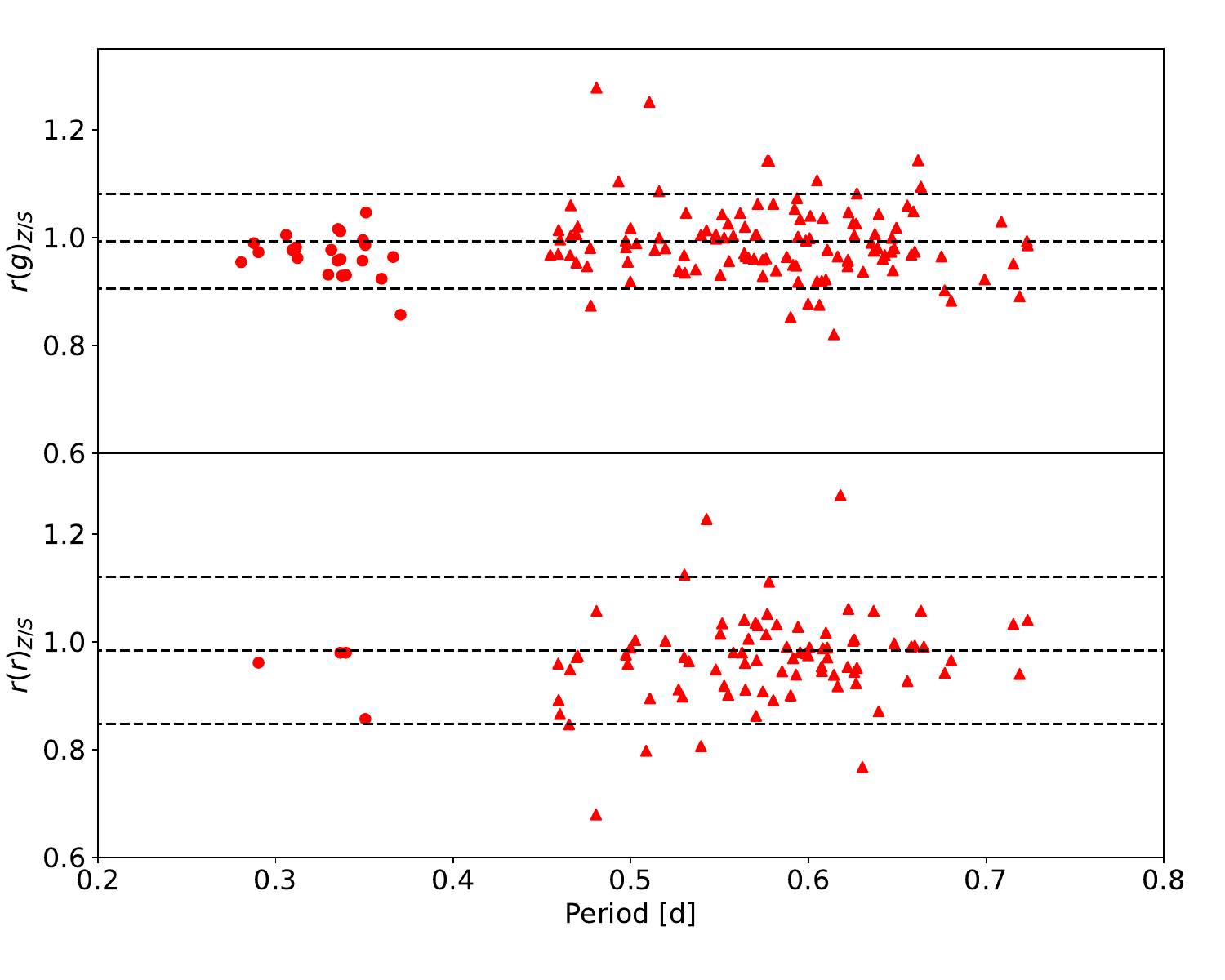}
\caption{Same as Fig.~\ref{fig:amplratio_ztf}, but for the ZTF over SDSS ratios.}
\label{fig:amplratio_sdss}
\end{figure}

The survey SDSS \citep{sdss_dr18} is one of the most extensive photometric surveys and the one that popularized the $ugriz(y)$ photometric system, therefore we derived also amplitude ratios between the LSST and the SDSS photometric system ($r(x)_{L/S}$). These ratios will be useful when applying the LCTs to RRLs with known amplitudes from SDSS. Also in this case, we must follow a two-step process: 1) we derive $r(x)_{Z/S}$ ratios; 2) we obtain $r(x)_{L/S} = r(x)_{L/Z} \cdot r(x)_{Z/S}$. The reference photometric system that we use is ZTF because it is the dataset for which we could find more matches with the RRLs found by \citet{sesar2010} in the Stripe 82 of SDSS. The results are displayed in Fig.~\ref{fig:amplratio_sdss} and Fig.~\ref{fig:amplratio_ogle}, and listed in Table~\ref{tab:amplratio}. The $r(x)_{Z/S}$ ratios are all consistent with 1 within the standard deviations, with the only, marginal exception of $r(r)_{Z/S}$ of the RRc. Finally, we derived also the $r(x)_{L/S}$ ratios by multiplying $r(x)_{L/Z}$ by $r(x)_{Z/S}$. The behavior of $r(r)_{L/S}$ is identical to that of $r(x)_{Z/S}$: all the values are 1 within the standard deviations, again with the only exception of $r(r)_{L/S}$ (see Table~\ref{tab:amplratio}).

\subsection{ZTF/DECam-over-$I$ ratios}
OGLE \citep{udalski92,udalski2015} is a still-ongoing, extensive optical survey, specifically conceived for micro-lensing and variability detections, that provides an outstanding combination of time, sky and depth coverage, on top of time series with 10$^3$-10$^4$ measurements per star in the Johnson-Cousins $V$ and $I$ band. It also has a far better pixel scale than ASAS-SN (0.25 vs 8.0 arcseconds). Since, to this point in the paper, we still miss an anchor to the Johnson system, we calculated the ZTF-over-OGLE ratios. In Fig.~\ref{fig:amplratio_ogle}, we note that the $\dfrac{Ampl(g_Z)}{Ampl(I)}$ and $\dfrac{Ampl(r_Z)}{Ampl(I)}$ ratios display a clear separation between ab and c type RRLs, the latter having higher ratios. On the other hand, the$\dfrac{Ampl(i_Z)}{Ampl(I)}$ ratios of RRab and RRc are similar. Differences in amplitude ratio between RRc and RRab were found by \citet{braga2018} in $\omega$ Cen. More specifically, they found that the NIR-to-$V$-band amplitude ratios of RRab are larger than RRc. This is consistent with our result because the difference between the RRc and RRab ratios progressively decreases from $g_Z$ to $r_Z$ to $i_Z$, where the difference is zero. Unfortunately, the overlap between our DECam sample and OGLE for the $i_D$ and $z_D$ bands is small and the dispersion is very large, not allowing to draw firm quantitative conclusions.

We also derived the ratios between $Ampl(z_D)$ and $Ampl(I)$, by matching the estimates of $Ampl(I)$ from \citet{thesis_neeley} and $Ampl(z_D)$ by \citet{vivas2017}. Despite the sample is very small and shows a large dispersion, it follows the quoted trend. Finally, we derived $\dfrac{Ampl(z_D)}{Ampl(I)}$ for Bulge RRab matched with OGLE (blue points in Fig.~\ref{fig:amplratio_ogle}). We found an average and standard deviation of 0.947$\pm$0.085, that are almost identical to the one obtained with M5 data. This is encouraging but also expected, because M5 $I$-band photometry is calibrated to the Kron-Cousins system, as OGLE. However, for homogeneity, in Table~\ref{tab:amplratio}, we provide only the ratios from M5 data, since these can be estimated for both RRab and RRc.

\begin{figure}[!htbp]
\centering
\includegraphics[width=9cm]{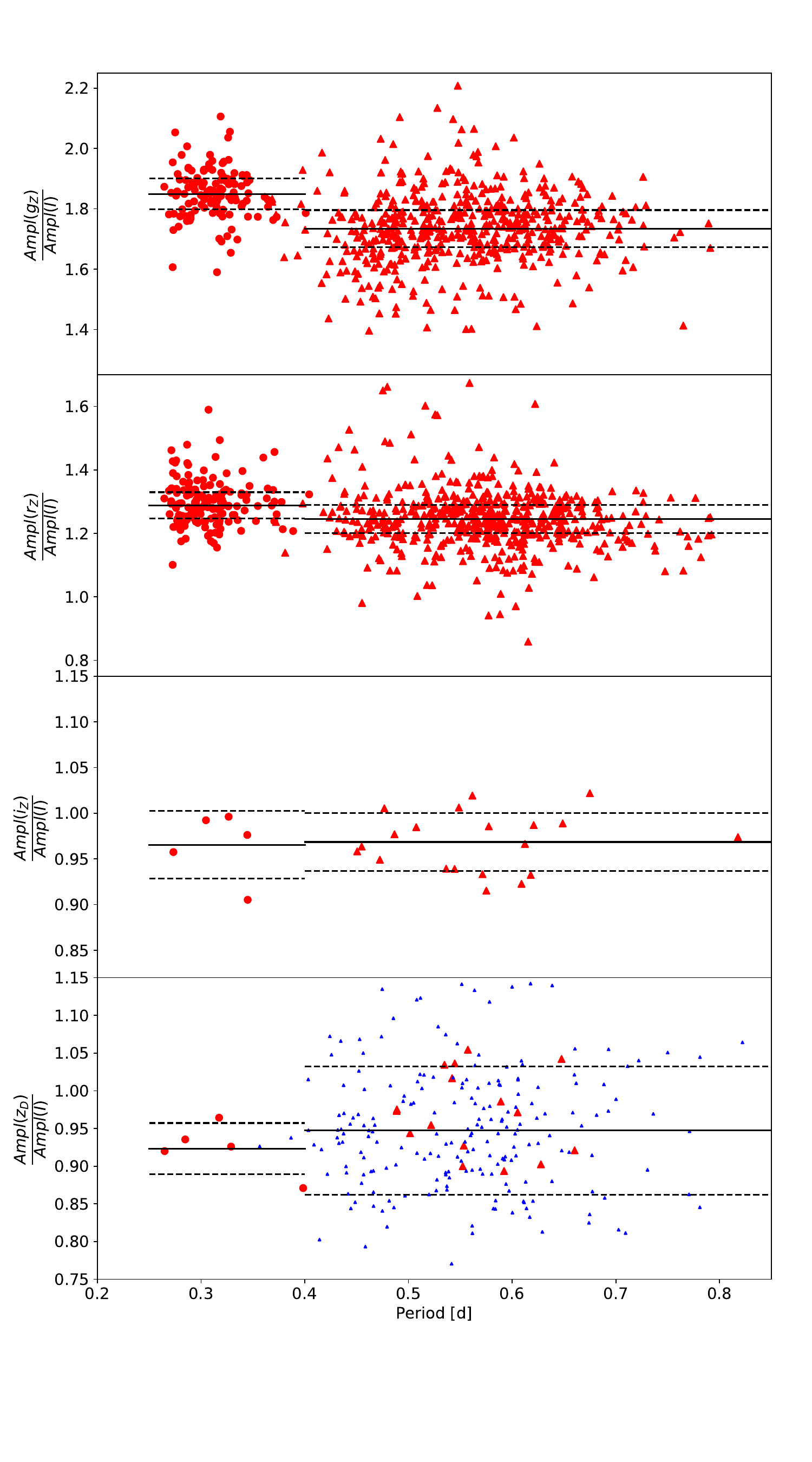}
\caption{From top to bottom: Amplitude ratios between $Ampl(g_Z)$, $Ampl(r_Z)$, $Ampl(i_Z)$ and $Ampl(z_D)$ over $Ampl(I)$. Blue points in the bottom panel represent the amplitude ratios derived from Bulge RRLs.}
\label{fig:amplratio_ogle}
\end{figure}

\subsection{ZTF/DECam-over-\gaia($G$) ratios}

The photometric system of \gaia ($G$, $Bp$, $Rp$) has become of primary importance for variable stars, since \gaia delivered catalogs and photometric properties for millions of variable stars, among which, the largest homogeneous catalog of known RRLs \citep{clementini2023}. Moreover, LSST brokers like Alerce \citep{alerce} and ANTARES \citep{antares} will provide cross-identification of LSST candidate variables with \gaia. Therefore, it is crucial to provide also amplitude ratios versus $Ampl(G)$ (as obtained from the \texttt{peak\_to\_peak\_g} column in the \texttt{gaiadr3.vari\_rrlyrae}) table within \gaia DR3, that will allow LCT users to adopt \gaia as a starting catalog. As described in Sect.~\ref{sect_ztf}, we already have \gaia-ZTF matching. By selecting only RRLs with photometric properties in \gaia DR3, we ended up with 23610 RRLs. The RRab and RRc display the same ratio in the $g_Z$ and $r_Z$ bands. In the $i_Z$ and $z_D$ bands, the ratio of the RRab is higher than that of RRc (but still within 1$\sigma$), following the same trend displayed by the ratios over $Ampl(I)$ displayed in Fig.~\ref{fig:amplratio_ogle}. We point out that, for the $z_D$ ratios, we used a match between \gaia DR3 and M5 photometry from \citet{vivas2017}, although the overlap of DECam Bulge sample with \gaia DR3 is larger. This is because Bulge RRLs are very faint in \gaia ($G$>18) and could be either affected by blending or by second-order extinction effects. In fact, we found that $\dfrac{Ampl(z_D)}{Ampl(G)}$ for Bulge RRLs is significantly higher (0.763$\pm$0.110) than that of M5 (0.679$\pm$0.076). We propose three possible explanations:  1) an underestimated $Ampl(G)$ in Bulge RRLs, as expected from blended sources; 2) an underestimated $Ampl(G)$ in Bulge RRLs due to foreground reddening. The latter hypothesis was tested, under suggestion of the referee, by checking the trend of $Ampl(G)$ with the $Bp-Rp$ color, revealing a decreasing trend for the high-amplitude RRLs. We ruled out the hypothesis that this difference is due to an insufficient number of $G$-band observations in the Bulge, since neither $Ampl(G)$ nor $\dfrac{Ampl(z_D)}{Ampl(G)}$ seem to depend on that.

\begin{figure}[!htbp]
\centering
\includegraphics[width=9cm]{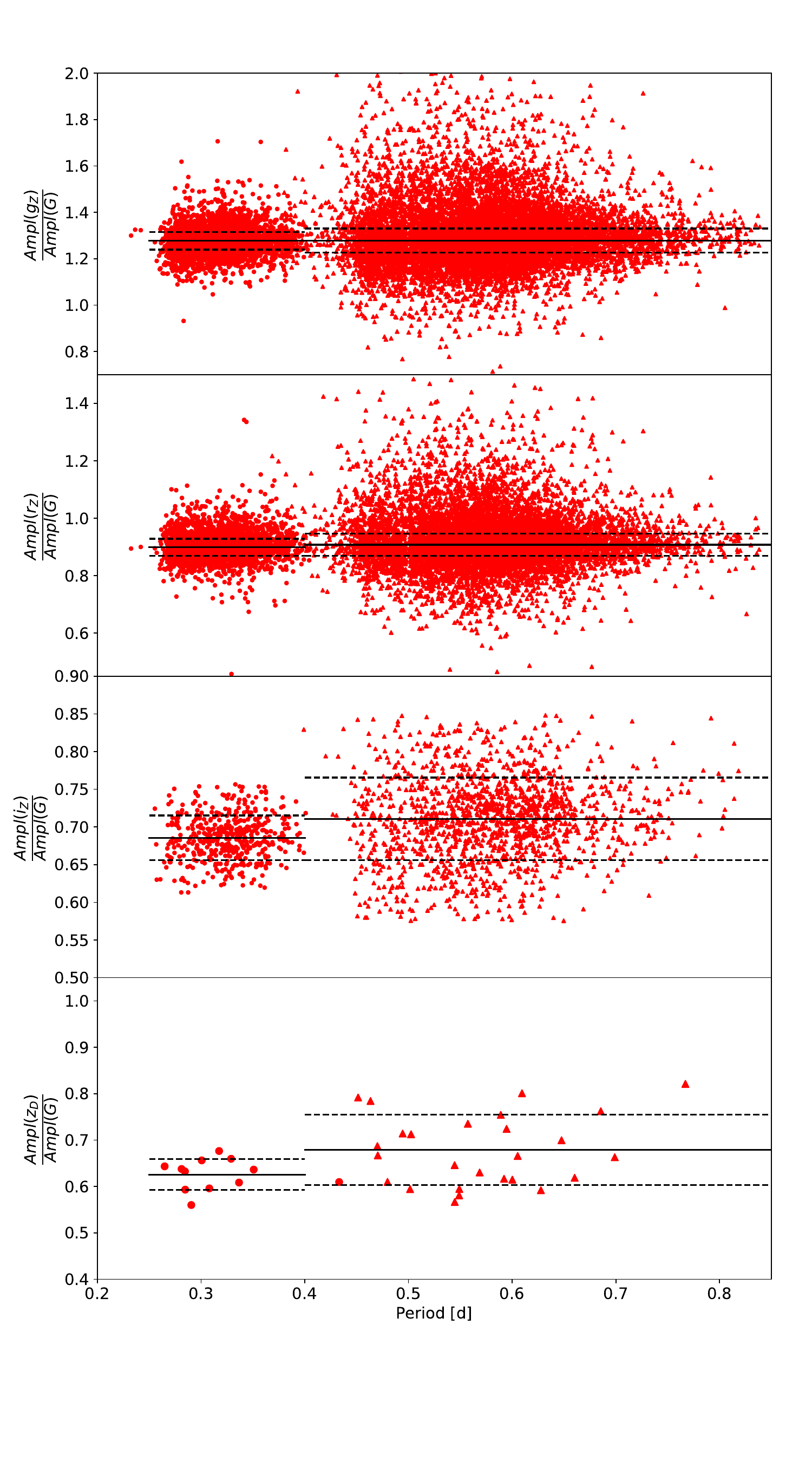}
\caption{From top to bottom: Amplitude ratios between $Ampl(g_Z)$, $Ampl(r_Z)$, $Ampl(i_Z)$ and $Ampl(z_D)$ over $Ampl(G)$.}
\label{fig:amplratio_ogle}
\end{figure}

\subsection{OGLE $I$-over-$V$ ratios}
Finally, we provide the amplitude ratios between the Johnson-Kron-Cousins $V$ and $I$ bands using OGLE data, in order to dispose of ratios allowing us to convert $Ampl$ from the historical standard Johnson $V$ to the LSST photometric systems. Unfortunately, the OGLE consortium does not provide $Ampl(V)$, so we fitted the $V$-band light curves available and calculated the amplitudes. For this task, we adopted the same Fourier series fitting method with an automatic selection of the degree, that we used for the ZTF data. To only keep the best estimates, we rejected all those light curves with too large average errors ($>$0.15 mag), those with less than 30 phase points and those with gaps in phase larger than 0.08, to avoid non-realistic ripples and bumps in the Fourier series fits.

We found a very notable and completely new behavior different behavior between the LMC/SMC and Bulge (BUL) sample of OGLE RRLs. In fact, while the LMC and SMC ratios display a flat behavior, the BUL sample clearly show that the ratios are strictly increasing with period, both for RRc and RRab. By leaving the slope free to change in the LMC sample, the maximum slope is around one quarter than that of the BUL sample. The SMC sample is too small to allow a similar, quantitatively meaningful investigation. Since the analysis of the data was performed homogeneously on the different datasets, we have to assume that this trend is an intrinsic feature of BUL RRLs. It is not the aim of this paper to delve deep into this feature, but we advance two preliminary hypotheses. The first is that the trend in the Bulge is due to blending. However, preliminary considerations tell us that, in the Bulge, redder wavelengths ($I$) should be more affected than blue wavelengths ($V$). This means that one should observe a decreasing trend because longer-period RRLs have, generally, redder colors instead of an increasing one. The other possibility is that the higher metallicity of Bulge RRLs \citep{walker1991} might be playing a role but to validate this hypothesis would need an extensive investigation, both with Globular Cluster data and pulsation models, that deserves a paper on its own.

\begin{figure}[!htbp]
\centering
\includegraphics[width=9cm]{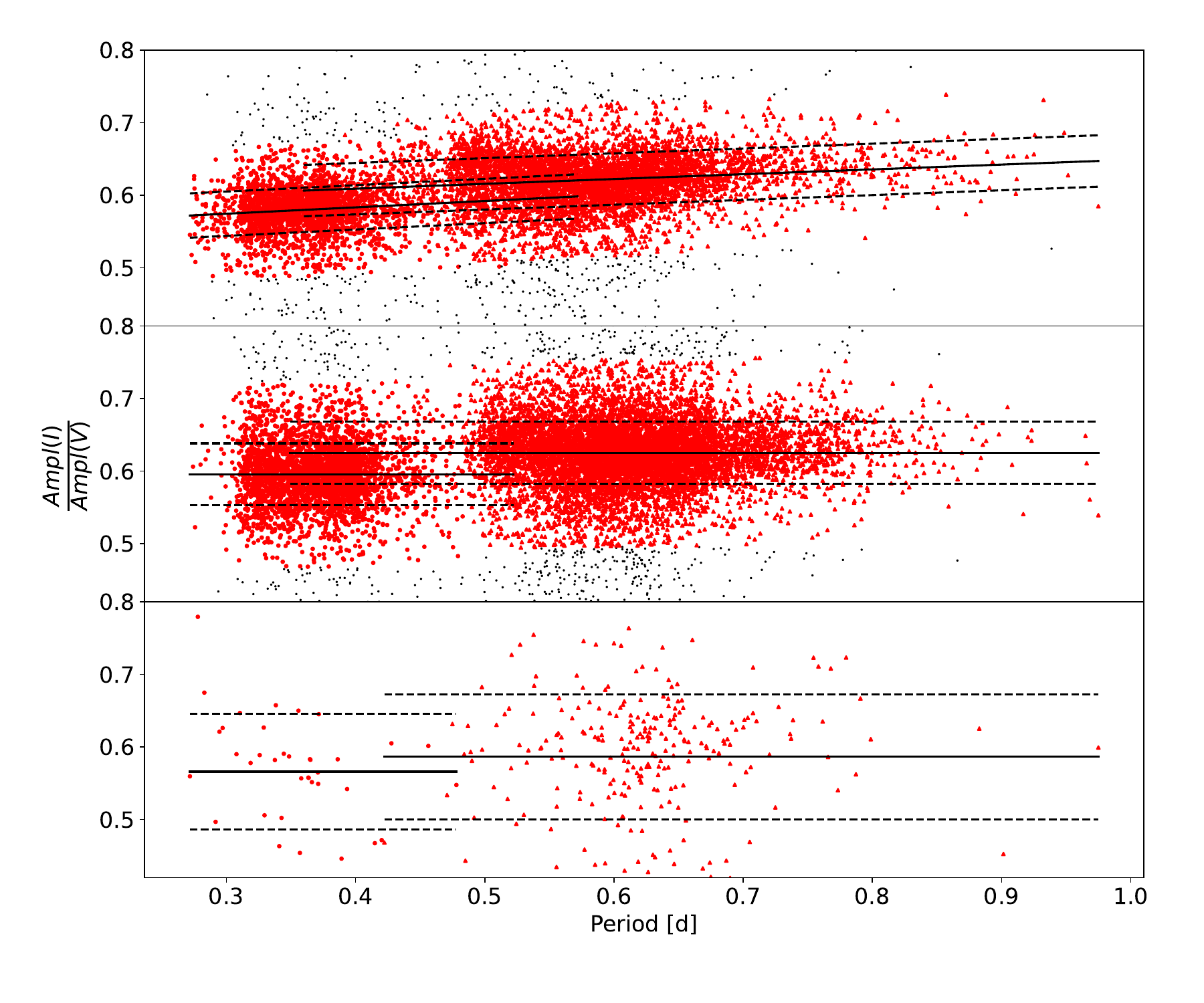}
\caption{Top: $\dfrac{Ampl(I)}{Ampl(V)}$ ratios for the RRLs of the Bulge. The red circles/triangles display the RRc/RRab that passed a recursive sigma clipping (cut at 2$\sigma$) for the linear fit. The black solid lines display the linear fits of the two samples. The black dashed lines display the standard deviation of the linear fit. Middle: same as top, but for the LMC sample; in this case, the fit is a constant. Bottom: same as middle but for the SMC sample.}
\label{fig:amplratio_ogle_ogle}
\end{figure}

\begin{table*}
\caption{Amplitude ratios between the LSST (suffix $L$), DECam (suffix $D$), ZTF (suffix $Z$), \gaia ($G$) and OGLE Johnson-Kron-Cousins ($VI$).}
\scriptsize
\begin{tabular}{llccc|ccc|ccc|ccc|ccc|ccc}\\
\hline
\hline

filter && \multicolumn{3}{c|}{$\dfrac{Ampl(x_L)}{Ampl(g_L)}(all)$} & \multicolumn{3}{c|}{$\sigma\dfrac{Ampl(x_L)}{Ampl(g_L)}(all)$} & \multicolumn{3}{c|}{$\dfrac{Ampl(x_L)}{Ampl(g_L)}(c)$} & \multicolumn{3}{c|}{$\sigma\dfrac{Ampl(x_L)}{Ampl(g_L)}(c)$} & \multicolumn{3}{c|}{$\dfrac{Ampl(x_L)}{Ampl(g_L)}(ab)$} & \multicolumn{3}{c}{$\sigma\dfrac{Ampl(x_L)}{Ampl(g_L)}(ab)$} \\
& Z & 0.0003 & 0.001 & 0.008 & 0.0003 & 0.001 & 0.008 & 0.0003 & 0.001 & 0.008 & 0.0003 & 0.001 & 0.008 & 0.0003 & 0.001 & 0.008 & 0.0003 & 0.001 & 0.008 \\
\hline
$u$ && 0.927 & 0.956 & 1.105 & 0.047 & 0.042 & 0.085 & 0.877 & 0.903 & 0.999 & 0.004 & 0.004 & 0.011 & 0.944 & 0.974 & 1.131 & 0.043 & 0.033 & 0.075 \\
$r$ && 0.747 & 0.746 & 0.725 & 0.014 & 0.011 & 0.024 & 0.754 & 0.752 & 0.734 & 0.010 & 0.011 & 0.008 & 0.745 & 0.743 & 0.723 & 0.014 & 0.011 & 0.026 \\
$i$ && 0.594 & 0.595 & 0.581 & 0.019 & 0.017 & 0.034 & 0.599 & 0.600 & 0.582 & 0.018 & 0.018 & 0.015 & 0.592 & 0.593 & 0.580 & 0.020 & 0.016 & 0.037 \\
$z$ && 0.519 & 0.523 & 0.512 & 0.025 & 0.022 & 0.041 & 0.526 & 0.529 & 0.513 & 0.019 & 0.020 & 0.016 & 0.516 & 0.521 & 0.511 & 0.027 & 0.022 & 0.045 \\
$y$ && 0.512 & 0.518 & 0.502 & 0.033 & 0.028 & 0.047 & 0.530 & 0.533 & 0.520 & 0.014 & 0.016 & 0.012 & 0.505 & 0.512 & 0.498 & 0.036 & 0.029 & 0.051 \\
\hline\\

 && \multicolumn{3}{c|}{$r(x)_{L/D}(all)$} & \multicolumn{3}{c|}{$\sigma r(x)_{L/D}(all)$} & \multicolumn{3}{c|}{$r(x)_{L/D}(c)$} & \multicolumn{3}{c|}{$\sigma r(x)_{L/D}(c)$} & \multicolumn{3}{c|}{$r(x)_{L/D}(ab)$} & \multicolumn{3}{c}{$\sigma r(x)_{L/D}(ab)$} \\
& Z & 0.0003 & 0.001 & 0.008 & 0.0003 & 0.001 & 0.008 & 0.0003 & 0.001 & 0.008 & 0.0003 & 0.001 & 0.008 & 0.0003 & 0.001 & 0.008 & 0.0003 & 0.001 & 0.008 \\
\hline
$u$ && 0.848 & 0.853 & 0.868 & 0.015 & 0.018 & 0.011 & 0.834 & 0.838 & 0.850 & 0.003 & 0.002 & 0.003 & 0.853 & 0.858 & 0.872 & 0.014 & 0.019 & 0.008 \\
$g$ && 1.010 & 1.010 & 1.012 & 0.001 & 0.001 & 0.002 & 1.009 & 1.009 & 1.010 & 0.000 & 0.000 & 0.000 & 1.010 & 1.010 & 1.013 & 0.001 & 0.001 & 0.002 \\
$r$ && 1.045 & 1.045 & 1.045 & 0.003 & 0.002 & 0.005 & 1.045 & 1.045 & 1.047 & 0.003 & 0.003 & 0.003 & 1.045 & 1.045 & 1.045 & 0.003 & 0.002 & 0.006 \\
$i$ && 1.054 & 1.053 & 1.050 & 0.004 & 0.005 & 0.009 & 1.056 & 1.055 & 1.055 & 0.005 & 0.005 & 0.004 & 1.053 & 1.052 & 1.048 & 0.004 & 0.004 & 0.009 \\
$z$ && 0.990 & 0.988 & 0.995 & 0.016 & 0.013 & 0.019 & 0.976 & 0.976 & 0.973 & 0.008 & 0.007 & 0.006 & 0.995 & 0.992 & 1.001 & 0.015 & 0.013 & 0.017 \\
$y$ && 1.022 & 1.021 & 1.019 & 0.002 & 0.003 & 0.005 & 1.021 & 1.020 & 1.019 & 0.002 & 0.002 & 0.001 & 1.022 & 1.021 & 1.019 & 0.002 & 0.004 & 0.006 \\
\hline\\

&& \multicolumn{3}{c}{$r(x)_{Z/D}(all)$} & \multicolumn{3}{c}{$\sigma r(x)_{Z/D}(all)$} & \multicolumn{3}{c}{$r(x)_{Z/D}(c)$} & \multicolumn{3}{c}{$\sigma r(x)_{Z/D}(c)$} & \multicolumn{3}{c}{$r(x)_{Z/D}(ab)$} & \multicolumn{3}{c}{$\sigma r(x)_{Z/D}(ab)$} \\
\hline
$g$ &&& 0.984 &&& 0.103 &&& 1.004 &&& 0.052 &&& 0.978 &&& 0.115 &\\
$r$ &&& 1.001 &&& 0.099 &&& 1.046 &&& 0.032 &&& 0.992 &&& 0.105 &\\
\hline\\

&& \multicolumn{3}{c}{$r(x)_{L/Z}(all)$} & \multicolumn{3}{c}{$\sigma r(x)_{L/Z}(all)$} & \multicolumn{3}{c}{$r(x)_{L/Z}(c)$}& \multicolumn{3}{c}{$\sigma r(x)_{L/Z}(c)$} & \multicolumn{3}{c}{$r(x)_{L/Z}(ab)$} & \multicolumn{3}{c}{$\sigma r(x)_{L/Z}(ab)$} \\
\hline
$g$ &&& 1.026 &&&0.107 &&&1.005 &&&0.103 &&&1.033 &&&0.121 &\\
$r$ &&& 1.044 &&&0.103 &&&0.999 &&&0.095 &&&1.053 &&&0.111 &\\
\hline\\

&& \multicolumn{3}{c}{$r(x)_{Z/S}(all)$} & \multicolumn{3}{c}{$\sigma r(x)_{Z/S}(all)$} & \multicolumn{3}{c}{$r(x)_{Z/S}(c)$} & \multicolumn{3}{c}{$\sigma r(x)_{Z/S}(c)$} & \multicolumn{3}{c}{$r(x)_{Z/S}(ab)$} & \multicolumn{3}{c}{$\sigma r(x)_{Z/S}(ab)$} \\
\hline
$g$ &&& 0.993 &&& 0.088 &&& 0.967 &&& 0.038 &&& 0.998 &&& 0.094 &\\
$r$ &&& 0.984 &&& 0.137 &&& 0.945 &&& 0.051 &&& 0.986 &&& 0.139 &\\
\hline\\

&& \multicolumn{3}{c}{$r(x)_{L/S}(tot)$} & \multicolumn{3}{c}{$\sigma r(x)_{L/S}(tot)$} & \multicolumn{3}{c}{$r(x)_{L/S}(c)$} & \multicolumn{3}{c}{$\sigma r(x)_{L/S}(c)$} & \multicolumn{3}{c}{$r(x)_{L/S}(ab)$} & \multicolumn{3}{c}{$\sigma r(x)_{L/S}(ab)$} \\
\hline
$g$ &&& 1.019 &&& 0.140 &&& 0.972 &&& 0.063 &&& 1.032 &&& 0.155 &\\
$r$ &&& 1.027 &&& 0.175 &&& 0.944 &&& 0.059 &&& 1.038 &&& 0.183 &\\
\hline\\

&& \multicolumn{3}{c}{$\dfrac{Ampl(x)}{Ampl(I)}(all)$} & \multicolumn{3}{c}{$\sigma \dfrac{Ampl(x)}{Ampl(I)}(all)$} & \multicolumn{3}{c}{$\dfrac{Ampl(x)}{Ampl(I)}(c)$} & \multicolumn{3}{c}{$\sigma \dfrac{Ampl(x)}{Ampl(I)}(c)$} & \multicolumn{3}{c}{$\dfrac{Ampl(x)}{Ampl(I)}(ab)$} & \multicolumn{3}{c}{$\sigma \dfrac{Ampl(x)}{Ampl(I)}(ab)$} \\
\hline
$g_Z$ &&& \ldots &&& \ldots&&& 1.849 &&& 0.051 &&& 1.734 &&& 0.062 &\\
$r_Z$ &&& \ldots &&& \ldots&&& 1.288 &&& 0.042 &&& 1.245 &&& 0.045 &\\
$i_Z$ &&& \ldots &&& \ldots&&& 0.965 &&& 0.037 &&& 0.968 &&& 0.032 &\\
$z_D$ &&& \ldots &&& \ldots&&& 0.923 &&& 0.033 &&& 0.971 &&& 0.054 &\\
\hline\\

&& \multicolumn{3}{c}{$\dfrac{Ampl(x)}{Ampl(G)}(all)$} & \multicolumn{3}{c}{$\sigma \dfrac{Ampl(x)}{Ampl(G)}(all)$} & \multicolumn{3}{c}{$\dfrac{Ampl(x)}{Ampl(G)}(c)$} & \multicolumn{3}{c}{$\sigma \dfrac{Ampl(x)}{Ampl(G)}(c)$} & \multicolumn{3}{c}{$\dfrac{Ampl(x)}{Ampl(G)}(ab)$} & \multicolumn{3}{c}{$\sigma \dfrac{Ampl(x)}{Ampl(G)}(ab)$} \\
\hline
$g_Z$ &&&  1.278 &&& 0.048 &&& 1.277 &&& 0.038 &&& 1.278 &&& 0.052 &\\
$r_Z$ &&&  0.905 &&& 0.034 &&& 0.899 &&& 0.029 &&& 0.908 &&& 0.039 &\\
$i_Z$ &&& \ldots &&& \ldots&&& 0.686 &&& 0.030 &&& 0.711 &&& 0.055 &\\
$z_D$ &&& \ldots &&& \ldots&&& 0.626 &&& 0.033 &&& 0.679 &&& 0.076 &\\
\hline\\

BUL/LMC && \multicolumn{3}{c}{$\dfrac{Ampl(I)}{Ampl(V)}(all)$} & \multicolumn{3}{c}{$\sigma \dfrac{Ampl(I)}{Ampl(V)}(all)$} & \multicolumn{3}{c}{$\dfrac{Ampl(I)}{Ampl(V)}(c)$} & \multicolumn{3}{c}{$\sigma \dfrac{Ampl(I)}{Ampl(V)}(c)$} & \multicolumn{3}{c}{$\dfrac{Ampl(I)}{Ampl(V)}(ab)$} & \multicolumn{3}{c}{$\sigma \dfrac{Ampl(I)}{Ampl(V)}(ab)$} \\
\hline
BUL &&& \ldots &&& \ldots && \multicolumn{3}{c|}{$0.557 + 0.078 \cdot P$} && 0.031 && \multicolumn{3}{c|}{$0.589 + 0.060 \cdot P$} && 0.035 &\\
LMC &&& \ldots &&& \ldots &&& 0.596 &&& 0.042 &&& 0.625 &&& 0.043 &\\
SMC &&& \ldots &&& \ldots &&& 0.596 &&& 0.080 &&& 0.587 &&& 0.086 &\\
\hline
\end{tabular}\\
\tablefoot{All amplitude ratios are averages and derived from empirical data, except the $\dfrac{Ampl(x_L)}{Ampl(g_L)}$ and $r(x)_{L/D}$, that are derived from pulsation models and, for this reason, are separated by three metallicity values (Z=0.0003, 0.0001, 0.008).}
\label{tab:amplratio}
\end{table*}

\end{appendix}

\end{document}